\documentclass{SCGE}
\usepackage{epsfig}

\begin{document}

%%%%%%新版式要加上这组
\begin{picture}(0,0){\rm
\put(0,-20){\makebox[160truemm][l]{\bf {\sanhao\raisebox{2pt}{.}}
Article  {\sanhao\raisebox{1.5pt}{.}}}}}
\put(0,-34){\jiuwuhao {\textcolor[rgb]{0.5,0.5,0.5}{\sf Progress of Projects Supported by NSFC
}}}%%(11月注释：调\textcolor[rgb]{x,x,x} 中的数字x越大越灰)
\end{picture}

\def\bm{\boldsymbol}

\def\dl{\displaystyle}
\def\du{\end{document}}
\def\d{{\rm d}}
\def\e{{\rm e}}
\def\i{{\rm i}}

\def\pi{{\uppi}}

% The author doesn't need fill in it.
\Year{2015} %
\Month{October} %
\Vol{xx} %  卷号
\No{x} %  期号
\BeginPage{1} % 起页码
\AuthorMark{{\rm A.G. Xu}, et al.}  %(11月注释：页眉上的作者)
\AuthorMarkCite{{\rm A.G. Xu, G.C. ZHANG, Y.Y. YING, C. WANG
}. } %(11月注释：citation中的作者)
\DOI{10.1007/s11433-014-5625-8} % The author doesn't need fill in it.
\ArtNo{000000}

% \title[short text for running head]{full title}{comments for title}
\title
%[Complex fields in heterogeneous materials under shock]
{Complex fields in heterogeneous materials under shock: modeling, simulation and analysis}

\author[1*,2,3]{XU Aiguo}{}
\author[1,3,4]{ZHANG Guangcai}{}
\author[1]{YING Yangjun}{}
\author[4]{WANG Cheng}{}
\address[{\rm1}]{Laboratory of Computational Physics, \\Institute of Applied
Physics and Computational Mathematics, P. O. Box 8009-26, Beijing 100088,
P.R.China}
\address[{\rm2}]{Center for Applied Physics and Technology, MOE Key Center for High Energy Density Physics Simulations, \\
College of Engineering, Peking University, Beijing 100871, P.R.China}
\address[{\rm3}]{State Key Laboratory of Theoretical Physics, Institute of Theoretical Physics, \\
Chinese Academy of Sciences,Beijing 100190, P.R.China }
\address[{\rm4}]{State Key Laboratory of Explosion Science and Technology, \\ Beijing Institute of Technology, Beijing 100081, P.R.China}

\maketitle \vspace{-3.5mm}{\footnotesize\begin{center} Received October xx, 2015; accepted November xx, 2015
\end{center}}\vspace*{-5mm}

%     Abstract is required.
\begin{center}
\rule{16.5cm}{0.4pt}
\parbox{16.5cm}
{\begin{abstract}

In this mini-review we summarize the progress of modeling, simulation and analysis of shock responses of heterogeneous materials in our group in recent years. The basic methodology is as below. We first decompose the problem into different scales. Construct/Choose a model according to the scale and main mechanisms working at that scale. Perform numerical simulations using the relatively mature schemes. The physical information is transferred between neighboring scales in such a way: The statistical information of results in smaller scale contributes to establishing the constitutive equation in larger one. Except for the microscopic Molecular Dynamics (MD) model, both the mesoscopic and macroscopic models can be further classified into two categories, solidic and fluidic models, respectively. The basic ideas and key techniques of the MD, material point method and discrete Boltzmann method are briefly reviewed. Among various schemes used in analyzing the complex fields and structures, the morphological analysis and the home-built software, GISO, are briefly introduced. New observations are summarized for scales from the larger to the smaller.

\end{abstract}}
\end{center}\vspace*{-0.6cm}

\begin{center}
\parbox{16.5cm}
{\bf\jiuhao complex fields,  heterogeneous material, molecular dynamics, material point method, discrete Boltzmann model.}%关键词
\end{center}

\begin{center}
{\PACS{\rm 05.40.-a, 62.50.Ef, 81.05.Rm, 81.05.Zx,62.20.mm}}%分类号
\CITA    %%(11月注释：Citation内容自动生成)
%\Cit{~~~???, et al. ???. Sci China-Phys Mech Astron, 2014, 57: 1--6, doi:}%%(11 月注释：Citation 内容需手动填写)
\end{center}

\textwidth=178truemm \textheight=236truemm%%%%%% 新版式要加上

%%%%%%%%%%%%%%%%%%%%%%%%%%%%%%%%%%%%%%%%%%%%%%%%%%%%%%%%%%%%
\wuhao\vspace*{1.5mm}

\begin{multicols}{2}

%%%%%%%%%%%%%%%%%%%%%%%%%%%%%%%%%%%%%%%%%%%%%%%%%%%%%%%%%%%%
%% Text of article.
%%%%%%%%%%%%%%%%%%%%%%%%%%%%%%%%%%%%%%%%%%%%%%%%%%%%%%%%%%%%
%    Section headings
\renewcommand{\baselinestretch}{1.08} \baselineskip 12.2pt\parindent=10.8pt

\renewcommand{\thefootnote}

\section{Introduction}
It has long been recognized that the properties of materials are not uniquely determined by their average chemical composition but also, to a large extent, influenced by their structures.
A heterogeneous material is a substance which is non-uniform in composition
or character. The heterogeneous materials are ubiquitous in natu-
%%%%%%%%%%%%%%%%%%%%%%%%%%%%%%%%
\linebreak
\vspace*{-4mm}

\noindent\rule{2.5cm}{0.4pt}\\[0.1mm]{\qihao *Corresponding author (email:
Xu\_Aiguo$@$iapcm.ac.cn)}% 手

\noindent
%%%%%%%%%%%%%%%%%%%%%%%%%%%%%%%%
ral and
industrial fields. In fact, nearly all the materials in
macroscopic scales
used in our daily life are heterogeneous.
When a heterogeneous material is
shocked, the morphology and distribution of the mesoscopic structures will
largely influence the fields of stress and temperature inside the
material, and consequently influence the mechanical properties of the
material, influence the processes of failure nucleation and phase transition, etc. The
behaviors of mesoscopic structures and the resulted  non-equilibrium
effects have been attracting more attention with time\cite{Heterogeneous-book1,Heterogeneous-book2}.
Such problems have been becoming an essential inter-discipline subject in the fields of modern mechanics, physics and material science, etc.\cite{Zhu-Review2010}.

From the experimental side, since the shocking process is
very quick, it is generally difficult to measure the details of the series actions occurred in the materials.
From the theoretical side, since related to strong nonlinearity and complex fields, a pure theoretical investigation on such a system is nearly impossible.
Therefore, numerical simulation plays a non-sustitutable
role in better understanding shocked heterogeneous materials.

Such problems generally show effects or influence on our life in macroscopic scale, but are originated from the microscopic scale. The scale for the series of actions spans  from $10^{-10}$m to $1$m, i.e., about 10 orders in magnitude.
How to model and simulate behaviors in such a wide scale has plagued the scientific community for a long time.
Currently, the studies under the terminology, multi-scale modeling and simulation, can be roughly classified into two categories. In the first category, the complex problem is decomposed into various scales. One chooses  the theory and method according to the specific scale and the dominant  mechanism working in that scale. The statistical results of simulations in the smaller scale contribute to formulate the constitutive equation used in the larger scale. In the second category, the mainly concerned problems are around how to bridge the neighboring scales in simulations. In this mini-review we focus mainly on studies in the first category.
Even in the first category there are too many problems to be studied in one decade. Therefore, under the topic of multi-scale modeling and simulations of heterogeneous materials, what we did in the past years are scattered and can only show a few aspects of the field.

In terms of the shear behavior heterogeneous materials can be classified into two kinds. The first kind is referred to as solid, and the second is referred to as fluid. A typical differences between solid and fluid
is that the solid show anisotropic behaviors in both the microscopic and macroscopic scales, while the fluid shows isotropic behavior in microscopic scale but anisotropic behavior in macroscopic scale. For solid heterogenous materials, what we concerned in the past years are mainly focused on the shocking responses of porous materials and microscopic mechanical behaviors of metal under various loadings. These studies present preliminary and fundamental references for modeling the elasto-plasticity, damage and  fracture of solid materials under shock loading and unloading conditions. For fluid heterogeneous materials, we mainly studied the mechanical behavior and non-equilibrium phenomena in the kinetic transportation, phase transition, chemical coupling, etc. These observations present fundamental insights for modeling and simulation of complex fluids.

A porous material is a material
containing voids or tunnels of different shapes and sizes. Such materials are
commonly found in natural and industrial materials. Examples are referred to  bricks, wood, carbon, ceramics, foams, explosives and metals, etc. To have an effective application, their  mechanical and thermodynamical behaviors must be understood in relating to their
mesoscopic structures.

There contains a large quantity of micro-structures like dislocations, grain boundaries, voids, cavities, second phase grains in macroscopic metal materials. These structures influence the strength of the materials. When model these materials, the morphology and evolution process of these micro-structures should be taken into account. In the past years, we investigated such systems via the MD simulation.

In fluid heterogeneous materials, around the shock-induced structures, material interfaces and structures induced by their stabilities, the thermodynamic non-equilibrium effects are pronounced. The traditional models based on Navier-Stokes equations are critical in treating with such problems. Under such cases, a kinetic model based on the Boltzmann equation, the Discrete Boltzmann Model (DBM)\cite{Review-FoP2012,Review-PiP2014,Review-APC2015,Succi-DBM2015,SoftMatter2015} , can be used to adaptively capture the various non-equilibrium behaviors.

When phase transition and/or chemical reaction exist, the creation of new phase or matter and their evolution result in heat creation/absorption and more complicated kinetic transportation processes. The DBM is an effective mesoscopic approach to access such a system.

The rest of the paper is organized as below. In Sect. 2 various models and simulation tools are introduced. Section 3 is for the analysis schemes for the complex fields and structures. The numerical experiments and observations are presented in Sect.4. Section 5 summarizes the paper and gives perspectives.

\section{Models and simulation tools}

Generally speaking, we can model the system in microscopic, mesoscopic and macroscopic scales. Since the matter can be divided infinitely, the delimitation of the scales is relative. In this review, the microscopic description is referred to as the description based on Molecular Dynamics (MD). The macroscopic scale is referred to the scale of the whole system or a scale which is comparable with the system dimension. Thus, the wide range of scales in between the microscopic and macroscopic scales are referred to as mesoscopic scale. It is clear that the so-called mesoscopic scale is generally not referred to a specific scale, but a scale series.

The macroscopic model is generally described by a set of partial differential equations corresponding to  the fundamental conservation laws. Because it uses the smallest number of the mechanical quantities, the macroscopic model is the simplest and frequently used in many engineering applications. It has been well-known that the macroscopic model is not sufficient to describe the complex behaviors occurring in heterogeneous materials under shock. Such behaviors are generally originated from the molecular scale and make effects in macroscopic via a series of complex interactions between various structures. Intuitively, the complete understanding of the whole story resorts to the MD simulation. But, practically, it is far from possible to use MD to simulate behaviors in macroscopic scale. Under such cases, the mesoscopic modeling technology which connecting the macroscopic and the most necessary microscopic behaviors is needed. Compared with the MD results, the mesoscopic modeling is some kind of coarse-grained  description of the microscopic details via some slow or conservative variables.

\subsection{Microscopic MD model}

Molecular dynamics model describes physical movements of particles (molecules or atoms) in the context of $N$-body interaction, where $N$ is the particle number in the system. In the most common MD simulations, the trajectories of particles are tracked via numerically solving the Newton's equations of motion for a system of interacting
particles, where forces between the particles are determined from the molecular mechanics force fields (or interatomic potentials). The MD method was originally proposed by theoretical physicists in the late 1950s\cite{MD1,MD2}, but now is extensively used in chemical physics, materials science and the modeling of bio-molecules, etc.
Due to the vast number of particles in the systems, the MD method resorts to numerical simulations.

The first important step in the MD simulation is to establish the inter-particle potential.  In principle, a molecule is influenced by all other surrounding molecules.
Fortunately, the strength of the interaction decreases quickly with the distance. Therefore, the second important step in the MD simulation is to  truncate of the inter-particle potential. The smaller the truncation radius, the less the computational cost. The validity of truncation position is determined by that the simulation results of known material parameters are correct.
To compute the force acting on a molecule, one has to search all the surrounding molecules located within the truncation radius.
Because the one-by-one searching is not affordable for a system with more than 1 million atoms, the third important step is to index the molecules in terms of a link-list corresponding to background grid mesh.

Based on the inter-particle potential we can obtain the summation of the external forces acting on any molecule and consequently the acceleration of it. Thus, from the current position and velocity, we can obtain its position and velocity at the next time step. The positions and velocities of all the molecules can be updated in the same way. Then the summation of external forces acting on any molecule is updated. via such iteration steps, we can track all the molecules in the system. The physical variables like the energy, temperature, pressure,density, flow velocity, etc, can be obtained from the MD data via appropriate statistical calculations. The micro-structures can be identified, described and tracked via some data post-treatment algorithms.
In our studies, the numerical MD simulations are performed using the
well-known LAMMPS software package. The interatomic interaction in each material is described by an embedded atom method (EAM) potential \cite{Daw,Harrison}. The dilative strain is applied uniformly through re-scaling the coordinates as in the Parrinello-Rahman approach. The data are analyzed by using our home-built softwares.

 Now, we can obtain a preliminary estimation on the scale of the material that the MD can be used to simulate. They molecule numbers used in current MD simulations are generally less than $10^{7}$. That is to say, molecule number in one dimension is only in the order $10^{2}$. For a general solid material, the distance in between two neighboring particles is about $10^{-10}$m. It is clear that, for a practical MD simulation in nowadays, the largest scale in one dimensional is smaller that $0.1 \mu$m. Since the time step is generally in the order of femto-second, i.e. $10^{-15}$s, the whole duration being simulated is generally in the order of pico-second, i.e., $10^{-12}$s.
At the same time, from the theoretical point of view, a long MD simulation is mathematically ill-conditioned. It generates cumulative errors in numerical integration which can be minimized via selecting proper algorithms and parameters, but can not eliminated entirely.

There are many phenomenological physical models for a heterogeneous solid materials.
In this review the material is assumed to follow an associative von Mises
plasticity model with linear kinematic and isotropic hardening\cite{CModel}.
The Material Point Method(MPM)\cite{ZhangX2006,MaS2009,ZhangXiong2009,MaPhDThesis,H1964,MPM1,MPM2,MPM3,MPM4,MPM5,JPCM2007,CTP2008,JPD2008}
is used to simulate the mesocopic and macroscopic behaviors in the shocked porous materials.

\subsection{Solid model and MPM}

\subsubsection{Physical model}

If we introduce a linear isotropic elastic relation and assume that the volumetric plastic
strain is zero, the deviatoric stress $\mathbf{s}$ or strain $\mathbf{e}$ can be decoupled from volumetric one, $-P\mathbf{I}$ or $\theta \mathbf{I}/3$, where $P$ and $\theta$ are scalars, $\mathbf{s}$ and $\mathbf{e}$ are tensors. The stress and strain tensors, $\boldsymbol{\sigma }$ and $\boldsymbol{\varepsilon }$, can be written as
\begin{eqnarray}
\boldsymbol{\sigma } &=&\mathbf{s}-P\mathbf{I},P=-\frac{1}{3}\verb|Tr|(%
\boldsymbol{\sigma })\mathtt{,}  \label{PMe1} \\
\boldsymbol{\varepsilon } &=&\mathbf{e}+\frac{1}{3}\theta \mathbf{I},\theta =%
\frac{1}{3}\verb|Tr|(\boldsymbol{\varepsilon })\mathtt{,}  \label{PMe2}
\end{eqnarray}%
Generally, the strain $\mathbf{e}$ can be
decomposed as $\mathbf{e}=\mathbf{e}^{e}+\mathbf{e}^{p}$, where $\mathbf{e}%
^{e}$ and $\mathbf{e}^{p}$ are the traceless elastic and plastic components,
respectively. Until the von Mises yield criterion,
\begin{equation}
\sqrt{\frac{3}{2}}\left\Vert \mathbf{s}\right\Vert =\sigma _{Y}\mathtt{,}
\label{PM1}
\end{equation}%
is reached, the material shows a linear elastic response, where $\sigma _{Y}$ is the plastic yield stress increasing linearly with the second invariant of the plastic strain tensor $\mathbf{e}^{p}$, i.e.,
\begin{equation}
\sigma _{Y}=\sigma _{Y0}+E_{\tan }\left\Vert \mathbf{e}^{p}\right\Vert
\mathtt{,}  \label{PM4}
\end{equation}%
where $\sigma _{Y0}$ is the initial yield stress and $E_{\tan }$ is the tangential module. The deviatoric stress $\mathbf{s}$ is calculated by $\mathbf{s}=[E/(1+\nu)]\mathbf{e}^{e}$,
where $E$ is the Yang's module and $\nu $ the Poisson's ratio.
 The pressure $P$ is
calculated by using the following Mie-Gr\"{u}neissen state of equation:
\begin{equation}
P-P_{H}=\frac{\gamma (V)}{V}[E-E_{H}(V_{H})] \text{,}  \label{eq-eos}
\end{equation}%
where $P_{H}$, $V_{H}$ and $E_{H}$ are pressure, specific volume and energy on the
Rankine-Hugoniot curve, respectively. The relation between $P_{H}$ and $V_{H} $ can be estimated by experiment. It can be written as
\begin{equation}
P_{H}=\left\{
\begin{array}{ll}
\frac{\rho _{0}c_{0}^{2}(1-\frac{V_{H}}{V_{0}})}{(\lambda -1)^{2}(\frac{%
\lambda }{\lambda -1}\times \frac{V_{H}}{V_{0}}-1)^{2}}, & V_{H}\leq V_{0}
\\
\rho _{0}c_{0}^{2}(\frac{V_{H}}{V_{0}}-1), & V_{H}>V_{0}%
\end{array}%
\right.
\end{equation}
We assume that the
initial material density and sound speed are $\rho _{0}$ and $c_{0}$,
respectively. The shock speed $U_{s}$ and the particle speed $U_{p}$ after
the shock front follows a linear relation,
\[
U_{s}=c_{0}+\lambda U_{p}\text{,}
\]
 where $\lambda $ is a characteristic coefficient of material.
Both the shock compression and the plastic work $W_{p}$ result in increasing of temperature. The temperature increase from shock compression is calculated by
\begin{equation}
\frac{\mathrm{d}T_{H}}{\mathrm{d}V_{H}}=\frac{c_{0}^{2}\cdot \lambda
(V_{0}-V_{H})^{2}}{c_{v}\big[(\lambda -1)V_{0}-\lambda V_{H}\big]^{3}}-\frac{%
\gamma (V)}{V_{H}}T_{H}.  \label{eq-eos-temprshock}
\end{equation}%
where $c_{v}$ is the specific heat. Eq.(\ref{eq-eos-temprshock}) can be obtained from
the thermal equation and the Mie-Gr\"{u}neissen state of equation%
\cite{explosion}. The temperature increase due to plastic work is calculated by
\begin{equation}
\mathrm{d}T_{p}=\frac{\mathrm{d}W_{p}}{c_{v}}  \label{eq-eos-temprplastic}
\end{equation}%
Equations (\ref{eq-eos-temprshock}) and (\ref{eq-eos-temprplastic}) can be written in the form of increment.

\subsubsection{Material-Point Method}

The MPM is a particle method. It was originally introduced in fluid dynamics by Harlow, et al\cite{H1964} and extended to solid mechanics by Burgess, et al \cite{MPM1}, then developed by various groups, including ours\cite{MPM2,MPM3,MPM4,MPM5,JPCM2007,CTP2008,JPD2008}.

 The MPM discretizes the continuum bodies with $N_{p}$ material particles, where $p$ is the index of particle. Each material particle carries the information of mass $m_{p}$, density $\rho _{p}$, position $\mathbf{x}_{p}$, velocity $\mathbf{v}_{p}$, strain tensor $\boldsymbol{\varepsilon }_{p}$, stress tensor $\boldsymbol{\sigma }_{p}$ and all other internal state variables necessary for the constitutive model.
At each time step, the calculations can be classified into two parts: a Lagrangian part and a convective one. At first, the computational mesh deforms with the body. It is used to determine the strain increment, and the stresses in the sequel. Then, a new position of the computational
mesh is chosen. Particularly, it may be the previous one. The velocity field is mapped from the particles to the mesh nodes. Nodal velocities are determined by using the equivalence of momentum calculated for the particles and for the computational grid. The MPM not only takes advantages of both the Lagrangian and Eulerian algorithms but also avoid their drawbacks as well.

At each time step, the mass and velocities of the material particles are mapped onto the background computational mesh. The mapped momentum at node $i$ is obtained by
\[
m_{i}\mathbf{v}_{i}=\sum_{p}m_{p}\mathbf{v}_{p}N_{i}(\mathbf{x}_{p}) \text{,}
\]
where $N_{i}$ is the element shape function and the nodal mass $m_{i} $ reads
\[
m_{i}=\sum_{p}m_{p}N_{i}(\mathbf{x}_{p}) \text{.}
\]
Suppose that a computational mesh is constructed of eight-node cells for three-dimensional problems, then the shape function reads
\begin{equation}
N_{i}=\frac{1}{8}(1+\xi \xi _{i})(1+\eta \eta _{i})(1+\varsigma \varsigma
_{i})\mathtt{,}  \label{MPMe1}
\end{equation}%
where $\xi $,$\eta $,$\varsigma $ are the natural coordinates of the material particle in the cell along the $x$-, $y$-, and $z$-directions, respectively, $\xi _{i}$,$\eta _{i}$,$\varsigma _{i}$ take corresponding nodal values $\pm 1$. The mass of each particle is equal and fixed, so the mass conservation equation,
\[
\mathrm{d}\rho /\mathrm{d}t+\rho \nabla \cdot \mathbf{v}=0 \text{,}
 \]
 is automatically satisfied. The momentum equation reads,
\begin{equation}
\rho \mathrm{d}\mathbf{v/}\mathrm{d}t=\nabla \cdot \boldsymbol{\sigma }+\rho
\mathbf{b}\mathtt{,}  \label{MPMeq2}
\end{equation}%
where $\rho $ is the mass density, $\mathbf{v}$ the velocity, $\boldsymbol{\sigma }$ the stress tensor and $\mathbf{b}$ the body force. Equation (\ref{MPMeq2}) is solved on a finite element mesh in a lagrangian frame. Its weak form reads
\begin{equation}
\begin{array}{ll}
& \int_{\Omega }{\rho \delta \mathbf{v}\cdot \mathrm{d}\mathbf{v/}\mathrm{d}t%
\mathrm{d}\Omega }+\int_{\Omega }{\delta (\mathbf{v}\nabla )\cdot
\boldsymbol{\sigma }\mathrm{d}\Omega }-\int_{\Gamma _{t}}{\ \delta \mathbf{v}%
\cdot \mathbf{t}\mathrm{d}\Gamma } \\
& -\int_{\Omega }{\ \rho \delta \mathbf{v}\cdot \mathbf{b}\mathrm{d}\Omega }%
=0\mathtt{.}\label{1}%
\end{array}%
\end{equation}%
Since the continuum bodies is described by a finite set of material particles, the mass density can be written as
 \[
 \rho (\mathbf{x})=\sum_{p=1}^{N_{p}}{\ m_{p}\delta (\mathbf{x}-\mathbf{x}_{p})} \text{,}
 \]
  where $\delta $ is the Dirac delta function with dimension of the inverse of volume. Substituting $\rho (\mathbf{x})$ into the weak form of the momentum equation converts the integral to the sums of quantities evaluated at the material particles. So,
\begin{equation}
m_{i}\mathrm{d}\mathbf{v}_{i}/\mathrm{d}t=(\mathbf{f}_{i})^{\mathrm{int}}+(%
\mathbf{f}_{i})^{\mathrm{ext}}\mathtt{,}  \label{MPMe2}
\end{equation}%
where the internal force vector is given by
\[
\mathbf{f}_{i}{}^{\mathrm{int}%
}=-\sum_{p}^{N_{p}}{m_{p}\boldsymbol{\sigma }}_{p}{\cdot (\nabla N_{i})/\rho
_{p}} \text{,}
\]
and the external force vector is given by
\[
\mathbf{f}_{i}{}^{\mathrm{ext}}=\sum_{p=1}^{N_{p}}{N_{i}\mathbf{b}_{p}+\mathbf{f}_{i}^{c}} \text{,}
\]
where the vector $\mathbf{f}_{i}^{c}$ is the contacting force between two bodies. In the case where all colliding bodies are composed of the same material, $\mathbf{f}_{i}^{c}$ is treated  in the same way as the internal force.

The nodal accelerations can be calculated by using Eq. (\ref{MPMe2}) with an explicit
time integrator. To have a stable simulation, the time step $\Delta t$ should be less than the critical value,
\[
\Delta t_{C} = \frac{\Delta x_{\min}}{\max(c_{p}+|\mathbf{v}_{p}|)} \text{,}
\]
where $\Delta x_{\min}$ is the smallest cell size, $c_{p}$ the sound speed at particle $p$.
Once the motion equations have been solved on the cell nodes, the new nodal values of acceleration
can be used to update the velocity of the material particles. The strain
increment for each material particle is determined by using the gradient of
nodal basis function evaluated at the position of the material particle. The
corresponding stress increment can be obtained from the constitutive model. The
internal state variables can also be completely updated. The computational
mesh may be the original one or a newly defined one, choose for convenience,
for the next time step. More details of the algorithm can be referred to Refs.\cite%
{JPD2008,CTP2008}.

In our studies the possible phase transitions are studied via MD and DBM, instead of MPM simulations.
%%%%%%%%%%%%%%%%%%%%%%%%%%%%%%%%%%%%%%%%%%%%%%%%%%%%%%%%%%%%%%%%%%%%%

\subsection{Fluid model and DBM}

Shock waves may occur in many different kinds of materials. However, in
parts of this review the discussion is restricted to situations where
the material may be described, to a good approximation, by the model of a
compressible, heat-conducting fluid.
The most frequently used hydrodynamic models in engineering applications are the Euler and Navier-Stokes equations.
The model described by Euler equations assumes that (i) the fluid is always at its local thermodynamic equilibrium, and consequently can be completely described by the thermodynamic quantities, (ii) the change or transition between various mechanical states is quasi-static and iso-entropic.
According to this model, the shock wave has no thickness.
The model by Navier-Stokes equations assumes that all the relevant non-equilibrium behaviors can be described by non-equilibrium linear responses of gradients of physical quantities. The linear response of momentum flux is the viscosity tensor which is proportional to the gradient of flow velocity. The linear response of energy flux is the heat conduction which is proportional to the gradient of local temperature. According to this model, the width of the shock wave depends on the viscosity and heat conductivity.
Around the various interfaces, such as the material interfaces, shock fronts, new phase boundaries, and within the chemical reaction zone, the non-equilibrium effects may be so pronounced that the linear response theory do not work any more. Under such conditions, the models by Euler and Navier-Stokes equations may lead to unreasonable results and the DBM is more  preferable.

\subsubsection{Brief review of DBM}

Historically, the DBM was developed from the well-known Lattice Boltzmann Method(LBM) which was developed from the lattice gas automaton model\cite{Succi-Book}.
Mathematically, the DBM can be regarded as a special discretization of the Boltzmann equation.

Roughly speaking, the discrete Boltzmann models can be further classified into two categories. The first category is the one generally referred to as LBM in literature. In the first category the DBM is regarded as a kind of new scheme to numerically solve partial differential equations, such as the Navier-Stokes equaitons, etc. In the second category the DBM works as a kind of novel mesoscopic and coarse-grained kinetic model for complex fluids.
The most important difference between the two kinds of DBMs are as below. In the first category the LBM must be faithful to the original physical model, while in the second category, the DBM must possess some points beyond the original physical model.
The second kind of DBM aims to probe the trans- and supercritical fluid behaviors\cite{Succi-DBM2015} or to study simultaneously the hydrodynamic non-equilibrium (HNE) and thermodynamic non-equilibrium (TNE) behaviors\cite{Review-FoP2012,Review-PiP2014,Review-APC2015,SoftMatter2015}. It has brought significant new physical insights into the systems and promoted the development of new methods in the fields. For example, new observations on fine structures of shock and detonation waves have been obtained\cite{XuLin-PRE2014,XuLin-PRE2015}; These new observations have been used to discriminate various interfaces\cite{XuLin-PRE2014,XuLin-PRE2015}; The intensity of TNE has been used as a physical criterion to discriminate the two stages, spinodal decomposition and domain growth, in phase separation\cite{SoftMatter2015}; Based on the features of TNE, some new front-tracking schemes have been designed\cite{XuLai2015}. In a recent study, the relation between TNE quantities and entropy production rate has been established\cite{XuZhang2016}.
Since the goals are different, the criteria used to formulate the two kinds of models are significantly different, even though there may be considerable overlaps between them.

Physically, the DBM can be regarded as a model being coarser-grained than the Boltzmann equation. It can be obtained via two important steps of coarser-grained modelings. The first step is the linearization of the collision operator in the Boltzmann equation. In this step, we obtain the Boltzmann-BGK-like equations. The second step is the special discretization of the particle velocity space. The DBM obtained in this way can be roughly regarded as hydrodynamic model supplemented by a coarse-grained model of the TNE behaviours.

From the side of physical modeling, the MD is a microscopic particle model which is independent of the continuum assumption. Consequently, it can be used to study the TNE behaviours, no matter the material is in solid or fluid state.
The MPM is based on the solid mechanics which is based on the continuum assumption.
The DBM is based on the Boltzmann equation. It includes and is beyond the hydrodynamic model, for example, the Navier-Stokes equations\cite{Review-FoP2012,Review-PiP2014,Review-APC2015,Succi-DBM2015,SoftMatter2015}.
Here we briefly review the recently developed DBMs for multiphase flows and for detonation systems.

\subsubsection{DBM for multiphase flows}

In 2007, Gonnella, Lamura and Sofonea (GLS)\cite{GLS} proposed a LBM for liquid-vapor two-phase flows, where the effects of interparticle force enter the force term of the lattice Boltzmann equation. In 2011, our group proposed to use the fast Fourier transform and its inverse to calculate the spatial derivatives in the GLS model\cite{Xu-PRE2011,Xu-EPL2012}. In this way,
 the total energy conservation can be better held and the spurious velocities can be refrained to a negligible scale in real simulations. Recently, our group further improved the model in two sides, inserted a more practical equation of state and supplemented a methodology to investigate the non-equilibrium features in the system\cite{SoftMatter2015,XZG-arXiv2014}.

The GLS LBM can be described by the following evolution equation,
\begin{equation}
\frac{\partial f_{ki}}{\partial t}+\mathbf{v}_{ki} \cdot \frac{\partial f_{ki}
}{\partial \mathbf{r}}=-\frac{1}{\tau} [f_{ki}-f_{ki}^{eq}]
+I_{ki}\text{,}  \label{GLS-LB}
\end{equation}%
where the subscript $ki$ are the indexes of the discrete velocity and $I_{ki}$ reads
\begin{equation}
I_{ki}=-[A+ \mathbf{B} \cdot (\mathbf{v}_{ki}-\mathbf{u})+(C+C_{q})(\mathbf{v}_{ki}
-\mathbf{u})^{2}]f_{ki}^{eq}\text{,}  \label{iki}
\end{equation}%
with \begin{equation}
A=-2(C+C_{q})T\text{,}  \label{AAA}
\end{equation}%
\begin{equation}
\mathbf{B}=\frac{1}{\rho T}[\mathbf{\nabla}(P^{\text{vdw}}-\rho T)+\mathbf{\nabla}\cdot
\boldsymbol{\Lambda }-\mathbf{\nabla}(\zeta \mathbf{\nabla}\cdot \mathbf{u})],
\label{BBB}
\end{equation}%
\begin{eqnarray}
C &=&\frac{1}{2\rho T^{2}}\{(P^{\text{vdw}}-\rho T)\mathbf{\nabla}\cdot \mathbf{u}+%
\boldsymbol{\Lambda \colon \mathbf{\nabla}u}-\zeta (\mathbf{\nabla}\cdot \mathbf{u})^{2}
\notag \\
&&+\frac{9}{8}\rho ^{2}\mathbf{\nabla}\cdot \mathbf{u}+K[-\frac{1}{2}(
\mathbf{\nabla}\rho \cdot \mathbf{\nabla}\rho )\mathbf{\nabla}\cdot \mathbf{u}  \notag \\
&&-\rho \mathbf{\nabla}\rho \cdot \mathbf{\nabla}(\mathbf{\nabla}\cdot \mathbf{u
})-\mathbf{\nabla}\rho \mathbf{\cdot \mathbf{\nabla}u\cdot }\mathbf{\nabla}\rho
]\}\text{,}  \label{Eq:CC}
\end{eqnarray}%
\begin{equation}
C_{q}=\frac{1}{2\rho T^{2}}\mathbf{\nabla}\cdot \lbrack 2q\rho T\mathbf{\nabla}T]%
\text{.}  \label{CCQQ}
\end{equation}%
Here $\rho $, $\mathbf{u}$, $T$ are the local density, velocity,
temperature, respectively. The tensor $\boldsymbol{\Lambda }$ is
contribution of density gradient to pressure tensor and read
\[
\boldsymbol{\Lambda }=K\mathbf{\nabla}\rho \mathbf{\nabla}
\rho -K(\rho \nabla ^{2}\rho +\left\vert \mathbf{\nabla}\rho \right\vert ^{2}/2)%
\mathbf{I}-[\rho T\mathbf{\nabla}\rho \cdot \mathbf{\nabla}(K/T)]\mathbf{I}
\text{,}
\]
where $\mathbf{I}$ is the
unit tensor, $K$  the surface tension coefficient and $\zeta$ the bulk viscosity.
The model is consistent with the thermodynamic relations proposed by Onuki\cite{Onuki-PRL}.

The original GLS model utilizes the van der Waals equation of state,
\[
P^{\text{vdw}}=\frac{3\rho T}{3-\rho }-\frac{9}{8}\rho ^{2}
\]
with fixed parameters. The density ratio $R$ between the liquid and vapor phases which can be stably simulated is generally less than $10$ due to the numerical instability problem.
Since the Carnahan-Starling equation of state \cite{CS} modifis the repulsive term of van der Waals equation of state so that it presents a
more accurate representation for hard sphere interactions.
The Carnahan-Starling equation of state
\begin{equation}
p^{\text{cs}}=\rho T\frac{1+\eta +\eta ^{2}-\eta ^{3}}{(1-\eta )^{3}}-a\rho ^{2},  \label{PCS}
\end{equation}%
with $\eta =b\rho /4$
can be applied via replacing the term $\frac{9}{8}\rho ^{2}\mathbf{\nabla}
\cdot \mathbf{u}$ in Eq. (\ref{Eq:CC}) by $a\rho ^{2}\mathbf{\nabla}\cdot \mathbf{u}$, where $a$ and $b$ are the attraction and repulsion parameters. Subsequently, the total energy density becomes
\[
e_{T}=\rho T-a\rho^{2}+K\left\vert \mathbf{\nabla}\rho \right\vert ^{2}/2+\rho u^{2}/2 \text{.}
\]

\subsubsection{DBM for system under detonation}

As for the
discrete Boltzmann modeling and simulation of combustion systems, the current studies can also be
classified into two categories. Most of existing studies belong to the first category where the DBM is used as a kind of alternative numerical scheme and are focused on cases with low Mach number where the incompressible models work. The first DBM for detonation system\cite{XuYan-FoP2013} appeared in 2013. It is also the
first study aiming to investigate both the HNE and TNE in the combustion system via the discrete kinetic modeling.
To model and simulate the non-equilibrium behaviors in axial symmetric implosion and explosion processes, a DBM for detonation system in polar-coordinates \cite{XuLin-CTP2014} was proposed in 2014. A multiple-relaxation-time version of DBM for detonation system was developed and some fundamental issues in formulating discrete kinetic models were reviewed in a recent study\cite{XuLin-PRE2015}. A double-distribution-function DBM for detonation system is referred to Ref.\cite{XuLin-DDF-DBM}.

Up to now, from the view of mathematical modeling, the only difference of the DBM from the traditional hydrodynamic model is that the Navier-Stokes or Euler equations for flow are replaced by the discrete Boltzmann equations(s). The phenomenological equation describing the reaction process is the same. But from the view of physical application, the DBM is roughly equivalent to a hydrodynamic model supplemented by a coarse-grained model of the TNE behaviors. Being able to capture various non-equilibrium effects and being easy to parallelize are two features of the second kind of DBM. The two pints are also the physical gain and computational gain from this replacement.
Some more realistic DBMs for detonation systems are in progress.

The hydrodynamic modeling and microscopic molecular dynamics have seen great achievements in detonation simulations. But for problems relevant to  the mesoscopic scales, where the hydrodynamic modeling is not enough to capture the non-equilibrium behaviors and the molecular dynamics simulation is not affordable, the modeling and simulation are still open and challenging. Roughly speaking, there are two research directions in accessing the mesoscopic behaviors. One direction is to start from the continuous description at macroscopic scale to kinetic descriptions at smaller scales, the other direction is to start from the particle description at microscopic scale to statistical descriptions at larger scales.
The idea of second kind of DBM belongs to that of the first direction. It will contribute more to the studies on the non-equilibrium behaviors in various complex fluids.

\subsubsection{Two kinds of non-equilibrium effects}

 If choose the Navier-Stokes model as the macroscopic counterpart, the DBM must be based on, at least, the following seven kinetic moments,
\begin{equation}
\mathbf{M}_{0}^{\text{eq}}=\sum_{ki}f_{ki}^{eq}=\rho \text{,}
\label{moment1}
\end{equation}
\begin{equation}
\mathbf{M}_{1}^{\text{eq}}=\sum_{ki}f_{ki}^{eq}\mathbf{v}
_{ki}=\rho \mathbf{u}\text{,}  \label{moment2}
\end{equation}
\begin{equation}
\mathbf{M}_{2,0}^{\text{eq}}=\sum_{ki}\frac{1}{2}f_{ki}^{eq}\mathbf{v}%
_{ki}\cdot \mathbf{v}_{ki}=\rho (T+\frac{u^{2}}{2})\text{,}
\end{equation}
\begin{equation}
\mathbf{M}_{2}^{\text{eq}}=\sum_{ki}f_{ki}^{eq}\mathbf{v}_{ki}\mathbf{v}%
_{ki}=\rho (T\mathbf{I}+\mathbf{uu})\text{,}  \label{moment4}
\end{equation}
\begin{eqnarray}
\mathbf{M}_{3}^{\text{eq}}=\sum_{ki}f_{ki}^{eq}\mathbf{v}_{ki}\mathbf{v}%
_{ki}\mathbf{v}_{ki}
=\rho \lbrack T(\mathbf{u}_{\alpha }\mathbf{e}_{\beta }\mathbf{e}_{\gamma
}\delta _{\beta \gamma }+\mathbf{e}_{\alpha }\mathbf{u}_{\beta }\mathbf{e}%
_{\gamma }\delta_{\alpha \gamma} \notag \\
+\mathbf{e}_{\alpha }\mathbf{e}_{\beta }
\mathbf{u}_{\gamma }\delta_{\alpha \beta })+\mathbf{uuu}]\text{,}
\label{moment5}
\end{eqnarray}
\begin{equation}
\mathbf{M}_{3,1}^{\text{eq}}=\sum_{ki}\frac{1}{2}f_{ki}^{eq}\mathbf{v}%
_{ki}\cdot \mathbf{v}_{ki}\mathbf{v}_{ki}=\rho \mathbf{u}(2T+\frac{1}{2}%
\mathbf{u}\cdot \mathbf{u})\text{,}  \label{moment6}
\end{equation}%
\begin{eqnarray}
\mathbf{M}_{4,2}^{\text{eq}} =\sum_{ki}\frac{1}{2}f_{ki}^{eq}\mathbf{v}%
_{ki}\cdot \mathbf{v}_{ki}\mathbf{v}_{ki}\mathbf{v}_{ki}=\rho \lbrack (2T+%
\frac{\mathbf{u\cdot u}}{2})T\mathbf{I}  \notag \\
+\mathbf{uu}(3T+\frac{\mathbf{u\cdot u}}{2})] \text{,} \label{moment7}
\end{eqnarray}%
where $\mathbf{M}_{m,n}^{\text{eq}}$ stands for that the $m$-th order tensor is contracted to a $n$-th order one.
Among the seven kinetic moment relations, only for the first three, the local equilibrium distribution function $f_{ki}^{eq}$ can be replaced by $f_{ki}$, which means that, when the system approaches or deviates from the thermodynamic equilibrium, the mass, momentum and energy are conserved. Replacing $f_{ki}^{eq}$ by $f_{ki}$ in Eqs. \eqref{moment4}-\eqref{moment7} results in the imbalance and the deviation described by
\begin{equation}
\boldsymbol{\Delta}_{n} =\mathbf{M}_{n}(f_{ki})-\mathbf{M}_{n}^{\text{eq}}(f_{ki}^{eq}),
\end{equation}
. The quantity
$\boldsymbol{\Delta}_{n}$ presents a simple, convenient and effective measure to the departure of the system from the local thermodynamic equilibrium.

If the shocking is so strong that the material can be regarded as ideal gas, we can consider that the HNE and TNE effects are only induced by gradients of macroscopic quantities, also referred to gradient force. When the inter-particle interaction potential can not be completely ignored and the system can be regarded as a multiphase flow system. The force term in the DBM equation works as the second driving force. Especially, the right-hand side of Eq. (\ref{GLS-LB}) can be rewritten as
\begin{equation}
\text{RHS}=-\frac{1}{\tau }[f_{ki}-(1+\tau \theta
)f_{ki}^{eq}]=-\frac{1}{\tau }[f_{ki}-f_{ki}^{eq,\text{NEW}}]\text{,}
\end{equation}%
where
\[
\theta=-[A+ \mathbf{B} \cdot (\mathbf{v}_{ki}-\mathbf{u})+(C+C_{q})(\mathbf{v}_{ki}
-\mathbf{u})^{2}]
\]
and
\[
f_{ki}^{eq,\text{NEW}}=(1+\tau \theta)f_{ki}^{eq}
\text{.}
\]
It can be considered as a new equilibrium state shifted by the interparticle force. Consequently,
\begin{equation}
\boldsymbol{\Delta }_{n}^{F}=\mathbf{M}_{n}(\tau \theta f_{ki}^{eq})=%
\mathbf{M}_{n}(\tau I_{ki})
\end{equation}%
describes the non-equilibrium effects induced by the interparticle force.
What we measure from $f_{ki}$ and $f^{eq}_{ki}$,
\begin{equation}
\boldsymbol{\Delta}_{n}=\mathbf{M}_{n}(f_{ki})-\mathbf{M}_{n}^{\text{eq}}(f^{eq}_{ki})=%
\boldsymbol{\Delta}_{n}^{F}+\boldsymbol{\Delta}_{n}^{G}  \label{Non_eq3e}
\end{equation}
are the combined or the net non-equilibrium effects, where
\begin{equation}
\boldsymbol{\Delta}_{n}^{G} =\mathbf{M}_{n}(f_{ki})-\mathbf{M}
_{n}^{\text{eq}}(f_{ki}^{eq,\text{NEW}})
\end{equation}
are the non-equilibrium effects induced by the gradient force.
It is clear that, when the interparticle force disappears, the net non-equilibrium effects are only from the gradient force, i.e., $\boldsymbol{\Delta}_{n}=\boldsymbol{\Delta }_{n}^{G}$, corresponding to an ideal gas system.
Note that, the kinetic moment $\mathbf{M}_{n}$ contains the information of $\mathbf{u}$, so do the non-equilibrium quantity $\boldsymbol{\Delta}_{n}$. They describe both the HNE and TNE effects. If we use the central kinetic moment $\mathbf{M}_{n}^{*}(f_{ki})=\sum f_{ki}(\mathbf{v}_{ki}-\mathbf{u})^{n}$,
then $\boldsymbol{\Delta}_{n}^{*}$ does not contain the effects of $\mathbf{u}$, describes only the TNE effects.
Because $\mathbf{M}_{n}^{*}(f_{ki})$  is only the representation of the thermo-fluctuations of molecules relative to $\mathbf{u}$.

\paragraph{Compromise}
The physically concerned hydrodynamic quantities are some kinetic moments of the distribution function, $f$. According to the Chapman-Enskog analysis, they can finally be roughly calculated from some kinetic moments of the local equilibrium distribution function, $f^{eq}$.
The calculation of any non-conserved quantity triggers the requirement of higher-order kinetic moments of $f^{eq}$.
When construct the discrete Boltzmann model, we must ensure the required kinetic moments of  $f^{eq}$, originally in integral form, can be calculated in discrete summation form.

All descriptions on the TNE based on finite number of kinetic moments are coarse-grained.
The more accurate the TNE is to be described, the more kinetic moments are required.
The more the required kinetic moments, the higher the computational cost. In practical applications, we
have to make compromise between what we want and what we can afford.

\section{Analysis schemes for complex fields and structures}

No matter which physical model and simulation tool are used, after the
simulation, how to analyze the data and pick out reliable information is of
key importance.
In our MD simulation studies, two methods are used.
(i) The atoms are distinguished by the Common Neighbor Analysis (CNA) method\cite{Pang-SciRep2014C,Pang-SciRep2014A,Pang-SciRep2014B}. In this method the signature of the local crystal structure of an atom is identified by computing three characteristic numbers for each of the $n$ neighbor bonds of the central atom; (ii) The
Burgers vectors of the evolved dislocations in the MD simulations are
calculated using our home-built code. In this code, dislocation lines and
their directions are first identified. Then, surrounding the dislocation
lines, appropriate Burgers circuits that cross stacking-fault planes or
perfect crystal are selected, and the atom-to-atom sequences corresponding
to the circuits are determined. Finally, after a summation over vectors of
the Thompson's tetrahedron and its mirrors that are most closest to the
atom-to-atom vectors, the Burgers vectors of the dislocations are obtained.

For complex system in the mesoscopic and macroscopic scales, nearly all the analysis methods are some sort of statistics. The most commonly used ones are the mean values of physical
variables and their corresponding fluctuations. The rheological description
provides helpful measurements like the spatial correlation, temporal
correlation, spatial-temporal correlation, structure factor, characteristic
length, etc\cite{XGL-PRE2003,XGL-EPL2005,XGL-PRE2006}. In our studies the morphological description is introduced to
describe the complex fields in heterogeneous materials under shock\cite{Xu-JPD2009,Xu-CTP2009B}. Several
new schemes, including the turbulence mixing, volume dissipation, entropy
increment, cluster identification, tracking of characteristic structures
were designed\cite{Xu-CTP2009A,MD-SHT-Book,Zhang-SciCN2010}.

\subsection{For fields and structures based on ordered points}

 To analyze fields and structures based on ordered points, a variety of schemes can be used.  For example, (i) common schemes of statistical physics, (ii) rheological descriptions, (iii) morphological characterization, etc.
  Scheme examples for (i) are referred to
 the mean value and fluctuations, turbulence dissipation, volume dissipation, entropy production.
 Scheme examples for (ii) are referred to
 spatial correlation, temporal correlation and spatial-temporal correlation, structure factor, characteristic length and time scales.
 Scheme examples for (iii) are referred to
 the Minkowskii functionals\cite{Serra1982}.
 All of them are some kinds of statistics.

 Here we concentrate only on the set of statistics known as Minkowski functionals\cite{Serra1982}.
A physical field can be described by $\Theta(\mathbf{x})$,
where $\mathbf{x}$ is the position, $\Theta$ a physical variable. The physical variable $\Theta$ can be a scalar state variable like temperature $T$, density $\rho$ and pressure $P$. It can also be the size of velocity $\mathbf{v}$, velocity component in one degree of freedom, as well as some specific stress component, etc.

According to a general theorem of integral geometry,
all properties of a $D$-dimensional convex set (or more generally, a finite union of
convex sets) satisfying the morphological properties (translational invariance and additivity),
 are contained in $D + 1$ numerical values \cite{PRE1996,Xu-JPD2009}.
The points with $\Theta(\mathbf{x}) \geq \Theta_{th}$ compose the two- or three-dimensional convex set and its morphological properties can be completely described by three or four functionals, where $\Theta_{th}$ is some threshold value. In the case of two or three dimensions, the Minkowski functionals have intuitive geometric interpretations.
For the two-dimensional case, the three Minkowski functionals
correspond geometrically to the total fractional area $A$, the total boundary length $L$, and the Euler characteristic $\chi $ which is equivalent
to the topological genus. In practical application, the Minkowski functions can be made dimensionless.
Such a morphological description has been successfully
applied in describing patterns in reaction-diffusion
system\cite{PRE1996}, phase separation\cite{JChemP2000,Sofonea,PR} and complex fields in porous materials under shock\cite{Xu-JPD2009}, etc.

\begin{figure*}[tbp]
\centering
\includegraphics*[ scale=0.3,angle=0,
bbllx=0pt,bblly=0pt,bburx=569pt,bbury=445pt
]{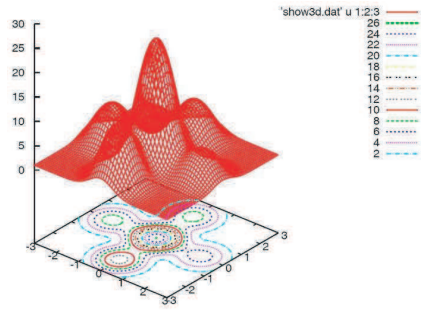}
\caption{(Color online) An example of two-dimensional patterns, where the $z$-axis corresponds to a physical quantity under consideration, $x$- and $y$- axes show the two-dimensional coordinates.}
\label{Fig:MorphologyFig1}
\end{figure*}

\begin{figure*}[tbp]
\centering
\includegraphics*[ scale=0.4,angle=0,
bbllx=0pt,bblly=0pt,bburx=575pt,bbury=543pt
]{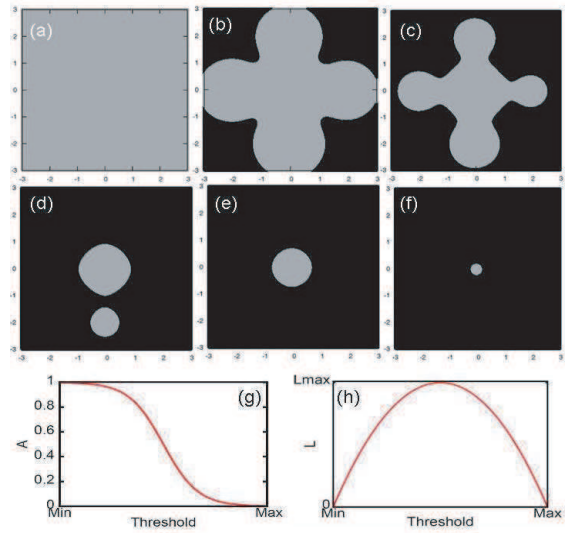}
\caption{(Color online)Schematics of morphological characterizations.
From (a) to (f) one can find that the white area $A$ decreases and the topology of the patten changes with increasing the threshold value $\Theta_{th}$.
(g) shows the schematic curve for $A$ versus $\Theta_{th}$.  (h) shows schematically the variation of boundary length $L$ with $\Theta_{th}$.
}
\label{Fig:MorphologyFig2}
\end{figure*}

For the two-dimensional square lattice, a lattice node possesses four vertices. A region with connected lattice nodes with $\Theta(\mathbf{x}) \geq \Theta_{th}$ or
connected lattice nodes with $\Theta(\mathbf{x})  <  \Theta_{th}$
is defined as a white or black domain, in the language of morphological description. Two neighboring white and black domains present a clear interface or boundary. When the threshold   $\Theta_{th}$ is increased from the lowest to the highest values of $\Theta$ in the system, the white area $A$ will decrease from $1$ to $0$; the boundary length $L$ first increases from $0$, then arrives at a maximum value, and finally decreases to $0$ again. There are several ways to define the Euler characteristic $\chi $. Two simplest ones are as below:
\begin{equation}
\chi =N_{W}-N_{B}\mathtt{,}  \label{Mink1}
\end{equation}%
or
\begin{equation}
\chi =\frac{N_{W}-N_{B}}{N}\mathtt{,}  \label{Mink2}
\end{equation}%
where $N_{W}$ ($N_{B}$) is the number of connected white (black) domains, $N $ is the total number of lattice nodes with $\Theta(\mathbf{x})  <  \Theta_{th}$.
The only difference of the two definitions is that the first keeps $\chi $ an integer. In contrast to the other two Minkowski functionals, white area $A$ and boundary length $L$, what the Euler characteristic $\chi $ describes is the connectivity of the domains in the lattice. It describes the pattern or structure in a purely topological way, i.e., without referring to any kind of metric. It is clear that it is negative (positive) if many disconnected black (white) regions dominate the pattern or structure. The smaller the Euler characteristic $\chi $, the higher the connectivity of the structure with
$\Theta(\mathbf{x}) \geq \Theta_{th}$ or $\Theta(\mathbf{x})  <  \Theta_{th}$.
Specifically, for the first definition, the integer $\chi = -1$ in the case with only  black drop in a large white lattice, and $\chi = +1$ vice versa,
since the surrounding white (black) region does conventionally not count. In our work, only the second definition is used. What the ratio,
$ \kappa = (N_{W}-N_{B})/ (NL) $,
describes is the mean curvature of the boundary line separating black and white
domains. Even though the Euler characteristic $\chi $ has a global meaning, it can
be calculated in a local way via the additivity relation\cite{PRE1996}.
When the number of white regimes dominates, $\chi > 0$; else, $\chi < 0$.
Figure \ref{Fig:MorphologyFig1} shows an example of two-dimensional patterns,  where the $z$-axis corresponds to a physical quantity $\Theta$ under consideration, $x$- and $y$- axes show the two-dimensional coordinates.
Figure \ref{Fig:MorphologyFig2} shows the white and back domains and schematic morphological characterizations.

\begin{figure*}[tbp]
\centering
\includegraphics*[ scale=0.2,angle=0,
bbllx=0pt,bblly=0pt,bburx=594pt,bbury=530pt
]{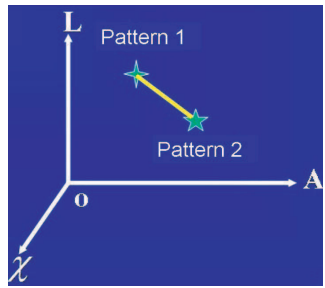}
\caption{(Color online)Morphological properties of two patterns in the space opened by the morphological quantities.}
\label{Fig:MorphologyFig3}
\end{figure*}

\begin{figure*}[tbp]
\centering
\includegraphics*[ scale=0.2,angle=0,
bbllx=0pt,bblly=0pt,bburx=593pt,bbury=506pt
]{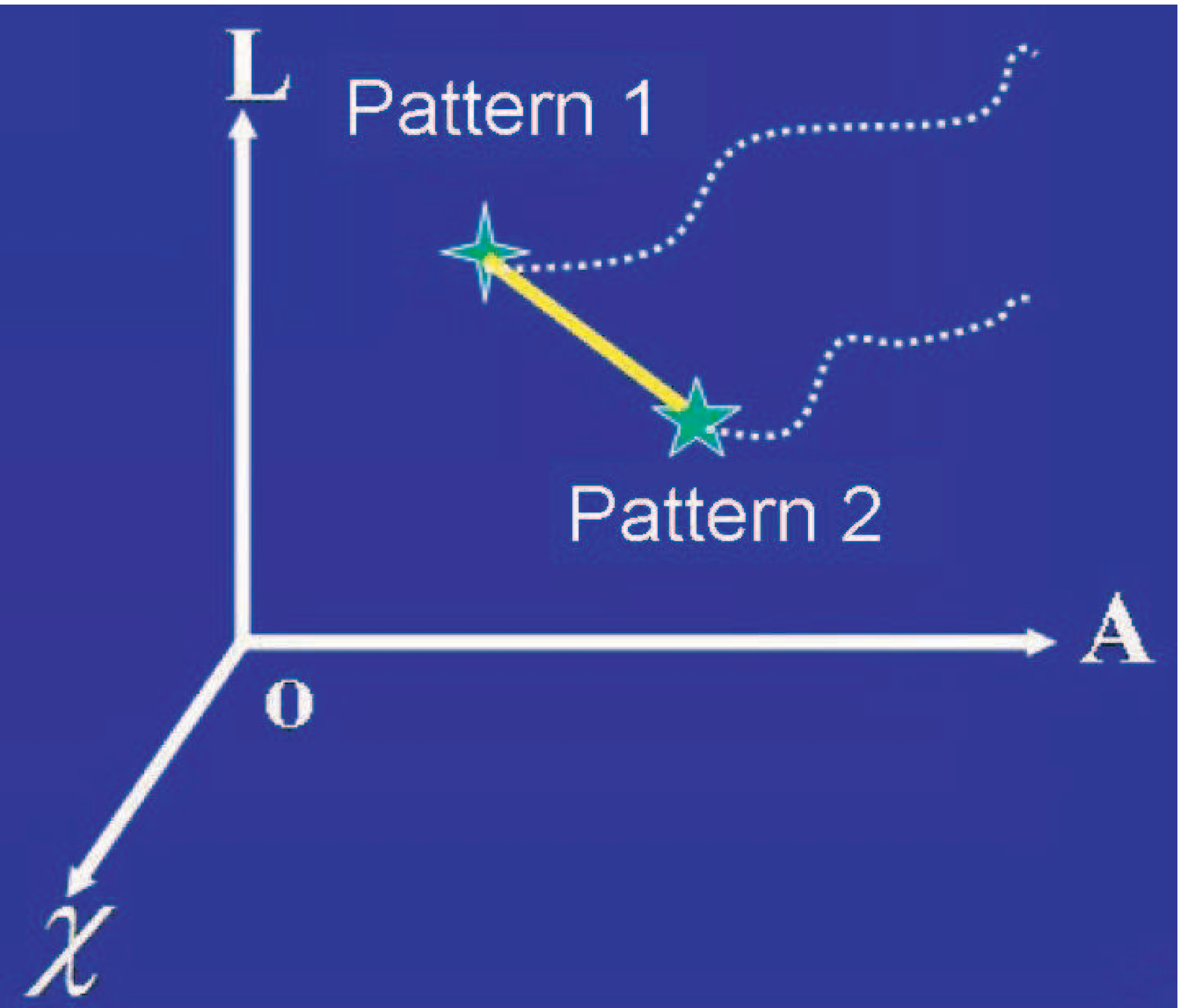}
\caption{(Color online)Evolution of the morphological properties with time.}
\label{Fig:MorphologyFig4}
\end{figure*}

The morphological characterizations of some physical fields, for example, the temperature field and density field, can be used to study the effects of material properties such as the porosity and effects of shocking strengths, etc. They can also be extended to investigate possible correlations and similarities occurred in various shocking processes.
Because all the morphological properties of a pattern in $D$-dimensional space are contained in the $D + 1$ morphological quantities, one can consider the morphological properties of the  pattern in a $D + 1$-dimensional space opened by the $D + 1$ morphological quantities. In this
$D + 1$-dimensional space one point corresponds to all the morphological behaviors of a pattern. The distance $d$ between two points in this space presents a coarse-grained description of the difference of the two patterns. The shorter the distance $d$, the higher the similarity between  the morphological properties of the two corresponding patterns. So, we can define a new quantity named structure similarity as  $S = 1/d$. (See Fig. \ref{Fig:MorphologyFig3} for a schematic.) If the two patterns evolve with time, then we can go a further step to define a dynamical similarity for the two pattern evolution processes from time $t_{1}$ to $t_{2}$,
\[
S_{D} = \int_{t_{1}}^{t_{2}}d\left( t\right) dt/\left( t_{2}-t_{1}\right) \text{.}
\]
 Specifically, for a pattern in the two-dimensional space, the difference of the morphological properties of pattern $1$ and pattern $2$ can be coarsely described by
 \[
 d = \sqrt{(A_2 - A_1)^2 + (L_2 - L_1)^2 + (\chi_2 - \chi_1)^2} \text{,}
  \]
  where the subscript is the index of the pattern.  (See Fig.\ref{Fig:MorphologyFig4} for a schematic.)

%%%%%%%%%%%%%%%%%%%%%%%%%%%%%%%%%%%%%%%%%%%%%%%%%%%%%%%%%%%%%%%%%%%%%%%%%%%%%%%%
%%%%%%%%%%%%%%%%%%%%%%%%%%%%%%%%%%%%%%%%%%%%%%%%%%%%%%%%%%%%%%%%%%%%%%%%%%%%%%%%

%%%%%%%%%%%%%%%%%%%%%%%%%%%%%%%%%%%%%%%%%%%%%%%%%%%%%%%%%%%%%%%%%%%%%%%%%%%%%%%%
%%%%%%%%%%%%%%%%%%%%%%%%%%%%%%%%%%%%%%%%%%%%%%%%%%%%%%%%%%%%%%%%%%%%%%%%%%%%%%%%

\
%%%%%%%%%%%%%%%%%%%%%%%%%%%%%%%%%%%%%%%%%%%%%%%%%%%%%%%%%%%%%%%%%%%%%%%%%%%%%%%%
%%%%%%%%%%%%%%%%%%%%%%%%%%%%%%%%%%%%%%%%%%%%%%%%%%%%%%%%%%%%%%%%%%%%%%%%%%%%%%%%

%%%%%%%%%%%%%%%%%%%%%%%%%%%%%%%%%%%%%%%%%%%%%%%%%%%%%%%%%%%%%%%%%%%%%%%%%%%%%%%%
%%%%%%%%%%%%%%%%%%%%%%%%%%%%%%%%%%%%%%%%%%%%%%%%%%%%%%%%%%%%%%%%%%%%%%%%%%%%%%%%

 \subsection{For fields and structures based on disordered points}

Structure analysis is the core issue in studies on material simulation and material dynamics. The micro-structures in metal material may be composed of defect atoms deviating from the crystal lattices. In principle, the defect atoms can be identified by analyzing the regularity of their neighboring atoms.
The distribution of the defect atoms in space is generally disordered. It is necessary to find an efficient algorithm for identifying and analyzing these structures. These complex algorithms include various searching schemes.
In scheme design and in coding, based on the defect atoms, the identification of high dimension structures like dislocations, grain boundaries and voids requires to construct the so-called line, surface and body.
A General Index of Spatial Objects (GISO) was designed in our group in the past years. In this section we introduce the GISO software and its applications in structure analysis on various micro-structures\cite{MD-SHT-Book,Zhang-SciCN2010}.

\subsubsection{Outline of GISO}

 Complex computations of relation between particles are inevitable in any elaborate defect identification methods. The computation time will dramatically increase with growing the system size in traditional methods.
 Without indexing the spatial objects, the computation quantity for searching object is generally very large. If a system contains $N$ objects, the computation complexity related to two objects is $N^2$ and that related to three bodies is $N^3$. In the case where the total number of objects is more than $10^4$, the computation complexity will not be acceptable. In such a case, schemes for effective storage and fast search of objects are crucial. To obtain such a scheme, it is necessary to design new data structure and indexing algorithm which significantly reduce the computation complexity. The computation complexity in defect identification methods can be greatly reduced by using background grid and linked list. The background grid index, together with the linked list data structure, is suitable for managing  uniform distributed points. It has been extensively used in computation and analysis of many simulation results. Complex structure in non-uniform system refers not only to points, but also to lines, surfaces and bodies. Their distributions in space are usually non-uniform. The background grid index cannot satisfy the needs for managing these objects, but a multi-level division of space is much more effective. The Space Hierarchy Tree (SHT) is a newly proposed data structure. It is a powerful dynamical management framework for any complex objects in any dimensional space.
 Based on the SHT, index of objects with complex structure can be created. Corresponding fast searching schemes can also be designed to satisfy various searching requirements.

\paragraph{SHT management structure}

\begin{figure*}[tbp]
\centering
\includegraphics*[scale=0.5,angle=0,
bbllx=0pt,bblly=0pt,bburx=595pt,bbury=320pt
]{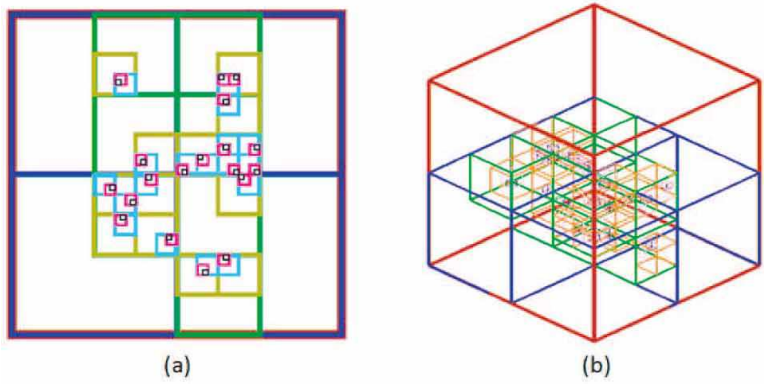}
\caption{(Color online)Scheme for management region of SHT of discrete points. (a) Two-dimensional points; (b) three-dimensional points.
(Adopted with permission from Ref. \cite{MD-SHT-Book}. The grey-level version is published in
Ref.\cite{Zhang-SciCN2010}.)
}
\label{Fig:ClusterFig1}
\end{figure*}

\begin{figure*}[tbp]
\centering
\includegraphics*[scale=0.7,angle=0,
bbllx=0pt,bblly=0pt,bburx=596pt,bbury=168pt
]{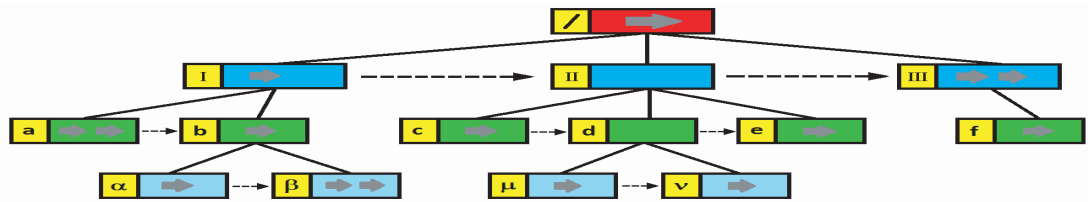}
\caption{(Color online)Scheme for object management by SHT. The rectangle in each row stands for cube, horizontal grey arrow stands for object, and vertical black arrow stands for the list which connects the sub-cubes belonging to a same cube.
(Adopted with permission from Ref. \cite{MD-SHT-Book}. )
}
\label{Fig:ClusterFig2}
\end{figure*}

The SHT data structure is similar to octree in the three-dimensional space. Go a further step, for a system in $n$-dimensional space, an $n$-dimensional cube is designed to contain the system. It is
a line segment in one-dimensional space, a rectangle in two-dimensional space, and a cube in three-dimensional space, and so on.
 Divide each dimension of this tube into two parts. $2^{n}$ sub-cubes are formed, but only retain the cubes with objects inside. Continue to decompose each cube until the required resolution is reached. Put the objects (points, lines, surfaces, bodies) into the appropriate cubes according to their locations and sizes (see Figure \ref{Fig:ClusterFig1}). Connect the retained cubes together according to their belonging relationships. Thus, a ``spatial hierarchy tree'' is constructed. (see Figure \ref{Fig:ClusterFig2}).

In practical applications, the SHT is constructed dynamically because the number of objects may be variable. The dynamic management procedure of SHT consists of the following three basic operations: (i) establishment of a tree, (ii) adding a new object to a tree, (iii) removing an object from a tree. The regimes managed by SHT is dynamically altered during these operations. In managing various objects with drastically different sizes and extremely scattered distributions, the SHT shows its effectiveness.

\subsubsection{Fast searching algorithms based on SHT}

 One generally needs a fast search of objects satisfying certain requirements in practical application. For  an ergodic search, the computational complexity is $N$. It is evidently not practical to treat with a huge number of objects. In  such cases, one needs to design fast searching algorithms. With the management of SHT, fast searchers with computational complexity $\ln N$ can be easily created. The basic idea is as follows: do not search the objects directly, but rather check cubes and  skip those cubes without objects. In this way, the searching is limited to a substantially smaller range. According to the requirements of applications, two fast searching algorithms are proposed. The first is referred to as conditional search, and the second is referred to as minimum search. The conditional search is to search for objects meeting certain conditions. For example, to find objects in a given area. The minimum search is to search for objects whose function values are minimum. For example, to find the nearest object to a fixed point.

\paragraph{Conditional search}

 The basic idea is as follows: From the largest cube to the smallest, hierarchically
check whether or not a cube contains objects meeting given conditions. If not, skip the cube (including all sub-cubes of it and corresponding objects).

In the searching process, only two operations are relevant to space dimension and type of object. The two operation are  as follows:  (i)to check whether or not an object is the needed one, or (ii) to check whether or not an cube is a candidate. Thus, the algorithm can be built in the abstract level. The conditional search is implemented via providing a conditional function and an identification function. The conditional function is used to check whether or not an object is needed. Assuming \emph{condition(o)} is the conditional function, the argument $o$ is object and the function value is a bool number. The identification function is used to assess whether or not a cube is a candidate. Assuming \emph{maycontain(b)} is identification function, the argument $b$ is cube and the function value is also a bool number. After defining the above two functions, conditional searching meeting any given conditions can be easily implemented.

\paragraph{Minimum search}

 One often needs to find objects satisfying some given extreme condition in programming related to spatial objects.
 For example, to find a point with the largest $z$ component from a set of three-dimensional points, or to search a point with the nearest distance to a given point, or to search a sphere closest to a plane, etc. Such searches can be classified to the minimum searching problem. For spatial objects, each one can be assigned a function value related to its location and size in such a way that the minimum search becomes to find the object with the minimum function value.

Corresponding fast searching scheme can be designed based on the SHT. The basic idea is as follows: Design a function to assess the range of the function values of all possible objects in a cube which has certain position and size. Via
 comparing the ranges of the function value of different cubes, some cubes can be excluded. For example,  $\Sigma$ is a set of discrete points in a region, one needs to search the nearest points to a given point, A. The fast searching is not to calculate the distance between each point in $\Sigma$ and the point A, but assess the range of distance between 'cubes' and point A to exclude unnecessary searching of cubes (including the point in them and their sub-cubes) with longer distance.

In minimum search, except for calculating the function value of an object and the range of a cube, other operations have nothing to do with space dimension and type of object, so that the algorithm can also be built in abstract level. Similar to case of conditional search, the minimum search is implemented via providing a value-finding function and range-evaluation function. The value-finding function is to calculate the function value of an object. Assuming \emph{value(o)} is value-finding function, the argument $o$ is an object and the function value is a real number. The range-evaluation function is used to compute the range of a cube. Assuming \emph{M(b)} is the upper limit and \emph{m(b)} is the lower limit of the range, the argument of function is cube $b$ and the function value is a real number. After defining the above two functions, various minimum searches can be easily implemented.

\subsubsection{Applications of GISO}

We first illustrate the algorithm of rolling-ball method to construct spatial surface. For other applications of GISO, only the basic ideas are briefly reviewed.

\paragraph{Rolling-ball method for finding interfaces}

On a regular grid, the most common method to find interface of physical domain is to use the contour of the corresponding physical field. This method works well for the case where the discrete points closed to interface are  uniformly distributed. When the distribution of discrete points is very complex, it is difficult to preserve the smoothness of the constructed interface. Consequently, the calculated interface will be significantly different from the actual one. A better means is to use the rolling-ball method. The basic idea of rolling-ball method is as follows: Roll a ball with fixed size over the discrete points;  each rolling goes through three points, and these points constitute a surface element of interface. After the rolling-ball goes through the overall region, the physical interface is constructed.

In the rolling-ball method, the initial localization needs two searching schemes, and the rolling process needs the other two searching schemes.
The four searching schemes are as follows.

\begin{figure*}[tbp]
\centering
\includegraphics*[scale=0.3,angle=0,
bbllx=0pt,bblly=0pt,bburx=594pt,bbury=480pt
]{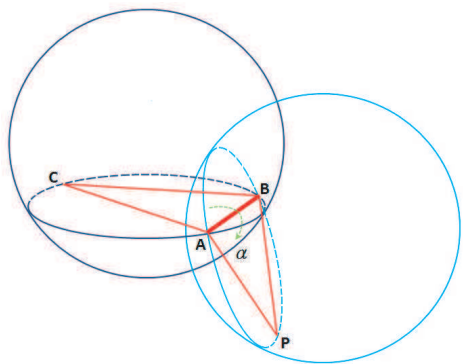}
\caption{(Color online)Scheme for the rotation of triangle ABC.
(Adopted with permission from Ref. \cite{MD-SHT-Book}.)
}
\label{Fig:ClusterFig3}
\end{figure*}

(I) Minimum searcher MS1: Given a triangle face ABC and one of its
edge AB, search in point tree for the first point met by the rolling-ball
above triangle ABC, where the radius of rolling-ball is $r$ and the rotation
axis is AB. To construct the value-finding function,
we first calculate the initial center $\mathbf{r}_{o}$ of the rolling-ball and the directions of local coordinate axes $\mathbf{\hat{x}}$,$\mathbf{\hat{y}}$,$\mathbf{\hat{%
z}}$ according to the following equations:
\[
r^{2}=\left( \mathbf{r}_{O}-\mathbf{r}_{A}\right) ^{2},
\]%
\[
r^{2}=\left( \mathbf{r}_{O}-\mathbf{r}_{B}\right) ^{2},
\]%
\[
r^{2}=\left( \mathbf{r}_{O}-\mathbf{r}_{C}\right) ^{2},
\]%
and \\
\[
\mathbf{\hat{x}}=\frac{\mathbf{P}_{xy}\cdot \left( \mathbf{r}_{O}-\mathbf{r}%
_{A}\right) }{\left\vert \mathbf{P}_{xy}\cdot \left( \mathbf{r}_{O}-\mathbf{r%
}_{A}\right) \right\vert },
\]%
\[
\mathbf{\hat{y}}=\mathbf{\hat{z}}\times \mathbf{\hat{x},}
\]%
\[
\mathbf{\hat{z}}=\frac{\mathbf{r}_{B}-\mathbf{r}_{A}}{\left\vert \mathbf{r}%
_{B}-\mathbf{r}_{A}\right\vert },
\]%
where the subscript ``o'' indicate ``old'' and
\[
{{\mathbf{P}}_{xy}=\mathbf{I}}-\mathbf{\hat{z}\hat{z}}.
\]
After the rotation,
calculate the new center $\mathbf{r}_{n}$ of the rolling ball and the corresponding local coordinates, $x$, $y$, $z$,
according to the following relations:
\[
{{r}^{2}}={{\left( {{\mathbf{r}}_{n}}-{{\mathbf{r}}_{A}}\right) }^{2}}
\]%
\[
\newline
{{r}^{2}}={{\left( {{\mathbf{r}}_{n}}-{{\mathbf{r}}_{B}}\right) }^{2}}
\]%
\[
\newline
{{r}^{2}}={{\left( {{\mathbf{r}}_{n}}-{{\mathbf{r}}_{P}}\right) }^{2}}
\]%
and\\
\[
x=\mathbf{\hat{x}}\cdot \left( {{\mathbf{r}}_{n}}-{{\mathbf{r}}_{A}}\right) ,
\]%
\[
y=\mathbf{\hat{y}}\cdot \left( {{\mathbf{r}}_{n}}-{{\mathbf{r}}_{A}}\right) ,
\]%
\[
z=\mathbf{\hat{z}}\cdot \left( {{\mathbf{r}}_{n}}-{{\mathbf{r}}_{A}}\right)
\]
where the subscript ``n'' means ``new''.
Calculate the rotation angle, i.e. the value of value-finding function.
	\[\text{value}(P)=\arctan2(y,x)\]	
Figure \ref{Fig:ClusterFig3} shows the scheme for the rotation of triangle ABC.
The procedure for constructing range-evaluation function is as follows: calculate the position $\mathbf{r}_T$ of the tangent point T of rolling-ball and the circumsphere of the cube b. The needed relations are as below:
\[
{{r}^{2}}={{\left\vert {{\mathbf{r}}_{T}}-{{\mathbf{r}}_{A}}\right\vert }^{2}%
}={{\left\vert {{\mathbf{r}}_{T}}-{{\mathbf{r}}_{B}}\right\vert }^{2}}\text{,%
}
\]%
\[
\left\vert {{\mathbf{r}}_{T}}-{{\mathbf{c}}_{b}}\right\vert =r+\sqrt{3}{{d}%
_{b}}\text{{.}}
\]
$\mathbf{r}_T$ has two roots, $\mathbf{r}_{ML}$ and $\mathbf{r}_{mL}$. The
corresponding tangent points are ML and mL. The range-evaluation functions are as follows:
\[
M\left( b\right) =\left\{
\begin{array}{cc}
\text{value}\left( \text{ML}\right) \text{,} & \text{if above equations have
real solutions} \\
\infty , & \text{else.}%
\end{array}%
\right.
\]%
\[
m\left( b\right) =\left\{
\begin{array}{cc}
\text{value}\left( \text{mL}\right) \text{,} & \text{if above equations have
real solutions} \\
\infty , & \text{else.}%
\end{array}%
\right.
\]

\begin{figure*}[tbp]
\centering
\includegraphics*[scale=0.3,angle=0,
bbllx=0pt,bblly=0pt,bburx=596pt,bbury=519pt
]{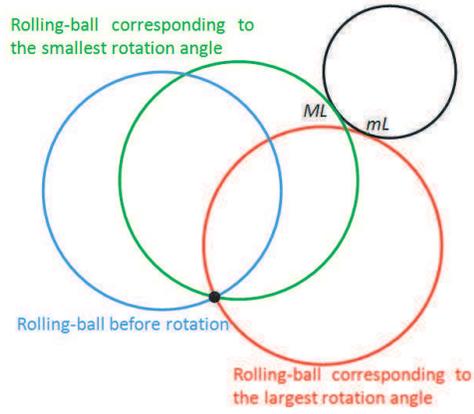}
\caption{(Color online)Scheme for the two tangent cases between the rolling-ball and the circumsphere of cube b.
(Adopted with permission from Ref. \cite{MD-SHT-Book}.)
}
\label{Fig:ClusterFig4}
\end{figure*}

Figure \ref{Fig:ClusterFig4} shows the cross section picture for two different cases. In each case, the circumspheres of cube b and the rolling-ball are tangent to each other. Here, the back circle is for the circumsphere of cube b. The blue, green and red circles are for rolling-balls. ML and mL are for corresponding tangent points. The rotation angle of rolling-ball takes its smallest value when the tangent point is ML. It takes its maximum when the tangent point is mL.

(II) Minimum searcher MS2: Given a point $\mathbf{r}_0$, search for its  nearest point P in point tree. The value-finding function is
	\[value(P)=\left| {{\mathbf{r}}_{P}}-{{\mathbf{r}}_{0}} \right|\]	
where $\mathbf{r}_P$ is coordinate of point P. The range-evaluation functions are
\[
M(b)=\left\vert {{\mathbf{c}}_{b}}-{{\mathbf{r}}_{0}}\right\vert +\sqrt{3}{{d%
}_{b}}\text{{,}}
\]%
\[
m(b)=\max (\left\vert {{\mathbf{c}}_{b}}-{{\mathbf{r}}_{0}}\right\vert -%
\sqrt{3}{{d}_{b}},0)\text{.}
\]

(III) Minimum searcher MS3: Given a point and a rotation axis, search for the first point met by the rolling-ball with fixed size in point tree. The algorithm is nearly the same as for MS1. We do not repeat here.

(IV) Conditional searcher CS1: Given two points, P1 and P2, search for segment BD, whose vertexes are P1 and P2, in segment tree. The conditional function is as follows:
\[
maycontain(b)=\left\{
\begin{array}{cc}
\text{true,} & {{P}_{1}},{{P}_{2}}\in S \\
\text{false,} & \text{else}%
\end{array}%
\right.
\]
The circumsphere S of cube b is used for identification. The identification function is as follows:
\[
s=\left\vert \sum\limits_{i\in \text{neighbour}}{({{\mathbf{r}}_{i}}-{{%
\mathbf{r}}_{0}})}\right\vert
\]

The rolling-ball algorithm is as follows: (I) Initialization: Generate a point tree, \emph{tp}, from given discrete points. Set the radius of rolling-ball as $r$ and the center as P0. Using the searcher MS2 to search for the nearest point P1 of P0 in tree \emph{tp}. Use searcher MS3 to search for a point P2 which is the first point met by the rolling-ball rotating around  $\mathbf{x}$ axis in tree \emph{tp}. Use searcher MS3 to search for a point P3 which is the first point met by the rolling-ball rotating around the direction of segment P1P2 in tree \emph{tp}. Generate a triangle from P1, P2 and P3. Construct a triangle tree \emph{tt} and a segment tree \emph{tb}.  Put the triangle P1P2P3 into \emph{tp} and put its three edges into \emph{tb}. (II) Interface construction: Check whether or not the tree \emph{tb} is null. If yes, exit. If not, cut down an edge AB of triangle ABC. Use the searcher MS1 to search, in tree \emph{tb}, for a point P to make smallest the rotation angle of circumsphere of triangle ABC. Here, AB is the rotation axis. Construct a triangle BAP, and put it into the triangle tree \emph{tt}. Use CS1 searcher to search, in tree \emph{tb}, for an segment L whose vertexes are point B and P.  If L exists, cut it down from \emph{tb}, and then delete it. If not, generate an segment PB and put it into \emph{tb}. Perform the same operations to points P and A. (III) Go back to step (II). The surface composed of triangles contained in the tree \emph{tt} is just the physical interface that we need.

\begin{figure*}[tbp]
\centering
\includegraphics*[scale=0.5,angle=0,
bbllx=0pt,bblly=0pt,bburx=593pt,bbury=300pt
]{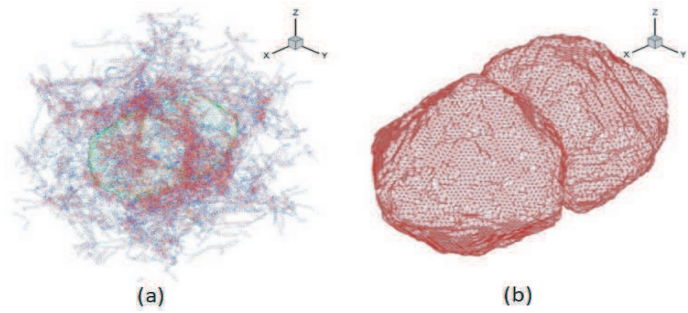}
\caption{(Color online)Interface of voids constructed from discrete points. (a) discrete points; (b)constructed interface.
(Adopted with permission from Ref. \cite{MD-SHT-Book}.)
}
\label{Fig:ClusterFig5}
\end{figure*}

\begin{figure*}[tbp]
\centering
\includegraphics*[scale=0.5,angle=0,
bbllx=0pt,bblly=0pt,bburx=595pt,bbury=322pt
]{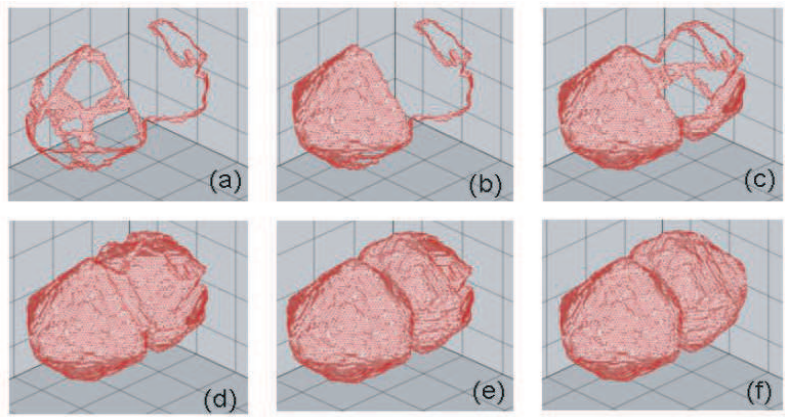}
\caption{(Color online)Process of constructing interface of voids from discrete points.
(Adopted with permission from Ref. \cite{MD-SHT-Book}.)
}
\label{Fig:ClusterFig6}
\end{figure*}

The interface of voids constructed from discrete points is shown Fig. \ref{Fig:ClusterFig5}. The process of constructing interface of voids from discrete points is shown in Fig. \ref{Fig:ClusterFig6}.

\paragraph{Delaunay division}

There have been a number of
 algorithms to construct Delaunay triangles in two-dimensional space and
 tetrahedrons in three-dimensional space. The complexities of most algorithms come from the searching procedures of disordered data. Here, we introduce an algorithm based on the GISO. The algorithm is simple and intuitive. It is convenient to extend to higher dimensional space. The algorithm for constructing Delaunay division from discrete points is as follows: Firstly, create a point tree \emph{tp}, put all the points into \emph{tp}. The largest cube of the \emph{tp} is centered at $r_0$ and has the size $a$. Construct a largest tetrahedron $T$ which contains all the points in the local region. This tetrahedron is just the most initial Delaunay tetrahedron. This tetrahedron can be chosen as regular tetrahedron centered at $r_0$ with enough large size, e.g. $20 a$. This ensue that all the points in the tree \emph{tp} are within $T$. Create a tetrahedron tree \emph{tt}, put the first tetrahedron  $T$ into the tree \emph{tt}.  Secondly, add each point to adjust the Delaunay division. Take off a point P from \emph{tp}, search the tetrahedrons in \emph{tt} whose circumsphere contains P. We use a set ,Q, to denote all the tetrahedrons checked out in above procedure. The tetrahedrons of Q forms a polyhedron. Remove these tetrahedrons from \emph{tt}. Construct new tetrahedrons by linking each triangle surface of the polyhedron and point P. Put these new tetrahedrons into \emph{tt}. Remove point P from \emph{tp}. Repeat the procedure until \emph{tp} is null. Finally, search for the tetrahedrons which share surface with T, and remove them from \emph{tt}. Then, the all the tetrahedron in the \emph{tt} construct the Delaunay division.

\begin{figure*}[tbp]
\centering
\includegraphics*[scale=0.6,angle=0,
bbllx=0pt,bblly=0pt,bburx=595pt,bbury=156pt
]{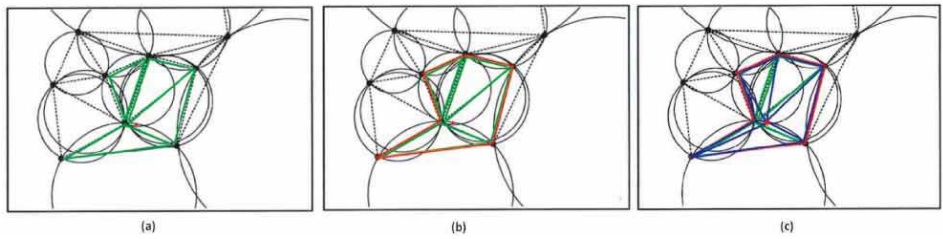}
\caption{(Color online)Three steps to add a new point to a two-dimensional Delaunay division. (a) Finding the triangles whose circumcircle contains the newly added point P; (b) Removing the internal lines of these triangles, retaining the external ones; (c) Connecting each left line with point P to form new triangles.
(Adopted with permission from Ref. \cite{MD-SHT-Book}. The grey-level version is published in
Ref.\cite{Zhang-SciCN2010}.)
}
\label{Fig:ClusterFig7}
\end{figure*}

\begin{figure*}[tbp]
\centering
\includegraphics*[scale=0.3,angle=0,
bbllx=0pt,bblly=0pt,bburx=595pt,bbury=637pt
]{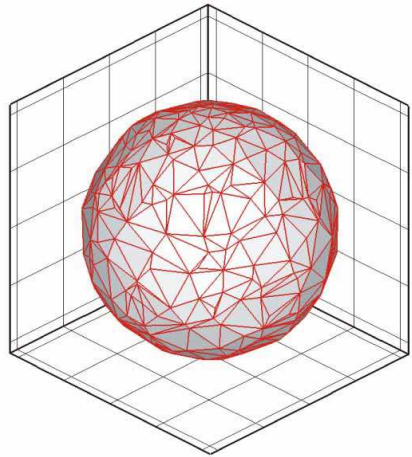}
\caption{(Color online)Delaunay division constructed from randomly distributed discrete points in a three- dimensional spherical region.
(Adopted with permission from Ref. \cite{MD-SHT-Book}. The grey-level version is published in
Ref.\cite{Zhang-SciCN2010}.)
}
\label{Fig:ClusterFig8}
\end{figure*}

 Figure \ref{Fig:ClusterFig7} shows the steps for adding a two-dimensional point and re-dividing the space, where the red point stands for the newly added point P, the green triangles in figure \ref{Fig:ClusterFig7}(a) are for the to-be-adjusted-triangles, the red segments in figure \ref{Fig:ClusterFig7}(b) are retained boundary segments, the blue segments in figure \ref{Fig:ClusterFig7}(c) are segments connecting point P and  vertexes of boundary. In the three-dimensional case, we need only to replace the triangle with a tetrahedron, replace the line with a triangular face, and replace the triangle with a tetrahedron. Figure \ref{Fig:ClusterFig8} shows the Delaunay division constructed from randomly distributed discrete points in a three-dimensional spherical region.

\paragraph{Cluster construction and analysis method}

 For discrete points, a cluster are defined as a group of points which have short distance. The critical distance is denoted as $d$, which is also the minimum distance between any two clusters. The algorithm to construct a cluster is as follows:
 Firstly, construct tetrahedron tree \emph{tt} containing Delaunay tetrahedrons using the Delaunay division algorithm.
  Search in \emph{tt} for the tetrahedron whose smallest edge is longer than $d$, and remove them from \emph{tt}. Divide the remaining tetrahedrons in \emph{tt} into different sets according to their connectivities. Create a cluster tree \emph{tcl} to contain all the clusters.  Secondly, create a tetrahedron tree \emph{tc} to contain all the tetrahedrons in the first cluster. For convenience of description, \emph{tc} is also referred to as a cluster.  Create a triangle tree \emph{ttr} to contain the inner surfaces of the clusters. Take off one tetrahedron T off \emph{tt}, put T into \emph{tc}, put each of its four triangle surfaces into \emph{ttr}. Take off triangle $tr$ from \emph{ttr},search in \emph{tt} for the tetrahedron, say T1, whose triangle surface coincides with $tr$. If find T1, remove it from \emph{tt} and put it into cluster \emph{tc}. Put all the surfaces except $tr$ into \emph{ttr}. Repeat the procedure until \emph{ttr} is null. Up to this step, the first cluster \emph{tc} is completely constructed. Put the cluster \emph{tc} into cluster tree \emph{tcl}.
  Then, construct a new cluster and put it into \emph{tcl} until \emph{ttr} is null.

\begin{figure*}[tbp]
\centering
\includegraphics*[scale=0.5,angle=0,
bbllx=0pt,bblly=0pt,bburx=580pt,bbury=263pt
]{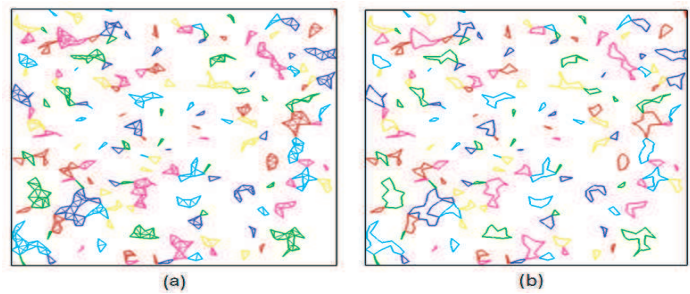}
\caption{(Color online) Cluster structure formed from 1000 random discrete points in a two-dimensional square area [0,1]$\times$[0,1]. (a) A cluster; (b) the corresponding cluster boundary.
(Adopted with permission from Ref. \cite{MD-SHT-Book}. The grey-level version is published in
Ref.\cite{Zhang-SciCN2010}.)
}
\label{Fig:ClusterFig9}
\end{figure*}

Figure \ref{Fig:ClusterFig9} shows the clusters constructed with random points in two-dimensional space.

\paragraph{Identification methods of defect atoms}

Here we introduce three methods.

(1) Excess energy method

In this method, the defect atoms are defined as those whose potential energies exceeds a critical value. This method requests that the MD simulation outputs not only atom positions but also the inter-atomic potentials.

(2) Centro-Symmetry Parameter method

In Centro-Symmetry Parameter(CSP) method\cite{Kelchner}, the geometrical symmetry of the collection of nearest atoms of an atom is used to identify defect atoms. All atoms in the perfect crystal are in the geometrical center of its nearest atoms, but the defect atoms are not. Therefore, an order parameter is defined as follows:
\[
s=\left\vert \sum\limits_{i\in \text{neighbour}}{({{\mathbf{r}}_{i}}-{{%
\mathbf{r}}_{0}})}\right\vert
\]
Atoms whose order parameter $s$ is greater than a critical value $s_c$ are defect atoms.

(3) Bond-pair analysis method

\begin{figure*}[tbp]
\centering
\includegraphics*[scale=0.6,angle=0,
bbllx=0pt,bblly=0pt,bburx=578pt,bbury=185pt
]{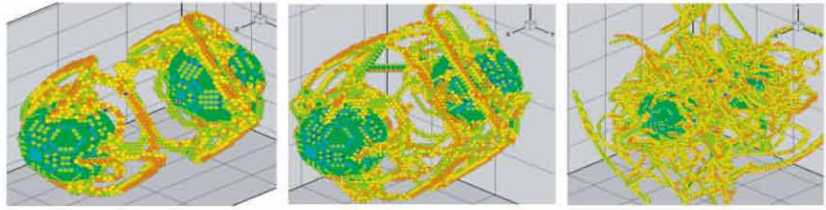}
\caption{(Color online)Voids surfaces and dislocations identified by bond-pair analysis.
(Adopted with permission from Ref. \cite{MD-SHT-Book}.)
}
\label{Fig:ClusterFig10}
\end{figure*}

The CSP and excess energy methods can be used to distinguish defect atoms, but can not be easily used to identify types of the defects atoms. The bond-pair analysis (BPA) \cite{Faken} based on local topological connections can be used to identify more accurately the atom type. The idea of BPA is as follows: The bond type is marked in terms of the connections among all the atoms bonding with the two atoms.
An atom type is marked in terms of all the bonds of itself. Figure \ref{Fig:ClusterFig10} shows the  voids surfaces and dislocations identified by bond-pair analysis.

\paragraph{Packing-sculpting method for constructing object surface}

In computational geometry, it is an important issue to construct object surface from disorder points. The current algorithms can be categorized into four groups \cite{Mencl}: space partition method \cite{Boissonant}, distance function method \cite{Hoppe}, deformation method \cite{Zhao}, and growth method \cite{Bernardini}. Space partition is generally based on Delaunay division. The outer surface is generated by removing some Delaunay mesh in the sculpting method. The packing-sculpting method presented below is an intuitive method. The out surface is constructed by dynamically sculpting the packing convex hull. As for the packing, the basic idea is as follows:
Firstly, create a point tree \emph{tp}, and put all the points into the tree \emph{tp}. Search in \emph{tp} for the point P1 which has largest $x$- coordinate; remove P1 from \emph{tp}. Define a plane passing P1 perpendicular to $x$- axis; rotate the plane around the axis which passes P1 and along the $y$ direction; search for the first point P2 it meets; remove it from \emph{tp}. Define a plane passing points P1 and P2; rotate the plane around P1P2; search for the first point P3 it meets. Remove P3 from \emph{tp}. Create triangle P1P2P3 by linking P1,P2,P3. That is the first triangle surface. Create a triangle tree \emph{tt}, and put triangle P1P2P3 into it. Create a boundary tree \emph{tb} and put P1P2,P2P3,P3P1 into it.  Secondly, take one boundary edge AB in \emph{tb}, define a half plane which is on the same plane as the triangle surface passing AB. This half plane includes the region opposite to the triangle. Rotate the half plane around AB, search in \emph{tp} for the point P it first meets. Create triangle BAP and put it into \emph{tt}. Find in \emph{tb} for each edge, AB,PA and BP. If find one,  remove it from \emph{tb}; if not, put its reverse edge, BA,AP or PB, into \emph{tb}. Repeat the procedure until \emph{tb} is null. In this way, all the packing surfaces are included in \emph{tp}.

As for the sculpting, the basic idea is as follows:
Define a size $s$ which represents the sculpting depth.  Firstly, create a triangle tree \emph{ts} to contain triangle surface. Take a triangle surface ABC from surface tree \emph{tt}. Define a sphere B which passes vertices A,B,C of triangle ABC and has a large enough radius,e.g.,$10^{100}$. Keep the sphere B passing the points, A,B and C, decrease the radius of B, search in \emph{tp} for the first point P that sphere B meets. If P can not be found before the radius shrink to be less than $s$, it means that the sculpting from triangle surface ABC can not be done any more, remove ABC from \emph{tt} and put it into \emph{ts}. If P exists, the sculpting from triangle surface ABC can be done, remove ABC from \emph{tt}. Find in \emph{tt} for each Triangle surface, ABC,CBP,BAP and ACP. If find one, remove it from \emph{tt}; if not, put its reverse triangle, BAC,CBP,ABP or CAP into \emph{tt}. Repeat the procedure until \emph{tt} is null. In this way, all the surfaces are included in \emph{ts}.

\begin{figure*}[tbp]
\centering
\includegraphics*[scale=0.5,angle=0,
bbllx=0pt,bblly=0pt,bburx=594pt,bbury=393pt
]{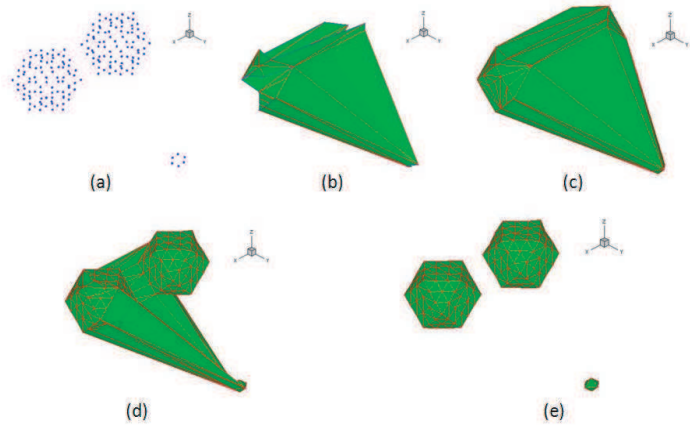}
\caption{(Color online)The procedure of packing-sculpting algorithm. (a) discrete points; (b) the mid of packing procedure; (c) packing convex hull; (d) sculpting procedure; (e) object surface.
(Adopted with permission from Ref. \cite{MD-SHT-Book}.)
}
\label{Fig:ClusterFig11}
\end{figure*}

 The procedure of packing-sculpting algorithm is shown in Fig. \ref{Fig:ClusterFig11}.

\paragraph{Calculation of the Burgers vector of dislocation loop}

\begin{figure*}[tbp]
\centering
\includegraphics*[scale=0.6,angle=0,
bbllx=0pt,bblly=0pt,bburx=593pt,bbury=251pt
]{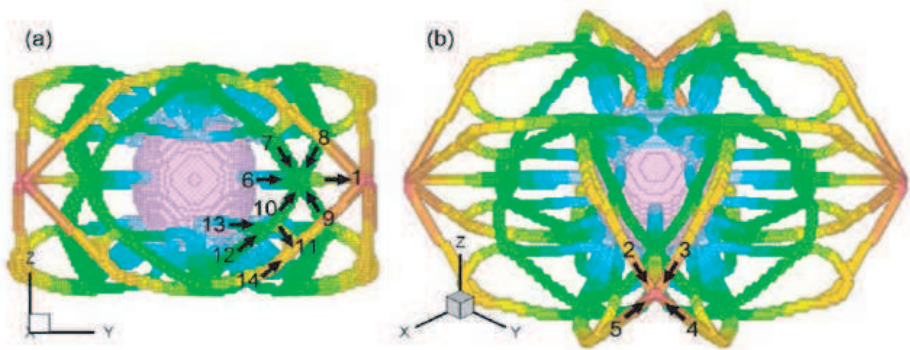}
\caption{(Color online)Burgers vectors for dislocations from a nanovoid in a deformed crystal under a uniaxial tensile loading. 
(a) Front view; (b) perspective view. The colors are based
on the distance to the void center. 
Specifically,
$\mathbf{b}_{1} =[0,0,0]$;
$\mathbf{b}_{2}=[1,2,-1]/6$;$\mathbf{b}_{3}=[-2,-1,1]$;
$\mathbf{b}_{4}=[2,1,1]/6$;$\mathbf{b}_{5}=[-1,-2,-1]/6$;
$\mathbf{b}_{6}=[0,0,0]$; $\mathbf{b}_{7} =[-1,1,-2]/6$;$\mathbf{b}_{8}=[-1,1,2]/6$;
$\mathbf{b}_{9}=[1,-1,2]/6$; $\mathbf{b}_{10} =[1,-1,-2]/6$;$\mathbf{b}_{11}=[-2,-1,1]/6$;
$\mathbf{b}_{12}=[1,-1,-2]/6$; $\mathbf{b}_{13} =[-2,-1,1]/6$;$\mathbf{b}_{14}=[1,-1,-2]/6$.
(Adopted with permission from Ref. \cite{WangZhang2013}.)
}
\label{Fig:ClusterFig12}
\end{figure*}

\begin{figure*}[tbp]
\centering
\includegraphics*[scale=0.5,angle=0,
bbllx=0pt,bblly=0pt,bburx=593pt,bbury=311pt
]{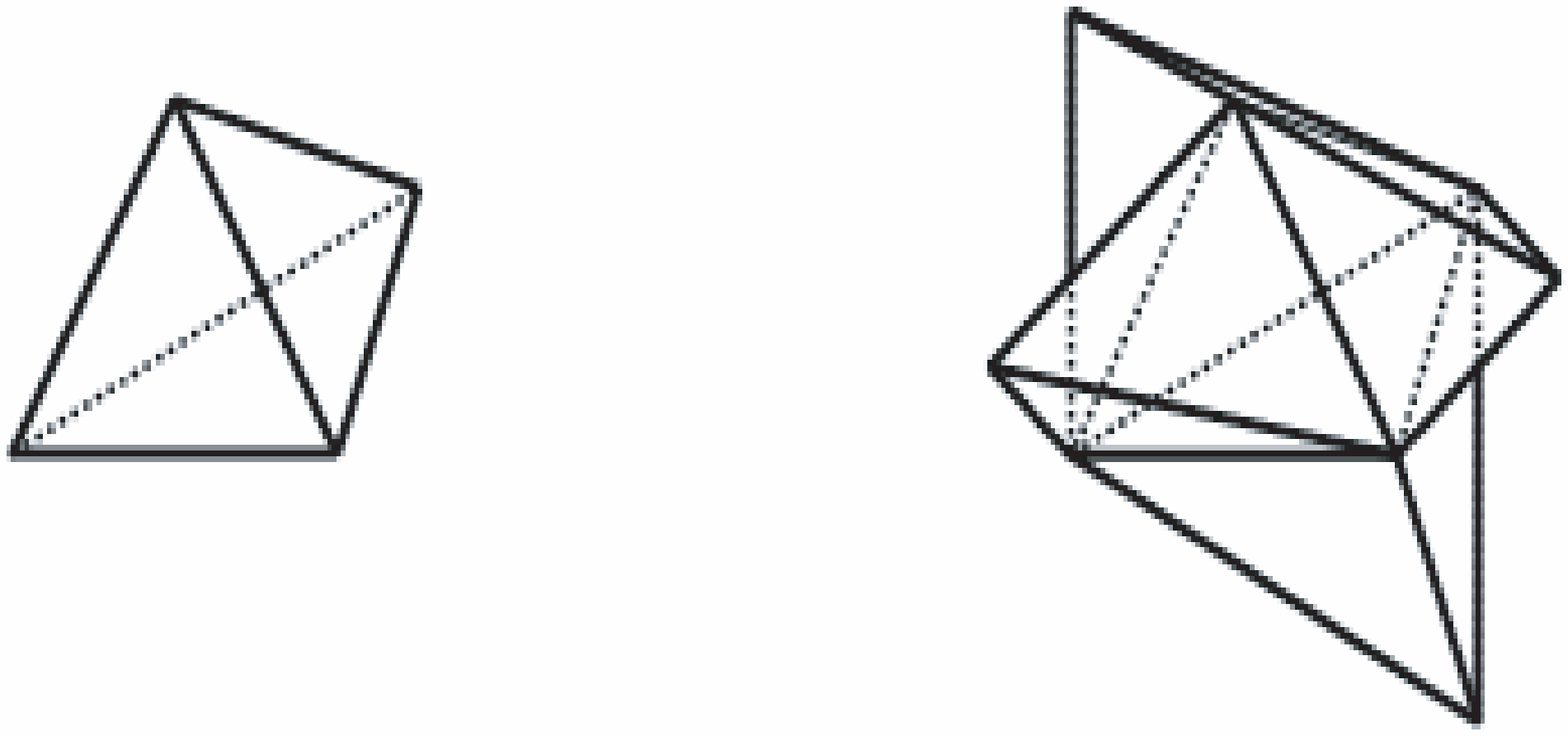}
\caption{Thompson's tetrahedron and its mirrors.
(Adopted with permission from Ref. \cite{MD-SHT-Book}.)
}
\label{Fig:ClusterFig13}
\end{figure*}

Based on the Thompson's tetrahedron, a Frank scheme is developed to calculate the Burgers vector of dislocations in a fcc crystal during its plastic deformation. A Burgers circuit is located firstly in a deformed crystal with a reference circle surrounding one or more dislocations. The atom-to-atom sequence, in a dislocation-free crystal, corresponding to the Burgers circuit is determined from the edge vectors of the Thompson's tetrahedron and its mirrors, instead of a local reference lattice. The Burgers vector can be calculated via summing over the vectors connecting neighboring atoms in the Burgers circuit. As long as the same dislocations are surrounded, the final Burgers vector obtained by its Frank definition is accurate. The present method is validated in determining the Burgers vectors for the dissociation of a perfect dislocation and for the complex reactions of the dislocations from a nanovoid in a deformed crystal under a uniaxial tensile loading\cite{WangZhang2013}. (See Figs. \ref{Fig:ClusterFig12}-\ref{Fig:ClusterFig13})

%%%%%%%%%%%%%%%%%%%%%%%%%%%%%%%%%%%%%%%%%%%%%%%%%%%%%%%%%%%%%%%%%%%%%%%%%%%%%%%%
%%%%%%%%%%%%%%%%%%%%%%%%%%%%%%%%%%%%%%%%%%%%%%%%%%%%%%%%%%%%%%%%%%%%%%%%%%%%%%%%

%%%%%%%%%%%%%%%%%%%%%%%%%%%%%%%%%%%%%%%%%%%%%%%%%%%%%%%%%%%%%%%%%%%%%%%%%%%%%%%%
%%%%%%%%%%%%%%%%%%%%%%%%%%%%%%%%%%%%%%%%%%%%%%%%%%%%%%%%%%%%%%%%%%%%%%%%%%%%%%%%

%%%%%%%%%%%%%%%%%%%%%%%%%%%%%%%%%%%%%%%%%%%%%%%%%%%%%%%%%%%%%%%%%%%%%%%%%%%%%%%%
%%%%%%%%%%%%%%%%%%%%%%%%%%%%%%%%%%%%%%%%%%%%%%%%%%%%%%%%%%%%%%%%%%%%%%%%%%%%%%%%

%%%%%%%%%%%%%%%%%%%%%%%%%%%%%%%%%%%%%%%%%%%%%%%%%%%%%%%%%%%%%%%%%%%%%%%%%%%%%%%%
%%%%%%%%%%%%%%%%%%%%%%%%%%%%%%%%%%%%%%%%%%%%%%%%%%%%%%%%%%%%%%%%%%%%%%%%%%%%%%%%

%%%%%%%%%%%%%%%%%%%%%%%%%%%%%%%%%%%%%%%%%%%%%%%%%%%%%%%%%%%%%%%%%%%%%%%%%%%%%%%%
%%%%%%%%%%%%%%%%%%%%%%%%%%%%%%%%%%%%%%%%%%%%%%%%%%%%%%%%%%%%%%%%%%%%%%%%%%%%%%%%

%%%%%%%%%%%%%%%%%%%%%%%%%%%%%%%%%%%%%%%%%%%%%%%%%%%%%%%%%%%%%%%%%%%%%%%%%%%%%%%%
%%%%%%%%%%%%%%%%%%%%%%%%%%%%%%%%%%%%%%%%%%%%%%%%%%%%%%%%%%%%%%%%%%%%%%%%%%%%%%%%

%%%%%%%%%%%%%%%%%%%%%%%%%%%%%%%%%%%%%%%%%%%%%%%%%%%%%%%%%%%%%%%%%%%%%%%%%%%%%%%%
%%%%%%%%%%%%%%%%%%%%%%%%%%%%%%%%%%%%%%%%%%%%%%%%%%%%%%%%%%%%%%%%%%%%%%%%%%%%%%%%

%%%%%%%%%%%%%%%%%%%%%%%%%%%%%%%%%%%%%%%%%%%%%%%%%%%%%%%%%%%%%%%%%%%%%%%%%%%%%%%%
%%%%%%%%%%%%%%%%%%%%%%%%%%%%%%%%%%%%%%%%%%%%%%%%%%%%%%%%%%%%%%%%%%%%%%%%%%%%%%%%

%%%%%%%%%%%%%%%%%%%%%%%%%%%%%%%%%%%%%%%%%%%%%%%%%%%%%%%%%%%%%%%%%%%%%%%%%%%%%%%%
%%%%%%%%%%%%%%%%%%%%%%%%%%%%%%%%%%%%%%%%%%%%%%%%%%%%%%%%%%%%%%%%%%%%%%%%%%%%%%%%

%%%%%%%%%%%%%%%%%%%%%%%%%%%%%%%%%%%%%%%%%%%%%%%%%%%%%%%%%%%%%%%%%%%%%%%%%%%%%%%%
%%%%%%%%%%%%%%%%%%%%%%%%%%%%%%%%%%%%%%%%%%%%%%%%%%%%%%%%%%%%%%%%%%%%%%%%%%%%%%%%

%%%%%%%%%%%%%%%%%%%%%%%%%%%%%%%%%%%%%%%%%%%%%%%%%%%%%%%%%%%%%%%%%%%%%%%%%%%%%%%%
%%%%%%%%%%%%%%%%%%%%%%%%%%%%%%%%%%%%%%%%%%%%%%%%%%%%%%%%%%%%%%%%%%%%%%%%%%%%%%%%

%%%%%%%%%%%%%%%%%%%%%%%%%%%%%%%%%%%%%%%%%%%%%%%%%%%%%%%%%%%%%%%%%%%%%%%%%%%%%%%%
%%%%%%%%%%%%%%%%%%%%%%%%%%%%%%%%%%%%%%%%%%%%%%%%%%%%%%%%%%%%%%%%%%%%%%%%%%%%%%%%

%%%%%%%%%%%%%%%%%%%%%%%%%%%%%%%%%%%%%%%%%%%%%%%%%%%%%%%%%%%%%%%%%%%%%%%%%%%%%%%%
%%%%%%%%%%%%%%%%%%%%%%%%%%%%%%%%%%%%%%%%%%%%%%%%%%%%%%%%%%%%%%%%%%%%%%%%%%%%%%%%

\section{Numerical experiments and observations}

Our investigations can be roughly classified into three groups, microscopic,
mesoscopic and macroscopic scales. As for the microscopic scale, what we
probed are limited to the cases which can be simulated by the MD
simulations. As for the mescoscopic scale, both the solid and fluid models are used. Based on the solid model, what we probed are limited to the cases with only one or a few cavities where the continuum theory works and the MPM can be used. The fluid model here is mainly referred to the DBM. By using the DBM, we can study both the hydrodynamic non-equilibrium and thermodynamic non-equilibrium behaviors, especially around the interfaces.
Both the MPM and DBM are also applied to simulate behaviors in the macroscopic scale.

\subsection{ MPM investigations: Global behaviors}

The global behaviors of shocked porous material based on MPM simulations are referred to those averaged or statistical behaviors\cite{Xu-JPD2009,Xu-CTP2009A,Xu-SciCN2010,Xu-CTP2009B,Xu-CAMWA2011,Xu-PhysScr2010}. Here ``global'' is relative to ``local''. The latter is referred to the case with only a single or a few cavities, while the former is referred to the case with several thousands or more.
In our numerical experiments the porous material is fabricated by a
solid material body with an amount of cavities randomly embedded. The particle feature of MPM makes easy the flexible setting of the initial configuration. We denote the mean density of the porous body as $\rho $ and the density of the solid portion as $\rho _{0}$. The porosity is defined as $\Delta = 1-\delta $, where $\delta = \rho/\rho_{0} $. The porosity $\Delta $ is controlled by the total number $N_{void}$ and mean size $r$ of voids embedded. In our numerical experiments, there are two kinds of equivalent shock loading schemes, colliding with a body with symmetric configuration or colliding with a rigid wall in the same material. In the studies on global behaviors, the shock is loaded via colliding with the rigid wall. In the studies on local behaviors, the shock is loaded via colliding with a body with symmetric configuration. The gravity effects are neglected. The rigid wall is located horizontally and keeps static at the bottom where $y=0$, the target porous body is on the upper side of the rigid wall and moves towards the rigid wall at a velocity with the size $v_{init}$. We start to count the time when the porous body begins to touch the rigid wall.  At the left and right boundaries we use periodic boundary conditions. This treatment means that the real system under consideration is composed of many of the simulated ones aligned periodically in the horizontal direction.

The sample material for MPM simulations in this paper is fixed at the metal aluminum. The corresponding parameters
are as follows: $E=69$ Mpa, $\nu =0.33$, $\rho_{0}=2700$ kg/m$^{3}$, $\sigma _{Y0}=120$
Mpa, $E_{\tan }=384$ MPa, $k=237$ W/(m$\cdot $K), $c_{0}=5.35$ km/s, $\lambda=1.34$, $c_{v}=880$
J/(Kg$\cdot $K) and $\gamma_0=1.96$ when the
pressure is below $270$ GPa. The initial temperature of the material is fixed at 300
K.

\subsubsection{Mean values and their fluctuations}

In this part of the studies, the effects of porosity and shock strength are the main concerns. For cases with the
same porosity, the effects of mean-cavity-size are further studied. Main observations are as follows: the local volume dissipation and turbulence mixing are two important mechanisms for transformation of kinetic energy to internal energy. In the cases with very small porosities, the shocked portion may arrive at a dynamical steady state; the cavities within the downstream portion reflect back rarefactive waves and make slight oscillations of mean density and pressure; in the cases with the same porosity, a larger mean-cavity-size results in a higher mean temperature. In the cases with high porosities, the hydrodynamic quantities vary with time during the whole shock-loading procedure: after the initial period, the mean density and pressure decrease, but the temperature increases with a higher rate. The distributions of local pressure, temperature, density and particle-velocity are generally non-Gaussian and vary with time. The changing rates are dependent on the shock strength, porosity value as well as the mean-cavity-size. The porosity effects becomes more pronounced with increasing shocking strength\cite{Xu-CTP2009A}.
We show some specific numerical results based on two-dimensional simulations below.

\begin{figure*}[tbp]
\centering
\includegraphics*[scale=0.7,angle=0,
bbllx=0pt,bblly=0pt,bburx=429pt,bbury=212pt
]{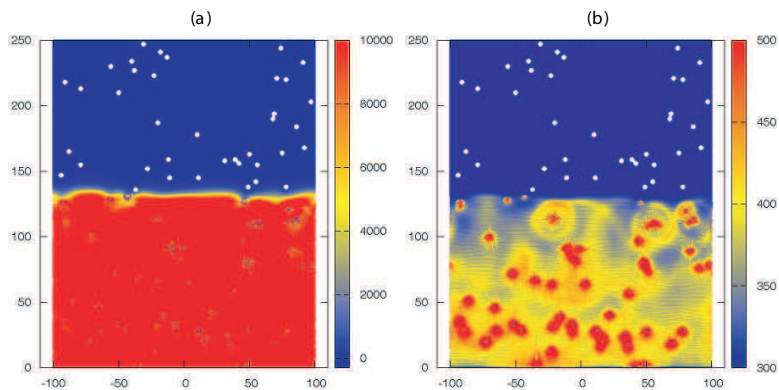}
\caption{(Color online) Snapshots of the shocked porous metal. $\protect%
\delta=1.03$, $\protect\Delta = 0.029$, t=250 ns. (a) Contour of pressure, (b) contour of temperature.
The unit of length in this figure is 10 $\protect\mu$m. From blue to red,
the contour value increases. The unit of contour is Mpa in (a) and is K in
(b). The initial velocities of the flyer and target are $\pm v_{init} = \pm
1000$ m/s in this case.
(The grey-scale version is published in Ref. \cite{Xu-CTP2009A}.)
}
\label{Fig:Figg1}
\end{figure*}

The computational unit here is $2$ mm in width, as shown in Fig.\ref{Fig:Figg1}. Since mainly interested in the loading procedure, the height of porous material is set a large enough value so that the rarefactive waves reflected from the upper free surfaces do not significantly influence the physical process within the time scale under investigation.

Two snapshots are shown in Fig. \ref{Fig:Figg1}, where Fig.(a) shows the contour of pressure and (b) shows the contour of temperature. Different from the cases with perfect solid material, no stable shock wave exists in the
porous materials. When the initial shock wave arrives at the first cavity, rarefactive
wave is reflected back and propagates within the compressed portion, which
destroys the original possible equilibrium state there. The shock waves at the two sides continue to propagate forward and meet again in front of the cavity. The waves begin to become complex. When a compressive wave meets a new cavity, similar behaviors occur. In this way, the waves in the porous material become very complex. For the
convenience of description, the concept, shock wave, is still used as a coarse-grained description. Correspondingly, the values of physical quantities, such as the pressure, the
particle velocity, temperature, density, etc, are corresponding
mean values calculated in a region $\Omega$ with $y_1 \leq y \leq y_2$.

\paragraph{Cases with low porosity}

\begin{figure*}[tbp]
\centering
\includegraphics*[ scale=0.7,angle=0,
bbllx=0pt,bblly=0pt,bburx=321pt,bbury=218pt
]{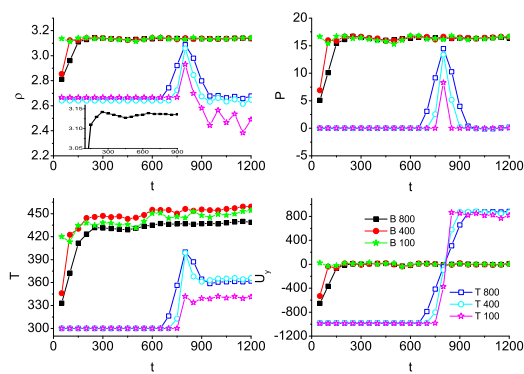}
\caption{(Color online) Variations of mean density, pressure, temperature
and particle velocity with time. The height of the measured domain are h=
800 $\protect\mu$m, 400$\protect\mu$m and 100 $\protect\mu$m, respectively,
as shown in the legends. ``B" and ``T" in the legends means the measured
domains are at the bottom and top of the target body, respectively. The
units of density, pressure, temperature, particle velocity and time are g/cm$%
^3$, Gpa, K, m/s and ns, respectively.
(The grey-scale version is published in Ref. \cite{Xu-CTP2009A}.)
}
\label{Fig:Figg2}
\end{figure*}

\begin{figure*}[tbp]
\centering
\includegraphics*[ scale=0.5,angle=0,
bbllx=0pt,bblly=242pt,bburx=595pt,bbury=607pt
]{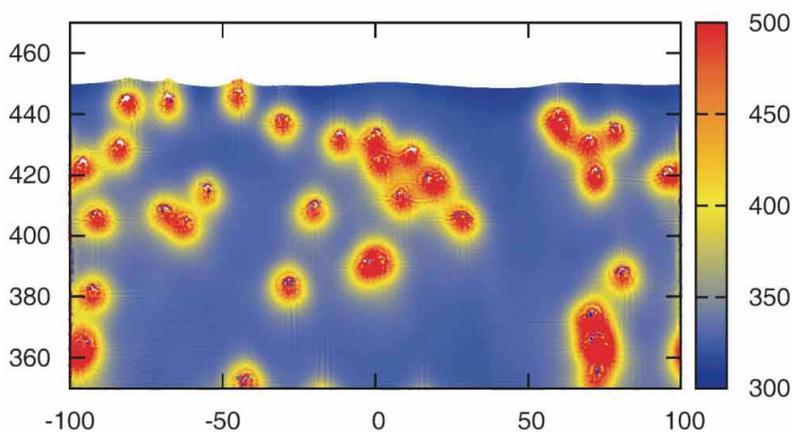}
\caption{(Color online) Configuration with temperature contour at time
t=1.15 $\protect\mu$s. Other parameters are referred to Fig.\protect\ref%
{Fig:Figg1}. and Fig.\protect\ref{Fig:Figg2}. The unit of temperature is K.
(The grey-scale version is published in Ref. \cite{Xu-CTP2009A}.)
}
\label{Fig:Figg3}
\end{figure*}

Figure \ref{Fig:Figg2} shows the mean density, pressure, temperature and particle velocity versus time for a case where  $\protect\Delta = 0.029$ ($\protect\delta =1.03$), $r =50\mu$m, $v_{init} = 1000$ m/s and the height of the porous material is 5 mm. These values are dynamically measured in a bottom and a top domains, respectively. For the bottom domain, we choose $y_1 = 100\mu$m, and for the top domain, $y_2$ takes the $y$-coordinate of the highest
material-particle. Three sets of measured results are shown. The heights of the measured domain are chosen as $h=800 \mu$m, $400 \mu$m and $100 \mu$m, respectively. The lines with solid symbols are for measured values from the
bottom domain, the lines with empty symbols show measured values from the top.
Simulation results show that, for the case of $h=800 \mu$m, when the shock waves propagate within the bottom domain $\Omega_b$, the measured mean density, pressure and temperature increase nearly linearly with time, up to about $t=150$ ns. After that the temperature further to increase with a much lower increasing rate. The three quantities arrive at their first maximum values, 3.14g/cm$^3$, 16.7GPa, and 432K, at about $t=250$ ns. At this time
the shock front has passed the downstream boundary, $y_2 = 810 \mu$m, of the
measured domain. (See Fig. \ref{Fig:Figg1}.) The concave regions in the $\rho$-,$P$-,$T$-curves at about $t = 450$ ns shows an unloading phenomenon of the compressive waves, i.e., rarefactive waves reflect back from the
cavities downstream neighboring to the measured domain. The values of $\rho$
and $P$ increase and recover to their (nearly) steady values after that, but the
temperature further to increase. The secondary loading-phenomenon is due to
the collisions of the upstream and downstream walls of cavities. During the following period the density and pressure keep nearly constants, while the temperature still increases very slowly. The weak
fluctuations in the density, pressure and temperature curves after $t=650$ ns result from the inputs of compressive and rarefactive waves from the two boundaries at the opposite sides of the measured domain $\Omega_b$. The visco-plastic work by these wave series makes the temperature increase slowly.
From the lines with empty symbols we can find that
the shock waves arrive at the top free surface at about $t= 800$ ns. After that, rarefactive waves come back into the shocked material. Within the time interval shown in the figure, for the cases with $h=800 \mu$m and $400 \mu$m,
the density (or pressure) recovers to a value slightly larger than its
initial one, but the temperature is about 60K higher than its initial value and still increases; for the case with $h=100 \mu$m, evident oscillations are found in the curve of density after $t=900$ ns. To
understand better this phenomena, we show in Fig.\ref{Fig:Figg3} the top portion of the
configuration with temperature contour for the time $t=1.15 \mu$s, from
which we can find jetting phenomena at the upper free surface.
From the same data used in Fig.\ref{Fig:Figg1}, we
can obtain the mutual dependences of these hydrodynamical quantities. The initial
transient stage and the final oscillatory steady state are clearly
observable. Due to existence of the randomly distributed voids, waves with
various wave vectors and frequencies propagate within the shocked material. When the measured domain becomes smaller, more detailed wave structures may be found. Figure \ref{Fig:Figg2} shows clearly this trend.

\begin{figure*}[tbp]
\centering
\includegraphics*[ scale=0.7,angle=0,
bbllx=0pt,bblly=0pt,bburx=320pt,bbury=220pt
]{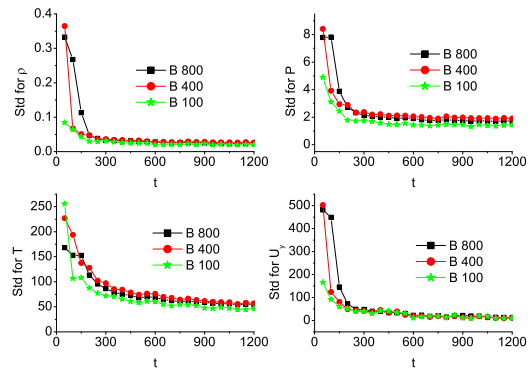}
\caption{(Color online) Standard deviations(Std) of the local quantities
averaged in various spatial scales. The heights of the measured domains are
shown in the legends where ``B" means the measured domains are at the bottom
of the target body. The length and time units are $\protect\mu$m and ns,
respectively.
(The grey-scale version is published in Ref. \cite{Xu-CTP2009A}.)
}
\label{Fig:Figg4}
\end{figure*}

Figure \ref{Fig:Figg4} shows the standard deviations of the above
four quantities measured in the bottom domains versus time. These quantities increase quickly with time at the initial stage, then decrease, nearly exponentially, to their steady values. The standard deviation of $u_y$
is larger than that of $u_x$, which means the system is out of the thermodynamic equilibrium and the internal energy in shocking degree of freedom is larger than in the transverse degrees of freedom. The non-zero values of these fluctuations confirm that the system is in a nearly steady state with local dynamical oscillations.

%%%%%%%%%%%%%%%%%%%
%% Fig. 1,2,3 uses data in /check/May5_08DN_V1000/
%%%%%%%%%%%%%%%%%%%

%%%%%%%%%%% Fig2

%%%%%%%%%%% Fig3
%%%%%%%%%%%%%%%%%%%%%%%%%%%%%%%%%%%%%%%%%%%%%%%%%%%%%%%%%%%
%%%%%%%%%% Fig4

%%%%%%%%% Fig5
\begin{figure*}[tbp]
\centering
\includegraphics*[ scale=0.7,angle=0,
bbllx=0pt,bblly=0pt,bburx=321pt,bbury=225pt
]{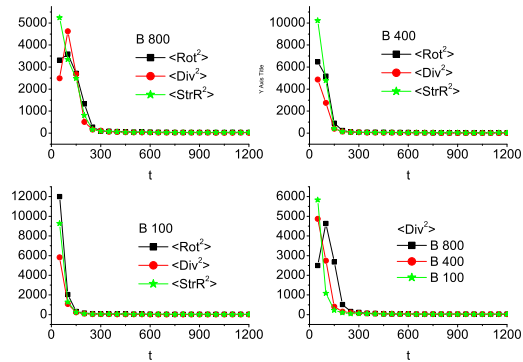}
\caption{(Color online) Variations of the mean values squared of local
rotation, divergence and strain rate with time. <...> in the legends denote
the mean value of the corresponding quantity and ``B" means the measured
domains are at the bottom of the target body. The length and time units are $\protect\mu$m and ns, respectively.
(The grey-scale version is published in Ref. \cite{Xu-CTP2009A}.)
}
\label{Fig:Figg5}
\end{figure*}

%%%%%%%%% Fig6
\begin{figure*}[tbp]
\centering
\includegraphics*[ scale=0.7,angle=0,
bbllx=0pt,bblly=0pt,bburx=444pt,bbury=387pt
]{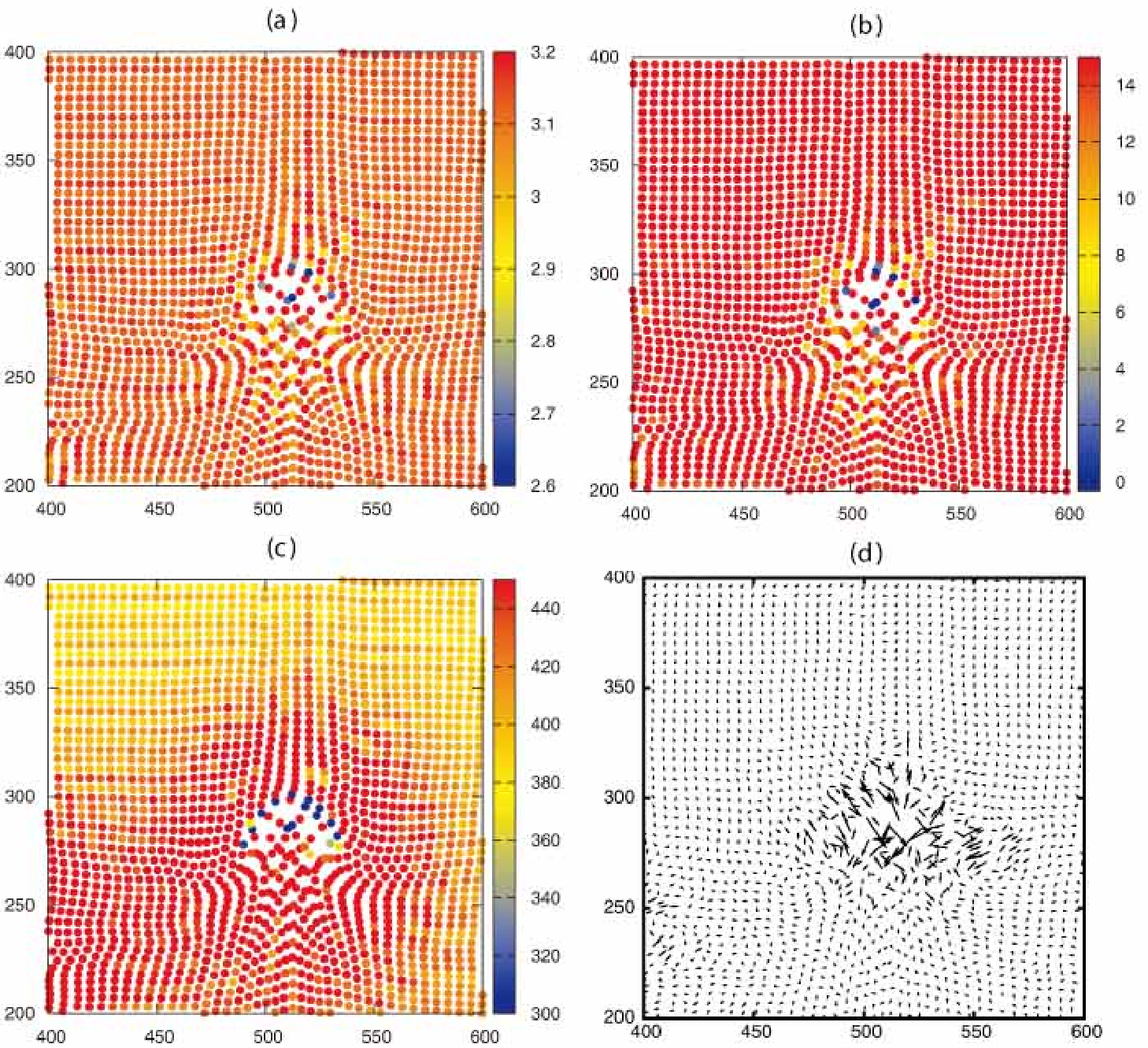}
\caption{(Color online) Configurations with density contour (a), pressure
contour (b), temperature contour (c) and velocity field (d) at time t=750
ns. The size of particle velocity is denoted by the length of arrow timed by
50. The units are the same as in Fig.\protect\ref{Fig:Figg2}.
(The grey-scale version is published in Ref. \cite{Xu-CTP2009A}.)
}
\label{Fig:Figg6}
\end{figure*}

For the case where a perfect crystal material is shocked, the entropy production occurs only in the non-equilibrium zone induced by the shock wave. In the case of porous material, the high plastic distortion of the materials
surrounding the collapsed cavities contributes extra entropy production. Therefore, we can roughly define a local rotation as
\[
\text{Rot}= |\nabla \times \mathbf{u}|, \text{,}
\]
and a local divergence as
\[
\text{Div}=  |\nabla \cdot \mathbf{u}| \text{.}
\]
Both the local rotation and divergence make significance sense in describing the dynamic process of porous material under shock. The local rotation, $|\nabla \times \mathbf{u}|$,
describes the circular flow and/or turbulence. The divergence,  $|\nabla \cdot \mathbf{u}|$, describes the changing rate of volume. Both of them work as important mechanisms of entropy production and temperature increase in dynamic responses of porous material.
The former indicates the turbulence dissipation, and the latter indicates the shock compression. Figure \ref{Fig:Figg5} shows their mean values squared versus time. As a comparison, the behavior of strain rate (``StrR'' in the figure) $\boldsymbol{\dot{\varepsilon}}$ is also shown.  All the three quantities decrease, nearly exponentially, to their steady state values as shock waves pass the measured domain $\Omega$. The amplitude of steady strain rate is very close to that of the rotation. The amplitude of the divergence is a little larger for this case. Cavity collapse and new
cavitation by the rarefactive waves are the main contributors to the local divergence. Figure \ref{Fig:Figg6} shows a portion of the configuration with density contour, pressure contour, temperature contour and velocity field at time $t=750$ns, from which one can understand better the fluctuations of the local density, pressure, temperature, particle velocity and the finite values of the rotation and divergence.

\begin{figure*}[tbp]
\centering
\includegraphics*[ scale=0.5,angle=0,
bbllx=0pt,bblly=0pt,bburx=332pt,bbury=245pt
]{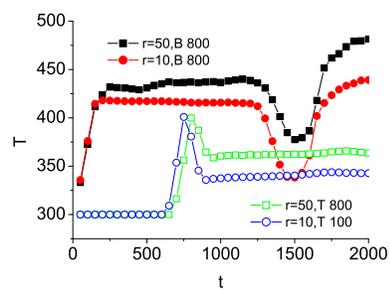}
\caption{(Color online) Effects of the mean void size on the mean
temperature. The mean void size $r$, position and height of the measured
domain are shown in the legend. ``B" and ``T" means the measured domains are
at the bottom and top of the target body, respectively. The length and time
units are $\protect\mu$m and ns, respectively.
(The grey-scale version is published in Ref. \cite{Xu-CTP2009A}.)
}
\label{Fig:Figg7}
\end{figure*}

\begin{figure*}[tbp]
\centering
\includegraphics*[ scale=1,angle=0,
bbllx=0pt,bblly=0pt,bburx=291pt,bbury=138pt
]{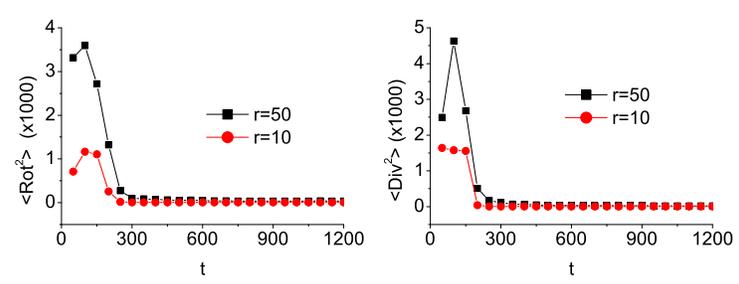}
\caption{(Color online) Effects of mean void size on the mean values squared
of local rotation, divergence. The mean sizes of void are shown in the
legends. The length and time units are $\protect\mu$m and ns, respectively.
(The grey-scale version is published in Ref. \cite{Xu-CTP2009A}.)
}
\label{Fig:Figg8}
\end{figure*}

There is a void around the position (510$\mu$m, 280$\mu$m) in this case.
To check the effects of the void size, results for different void sizes are shown and compared.
The mean density, pressure and particle velocity in the steady state do not show evident differences.
But the temperature shows significant dependence on the void size. Larger voids result in higher mean temperature. (See Fig.\ref{Fig:Figg7}.) As for influences of void size on the mean value squared of the local rotation and divergence, the void size make effects only in the transient period. See Fig.\ref{Fig:Figg8},
where the two cases correspond to different mean-void-sizes but the same
value of porosity, $\Delta=0.029$, are shown.

%%%%%%%%%%%%%
%%%% nr effects
%%%%% Fig.7

%%%%%% Fig.8

\paragraph{Cases with higher porosity}

\begin{figure*}[tbp]
\centering
\includegraphics*[ scale=0.7,angle=0,
bbllx=0pt,bblly=0pt,bburx=317pt,bbury=219pt
]{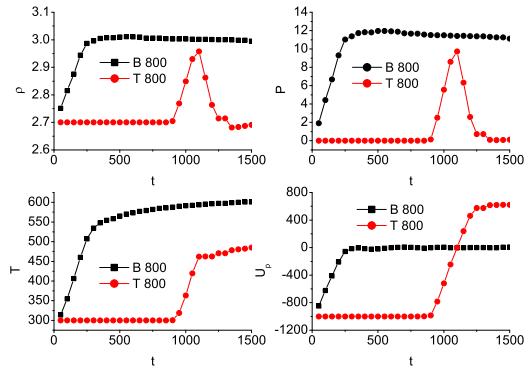}
\caption{(Color online) Variations of mean density, pressure, temperature
and particle velocity with time. Here $\protect\delta = 1.4$, $\protect\Delta = 0.286$, $v_{init} = 1000$
m/s. The meanings of ``B", ``T" and units are the same as in Fig.\protect\ref%
{Fig:Figg2}.
(The grey-scale version is published in Ref. \cite{Xu-CTP2009A}.)
}
\label{Fig:Figg9}
\end{figure*}

\begin{figure*}[tbp]
\centering
\includegraphics*[ scale=0.5,angle=0,
bbllx=0pt,bblly=0pt,bburx=296pt,bbury=223pt
]{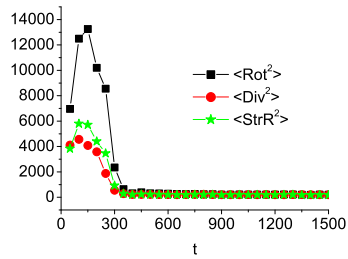}
\caption{(Color online) Variations of the mean values squared of local
rotation, divergence and strain rate with time. The unit of time is ns.
(The grey-scale version is published in Ref. \cite{Xu-CTP2009A}.)
}
\label{Fig:Figg10}
\end{figure*}

For the cases with higher porosity, we show the variations of mean density, pressure, temperature and particle velocity with time for the case with  $\protect\Delta = 0.286$, $\delta=1.4$, $r=10 \mu$m and $v_{init} = 1000$m/s in Fig.  \ref{Fig:Figg9}. Here only results averaged in the upper and bottom domains with the same hight, $h=800 \mu$m,  are  shown. Different from the low-porosity case with $\delta=1.03$, the mean density and pressure decrease with time, while the mean temperature increase with a higher rate after the initial stage. This is due to the rarefactive
waves reflected back from the cavities in the downstream region. The rarefactive waves make looser the shocked material and result in a relatively higher local divergence. Consequently, more kinetic energy into heat. At the same time, a higher porosity means more cavities embedded in the material, more jetting phenomena may occur under shock. Both the  jetting phenomena and the collisions of jetted materials with the downstream walls of cavities result in a significant increase of local temperature, local divergence and local rotation. Figure \ref{Fig:Figg10} shows the mean values squared of the local rotation, divergence and strain rate. During the initial transient period,
the turbulence dissipation is the main mechanism for the temperature increase in this case. In the later steady state, all the three kinds of dissipations make nearly the same contributions.

\begin{figure*}[tbp]
\centering
\includegraphics*[ scale=0.7,angle=0,
bbllx=0pt,bblly=0pt,bburx=321pt,bbury=218pt
]{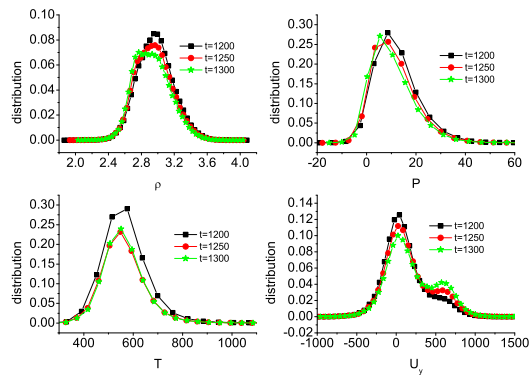}
\caption{(Color online) Distribution of local density, pressure,
temperature, particle velocity at various times. The units are the same as
in Fig.\protect\ref{Fig:Figg2}.
(The grey-scale version is published in Ref. \cite{Xu-CTP2009A}.)
}
\label{Fig:Figg11}
\end{figure*}

\begin{figure*}[tbp]
\centering
\includegraphics*[ scale=1,angle=0,
bbllx=0pt,bblly=0pt,bburx=314pt,bbury=130pt
]{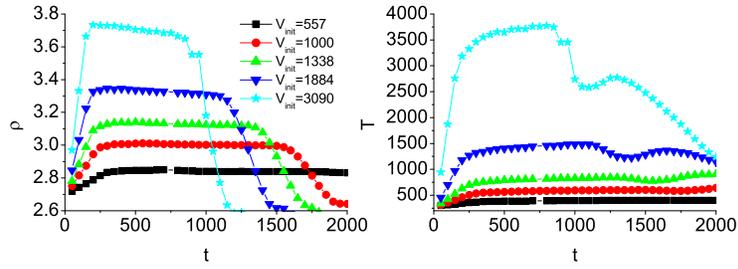}
\caption{(Color online) Mean density and temperature versus time for various
shock strengths. The initial velocity $v_{init}$ are shown in the legend.
The units are the same as in Fig.\protect\ref{Fig:Figg2}.
(The grey-scale version is published in Ref. \cite{Xu-CTP2009A}.)
}
\label{Fig:Figg12}
\end{figure*}

To further clarify the inhomogeneity effects in the shocked regime, in Fig.\ref{Fig:Figg11}, we show the distributions of density, pressure, temperature and particle velocity at three times, $t=1200$ns, $1250$ns and $1300$ns.
Their distributions generally deviate from the Gaussian distribution and vary with time. The effects of
initial impact velocity on the mean density, pressure and temperature are shown in Fig.\ref{Fig:Figg12}. It is clear that the decreasing rate of the mean density and the increasing rate of mean temperature becomes larger as increasing the strength of the initial shock.

\begin{figure*}[tbp]
\centering
\includegraphics*[ scale=0.95,angle=0,
bbllx=88pt,bblly=382pt,bburx=470pt,bbury=520pt
]{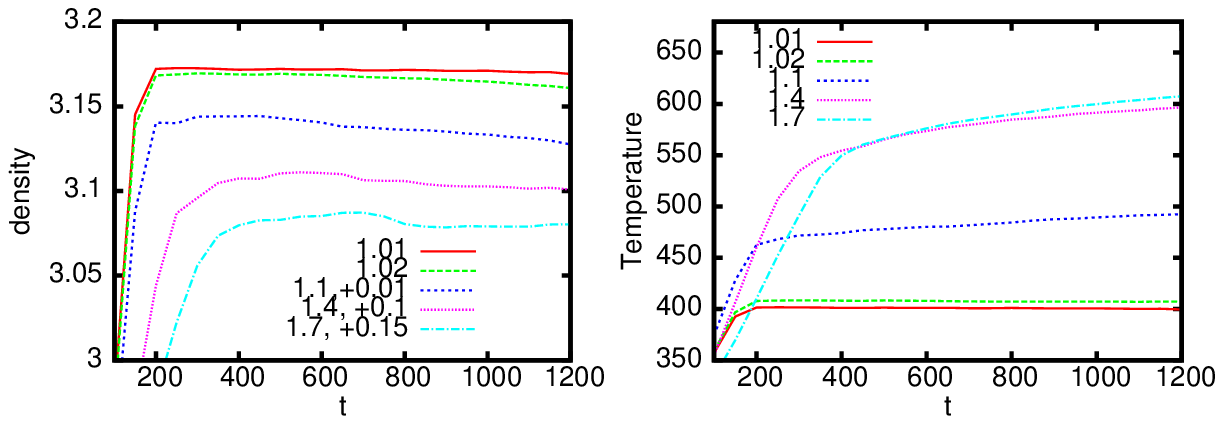}
\caption{(Color online) Mean density and temperature versus time for various
porosities. The values of $\delta$, 1.01,1.02,1.1,1.4,1.7 are shown in the
legends. In the left figure, the lines for cases with $\protect\delta = 1.1$,
$1.4$ and $1.7$ are moved upwards by 0.01,0.1, and 0.15, respectively. The
units are the same as in Fig.\protect\ref{Fig:Figg2}.
(The grey-scale version is published in Ref. \cite{Xu-CTP2009A}.)
}
\label{Fig:Figg13}
\end{figure*}

When study the porosity effects, we fix the shock strength. Figure \ref{Fig:Figg13}
shows the mean density, and temperature versus time for various porosities. Here initial velocity  $v_{init} = 1000$m/s.
When the porosity is very small, the mean density decreases more quickly with increasing the porosity. But when the porosity is high, the mean density show more complex behaviors.

%%%%%%%%%%%%%%%%%%%
%%%%% Fig9
%%%%%%%%%%%%%%%%%%%
%%%%% Fig10

%%%%% Fig11

%%%%%%%%%%%%% Up effects
%%%%% Fig12

%%%%% Fig13

\subsubsection{Morphological analysis}

Morphological analysis describes the geometrical and topological properties of the fields of temperature, pressure, density, etc. Shock wave results in complicated series of compressions and rarefactions in the porous material. In the case of temperature field, $A$ describes the fraction of high temperature particles. It increasing rate roughly gives the velocity $D$ of a compressive-wave series. The velocity $D$ decreases with increasing the threshold value $T_{th}$ of temperature. The fraction $A$ increases, nearly parabolically, with time $t$ during the initial period. The $A(t)$ curve recover to be linear in the following three cases: (i) when the porosity $\Delta$ approaches $0$, (ii) when the initial shock becomes much stronger, and (iii) when the threshold value approaches the minimum value of the temperature.
The fraction $A$ of high temperature particles may continue to increase even after the early compressive waves have arrived at the downstream free surface and some rarefactive waves have come back into the material. In the case of energetic material needing a higher temperature for ignition, a higher porosity is preferred and the material may be ignited after the precursory compressive waves have scanned the entire material.
 In morphology analysis, the result dependence on experimental conditions is reflected simply by a few coefficients.
Here we show some observations for the temperature field\cite{Xu-JPD2009}.

%%%%%%%%%%%%%%%%%%%%%%%%%%%%%%%%%%%%%%%%%%%%%%%%%%%%%%
\begin{figure*}[tbp]
\centerline{\epsfig{file= 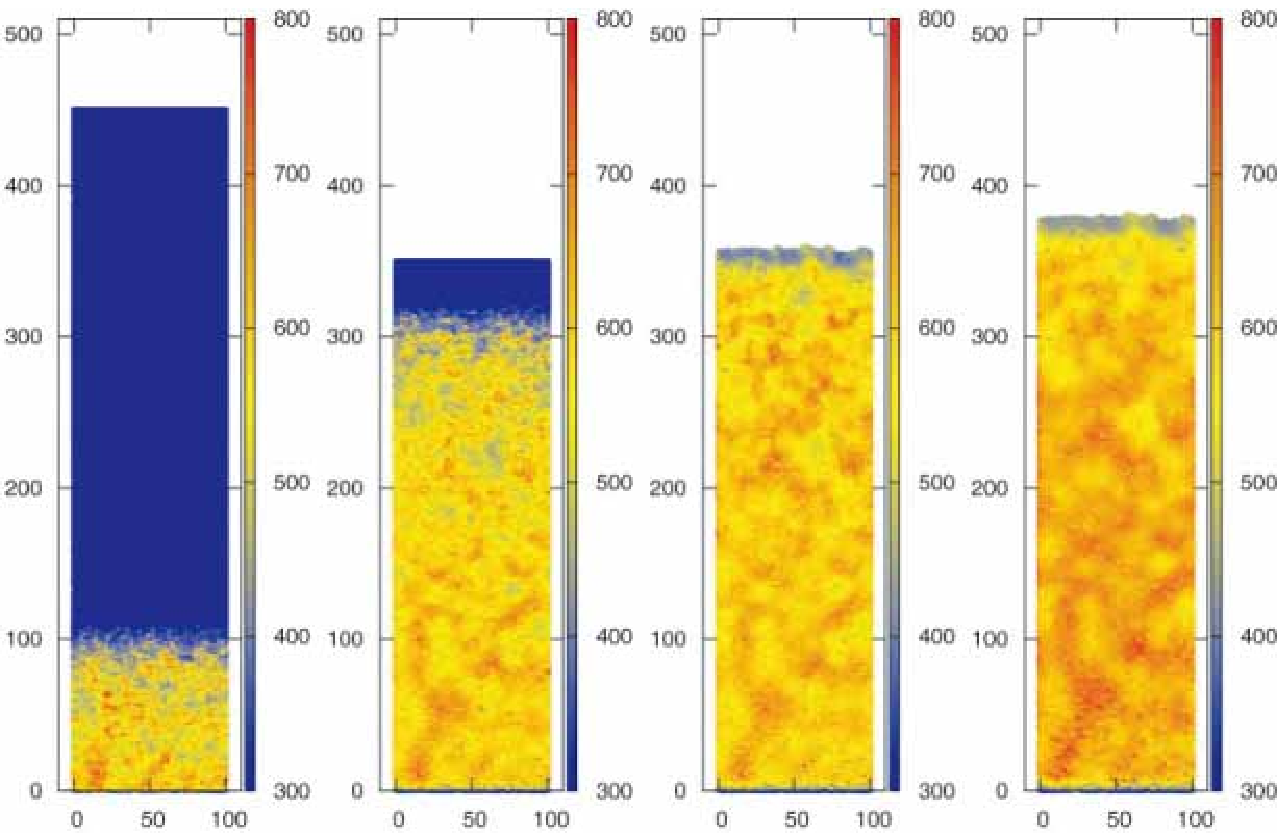, width=0.42\textwidth,clip=}}
\caption{(Color online) Configurations with temperature contours.
$\protect\Delta=0.5$ and $v_{init}=1000$m/s. From left to right,
t=500ns, 1500ns, 2000ns, and 2500ns, respectively. The length unit here is 10 $%
\protect\mu $m.
(Adopted with permission from Ref. \cite{Xu-JPD2009}.)
}
\label{Fig:Figu1}
\end{figure*}
%%%%%%%%%%%%%%%%%%%%%%%%%%%%%%%%%%%%%%%%%%%%%%%%%%%%%%%%%%%%%%%

%%%%%%%%%%%%%%%%%%%%%%%%%%%%%%%%%%%%%%%%%%%%%%%%%%%%%%%%%%%%%%%
\begin{figure*}[tbp]
\centerline{\epsfig{file= 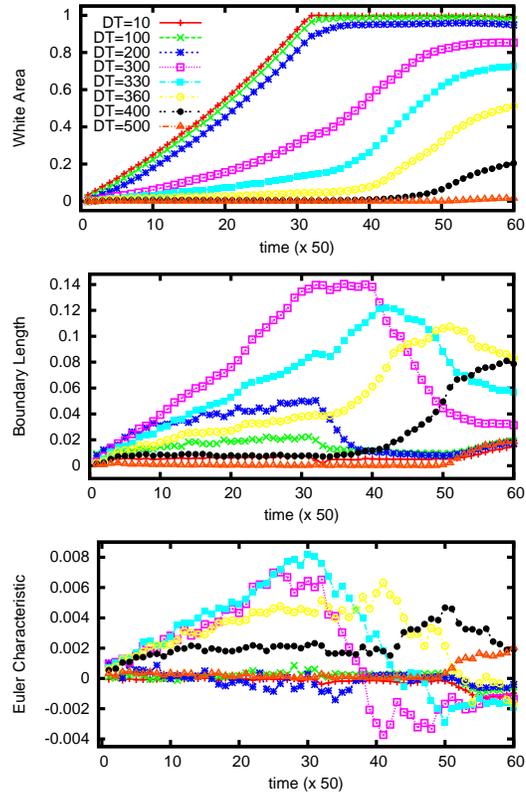, bbllx=103 pt, bblly=99
pt,bburx=509 pt,bbury=701 pt, width=0.4\textwidth,clip=}}
\caption{(Color online) Minkowski measures for the procedure shown in Fig.\ref{Fig:Figu1}.
The contour levels of the temperature increment are shown in the legend.
(Adopted with permission from Ref. \cite{Xu-JPD2009}.)
}
\label{Fig:Figu2}
\end{figure*}
%%%%%%%%%%%%%%%%%%%%%%%%%%%%%%%%%%%%%%%%%%%%%%%%%%%%%%%%%%%%%%%

%%%%%%%%%%%%%%%%%%%%%%%%%%%%%%%%%%%%%%%%%%%%%%%%%%%%%%%%%%%%%%%
\begin{figure*}[tbp]
\centerline{\epsfig{file= 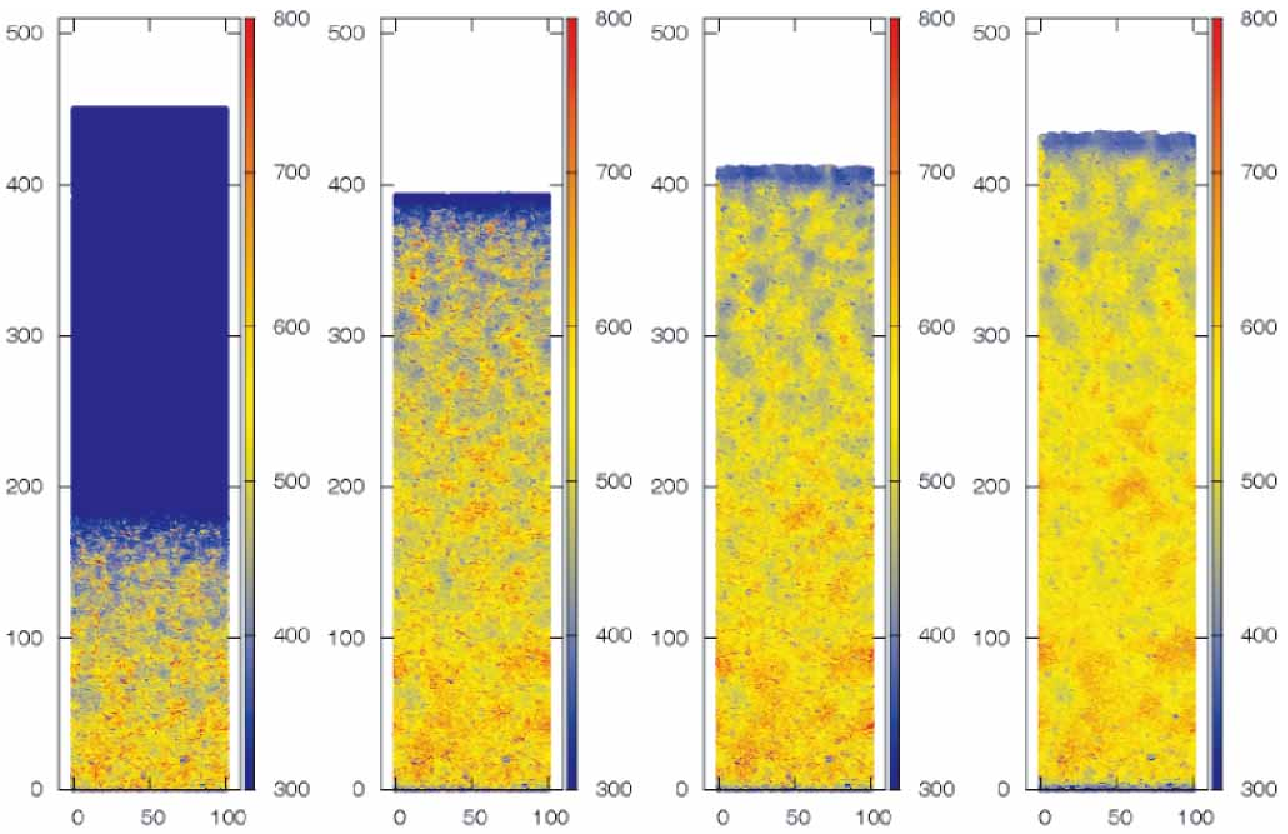, width=0.42\textwidth,clip=}}
\caption{(Color online) Configurations with temperature contours.
$\protect\Delta=0.286$ and $v_{init}=1000$m/s. From left to right,
t=500ns, 1100ns, 1400ns, and 1700ns, respectively. The length unit here is 10 $%
\protect\mu $m.
(Adopted with permission from Ref. \cite{Xu-JPD2009}.)
}
\label{Fig:Figu3}
\end{figure*}
%%%%%%%%%%%%%%%%%%%%%%%%%%%%%%%%%%%%%%%%%%%%%%%%%%%%%%%%%%%%%%%

%%%%%%%%%%%%%%%%%%%%%%%%%%%%%%%%%%%%%%%%%%%%%%%%%%%%%%%%%%%%%%%
%%%%%% file= MinT100_d.ps
\begin{figure*}[tbp]
\centerline{\epsfig{file= 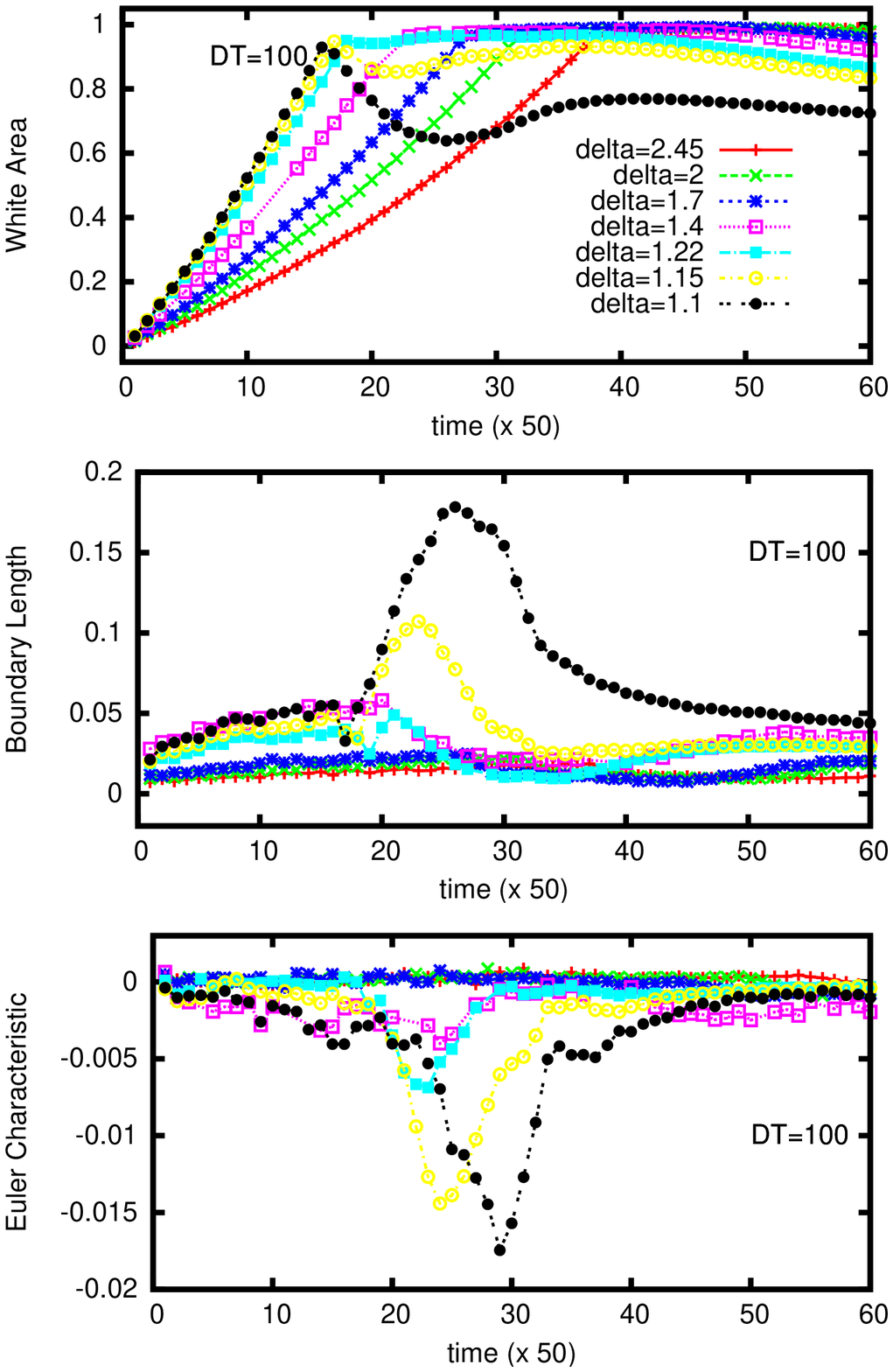, bbllx=103 pt, bblly=99
pt,bburx=509 pt,bbury=701 pt, width=0.4\textwidth,clip=}}
\caption{(Color online) Minkowski measures for  cases with various
porosities. $T_{th}=400$ K. The values of $\delta$ are shown in the legend.
(Adopted with permission from Ref. \cite{Xu-JPD2009}.)
}
\label{Fig:Figu4}
\end{figure*}
%%%%%%%%%%%%%%%%%%%%%%%%%%%%%%%%%%%%%%%%%%%%%%%%%%%%%%%%%%%%%%%

%%%%%%%%%%%%%%%%%%%%%%%%%%%%%%%%%%%%%%%%%%%%%%%%%%%%%%%%%%%%%%%
%%%%% file= MinT200_d.ps
\begin{figure*}[tbp]
\centerline{\epsfig{file= 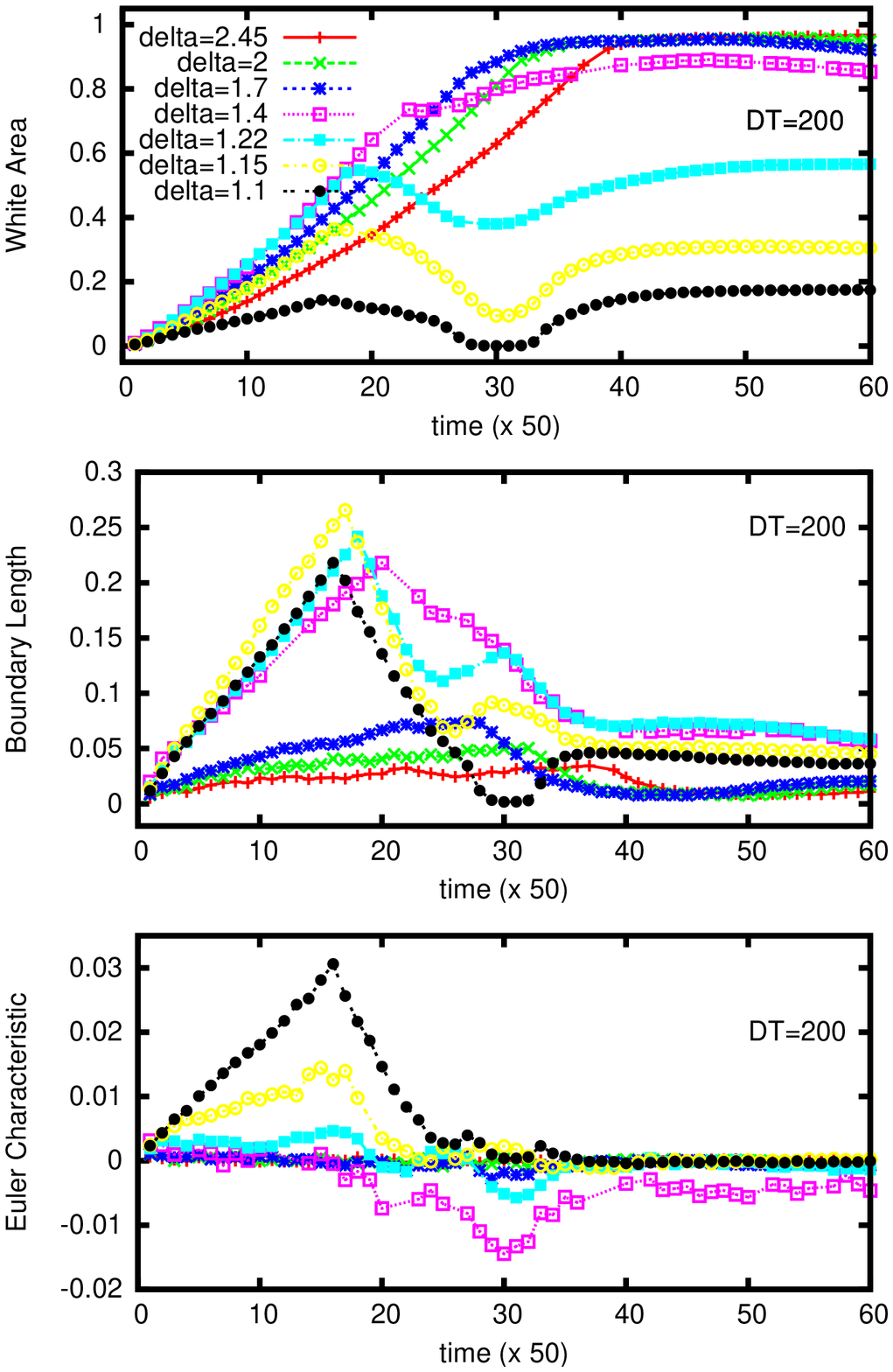, bbllx=103 pt, bblly=99
pt,bburx=509 pt,bbury=701 pt, width=0.4\textwidth,clip=}}
\caption{(Color online) Minkowski measures for  cases with various porosities.
$T_{th} = 500$ K. The values of $\delta$ are shown in the legend.
(Adopted with permission from Ref. \cite{Xu-JPD2009}.)
}
\label{Fig:Figu5}
\end{figure*}
%%%%%%%%%%%%%%%%%%%%%%%%%%%%%%%%%%%%%%%%%%%%%%%%%%%%%%%%%%%%%%%

%%%%%%%%%%%%%%%%%%%%%%%%%%%%%%%%%%%%%%%%%%%%%%%%%%%%%%%%%%%%%%%
%%%%% file= MinT300_d.ps
\begin{figure*}[tbp]
\centerline{\epsfig{file= 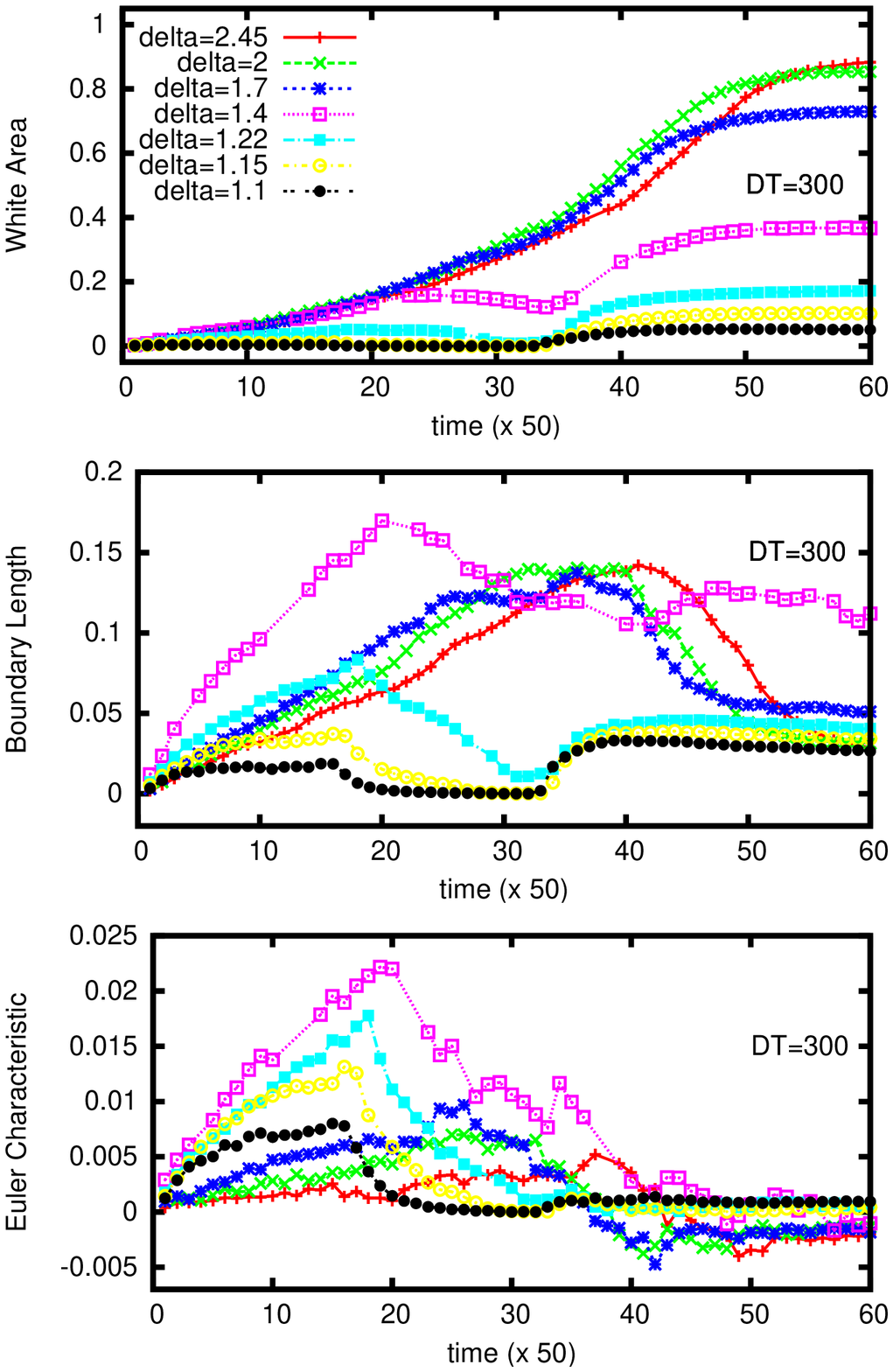, bbllx=103 pt, bblly=99
pt,bburx=509 pt,bbury=701 pt, width=0.4\textwidth,clip=}}
\caption{(Color online) Minkowski measures for cases with various porosities.
 $T_{th} = 600$ K. The values of $\delta$ are shown in the legend.
 (Adopted with permission from Ref. \cite{Xu-JPD2009}.)
 }
\label{Fig:Figu6}
\end{figure*}
%%%%%%%%%%%%%%%%%%%%%%%%%%%%%%%%%%%%%%%%%%%%%%%%%%%%%%%%%%%%%%%
%%%%%%%%%%%%%%%%%%%%%%%%%%%%%%%%%%%%%%%%%%%%%%%%%%%%%%%%%%%%%%%%%
\begin{figure*}[tbp]
\centerline{\epsfig{file= 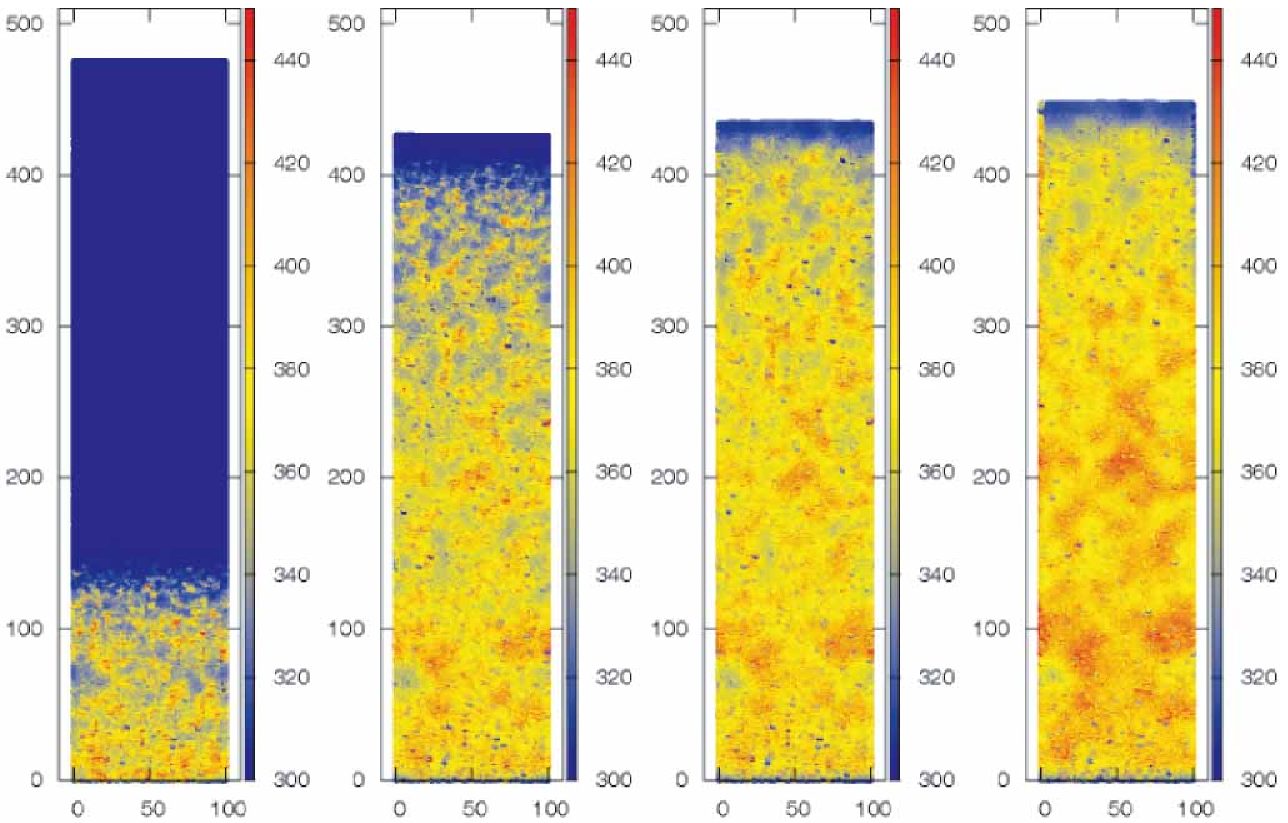, width=0.42\textwidth,clip=}}
\caption{(Color online) Configurations with temperature contours.
$\protect\Delta=0.286$ and $v_{init}=500$m/s. From left to right, t =
500 ns, 1500 ns, 2000 ns, and 2500 ns, respectively. The length unit here is 10 $%
\protect\mu $m.
(Adopted with permission from Ref. \cite{Xu-JPD2009}.)
}
\label{Fig:Figu7}
\end{figure*}
%%%%%%%%%%%%%%%%%%%%%%%%%%%%%%%%%%%%%%%%%%%%%%%%%%%%%%%%%%%%%%%

%%%%%%%%%%%%%%%%%%%%%%%%%%%%%%%%%%%%%%%%%%%%%%%%%%%%%%%%%%%%%%%
%%%%% file= d0.286V500Min.ps
\begin{figure*}[tbp]
\centerline{\epsfig{file= 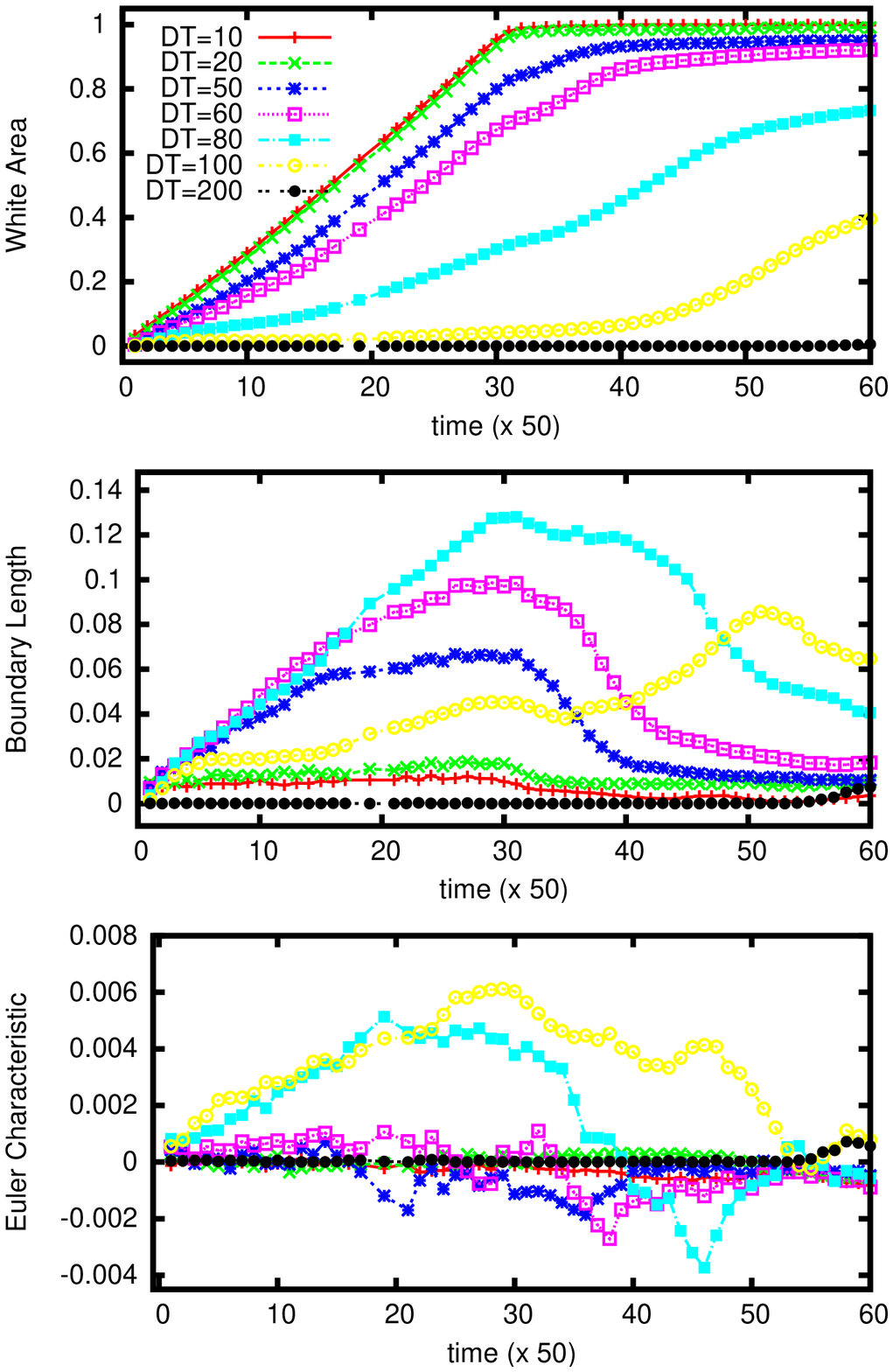, bbllx=103 pt, bblly=99
pt,bburx=509 pt,bbury=701 pt, width=0.4\textwidth,clip=}}
\caption{(Color online) Minkowski measures for the case of $\protect%
\Delta=0.286$ and $v_{init}=500$m/s. The values of contour level are shown in
the legend.
(Adopted with permission from Ref. \cite{Xu-JPD2009}.)
}
\label{Fig:Figu8}
\end{figure*}

%%%%%%%%%%%%%%%%%%%%%%%%%%%%%%%%%%%%%%%%%%%%%%%%%%%%%%%%%%%%%%%
%%%%% file= d0.286V400Min.ps
\begin{figure*}[tbp]
\centerline{\epsfig{file= 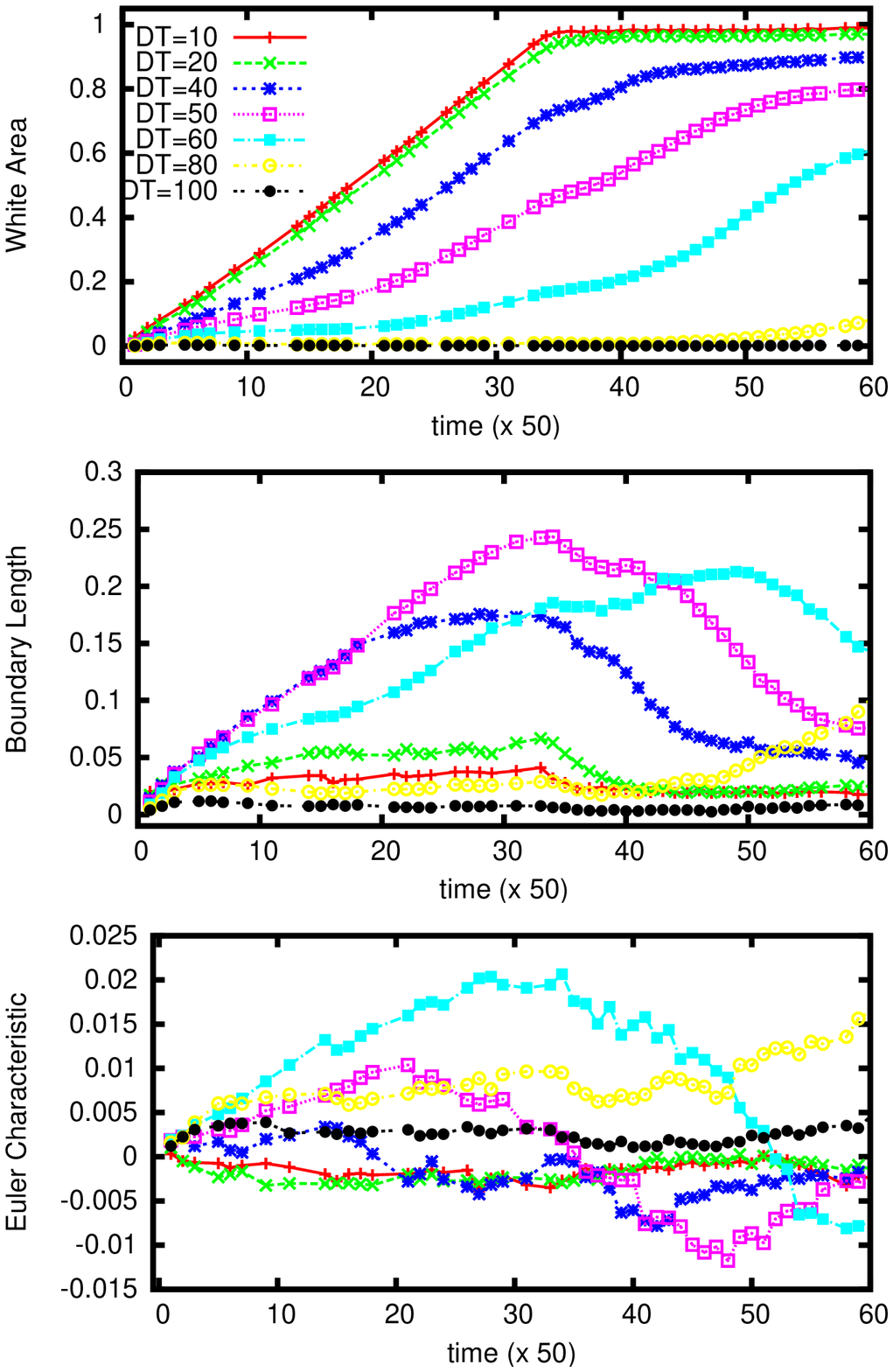, bbllx=103 pt, bblly=99
pt,bburx=509 pt,bbury=701 pt, width=0.4\textwidth,clip=}}
\caption{(Color online) Minkowski measures for the case of $\protect%
\Delta=0.286$ and $v_{init}=400$m/s. The values of contour level are shown in
the legend.
(Adopted with permission from Ref. \cite{Xu-JPD2009}.)
}
\label{Fig:Figu9}
\end{figure*}

%%%%%%%%%%%%%%%%%%%%%%%%%%%%%%%%%%%%%%%%%%%%%%%%%%%%%%%%%%%%%%%
%%%%% file= d0.286V300Min.ps
\begin{figure*}[tbp]
\centerline{\epsfig{file= 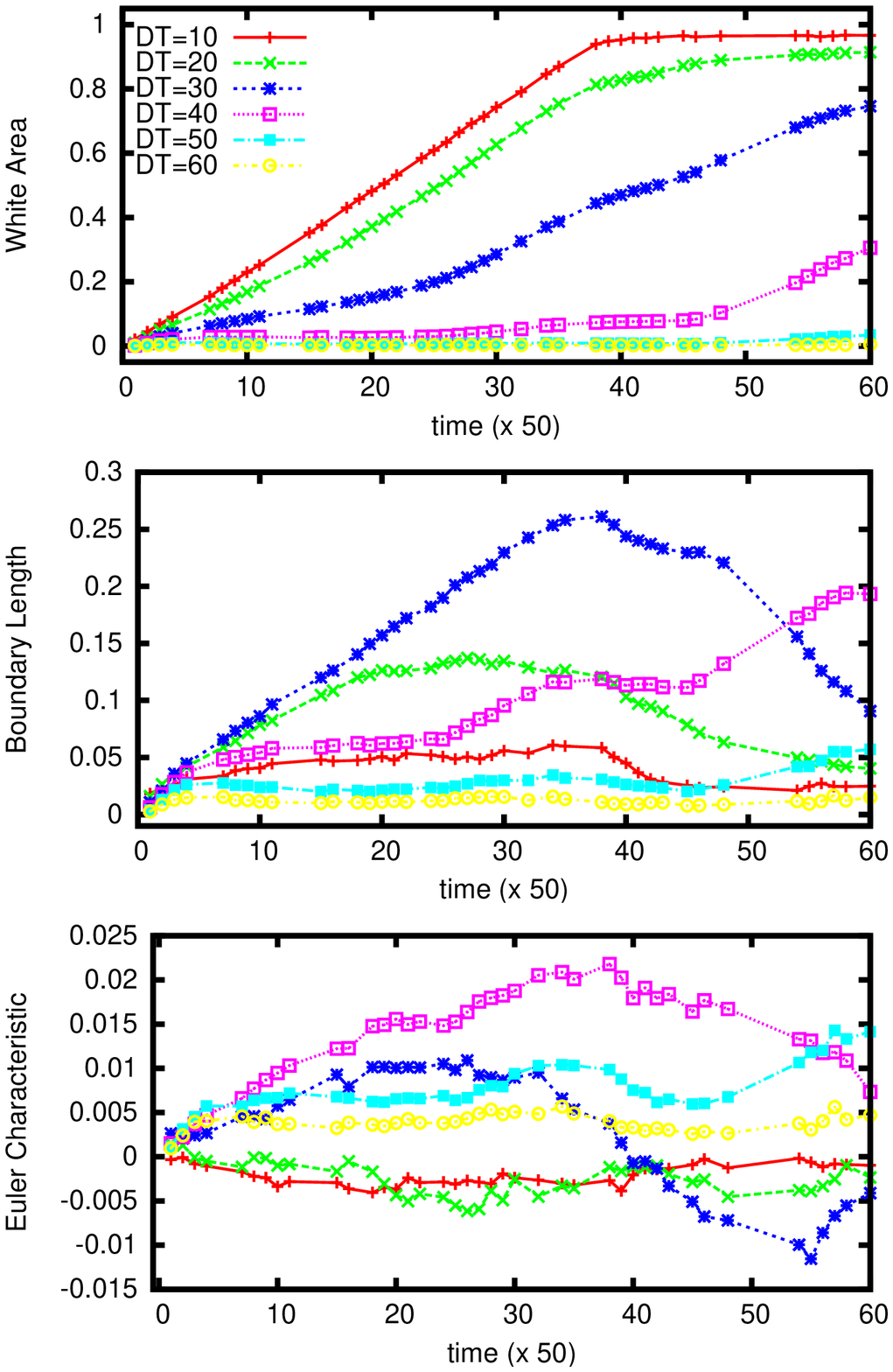, bbllx=103 pt, bblly=99
pt,bburx=509 pt,bbury=701 pt, width=0.4\textwidth,clip=}}
\caption{(Color online) Minkowski measures for the case of $\protect%
\Delta=0.286$ and $v_{init}=300$m/s. The values of contour level are shown in
the legend.
(Adopted with permission from Ref. \cite{Xu-JPD2009}.)
}
\label{Fig:Figu10}
\end{figure*}

%%%%%%%%%%%%%%%%%%%%%%%%%%%%%%%%%%%%%%%%%%%%%%%%%%%%%%%%%%%%%%%%%%%

\paragraph{Basic observations}
 A set of snapshots for a shock process are shown in Fig.\ref{Fig:Figu1}, where the contours are for the temperature.
From blue to red, the temperature increases. The first two show the loading process.  The last two are for the unloading process of the compressive waves. As mentioned in previous part, rarefactive waves are reflected back into the material when compressive waves reaches the upper free surface. Under the tensional action of rarefactive wave, the height of the porous material increases with time. In fact,  a large number of local unloading phenomena have occurred within the material before the compressive waves arrive at
the upper free surface. The details of wave series are very complex, we use the Minkowski functionals to characterize the physical fields inside the material.

In this review, the morphological analysis is mainly for the temperature
field. To use the Minkowski functionals, we first choose a threshold temperature $T_{th}$ and condense the
temperature field $T(\mathbf{x})$ into high temperature regions (with $T(\mathbf{x})\geq T(\mathbf{x})_{th}$) and
low temperature regions (with $T(\mathbf{x})<T(\mathbf{x})_{th}$ ). Figure \ref{Fig:Figu2} shows the several sets of Morphological analysis for the shocking process shown in Fig.\ref{Fig:Figu1}. \textquotedblleft $DT$ \textquotedblright\ in the legend means $T_{th}-300$. The unit of temperature is K. The
time unit is ns. One can find that, when $DT$ is very small, the wave front is nearly a plane, which is similar to the case of uniform solid material. When $DT=10$K, the total fractional high temperature area $A$
increases up to be nearly $1$ at about the time $t=1600$ ns and keeps this saturation value until the time $t=2600$ns, then shows a slight decreasing. This indicates that (i) the early compressive wave reaches the upper free surface
at about the time $t=1600$ ns, (in fact before that), (ii) nearly all material particles obtain a temperature higher than $310$ K during the following $1000$ns period. In the unloading process the
 a very small fraction of material particles decrease their temperature to below $310$ K due to the action of rarefactive waves. The high temperature area $A$ decreases with increasing the temperature threshold.
 In the case where $DT=100$ K, at the time $t=1900$ ns, the high temperature area gets a (nearly) steady value $0.96$%
, which indicates that $4\%$ of the material particles could not have a temperature higher than $400$ K in the whole process during shown period. If compare with the case with $DT=10$K, it will be easy to find that the temperature increase in shocked porous material is much slower than in shocked uniform solid material.
When the compressive wave arrives at a cavity, it is decomposed of many components. The components in the
solid portion propagate forwards more quickly, while the portion facing cavity may result in jet phenomenon. When jetted material particles hit the downstream wall of the cavity, new compressive waves occur. At the meanwhile, the cavity  reflects rarefactive wave back to the compressed regime. A large number of similar processes occur inside the shocked porous material. Thus, the shock loading process is manifested as successive actions of many compressive and rarefactive waves. The effects of compressive waves dominate during the shock-loading process. All the plastic deformations contribute to the temperature increase. Similarly, one can interpret the curve for $DT=200$ K. When $DT$ increases from $200$K to $300$K, the curve of high temperature area shows a significant variation. For the case of $DT=400$K, the high temperature area reach $0.2$ at about $t=3000$ns, which indicates $80\%$ of material particles could not obtain a
temperature higher than $700$ K up to this time. When $DT=500$K, the high temperature area keeps nearly zero during the whole period shown here, which indicates all the local temperature is lower than $800$K up to $t=3000$ns. For cases with $DT=300$K, $330$K, $360$K and $400$K, after the initial slow increasing period, the high temperature area shows a quick increasing period. The latter means that a large number of high temperature regimes in the previously compressed region
coalesce during this period. After that the increasing  rate of $A$ shows a slowing-down phenomenon. The slope of the $A(t)$ curve roughly indicates a mean propagation speed of some components of the compressive waves. When a velocity $D$ of the profile front of high temperature domains is mentioned, the corresponding temperature should also be claimed. It is clear that $D(T_{th})$ decreases when $T_{th}$ becomes higher. The total fractional high temperature area $A(t)$ shows roughly a parabolic behavior during the initial period. When $DT$ approaches $0$, the curve for high temperature area $A(t)$ goes back to be linear.

Now we analyze the second Minkowski measure, the boundary length $L$.
For the case of $DT=10$K, after the initial increase period, $L$ keeps a small constant up to about the time $t=2600$%
ns. The phenomenon that boundary length $L$ keeps constant while the high temperature area $A$ increase indicates also that the compressive wave is propagating towards the upper free surface and the interface is nearly a plane. The increasing of $L$ after $t=2600$ns is companied by a decreasing of high temperature area $A$, which indicates some small low temperature spots occur inside the background of high temperature area.
The curves for $DT=100$ K and $DT=200$K present similar information. They first increase with time due to the creation   of more spots with high temperature, then decreases due to the coalesce of
high temperature areas, and finally increase, accompanied by a slight decrease of the total fractional high temperature  area. For the case with $DT=300$K, during the period with $1500 $ ns $ < t < 2500$ ns, the high temperature area $A$ increases, while the total fractional boundary length $L$ is nearly a constant.
This phenomenon indicates that, during this period, the
compressive waves propagate forwards, more scattered high temperature spots appear in the newly compressed regime; at the same time, some previous scattered spots with high temperatures coalesce. From the time $2500
$ ns to $3000$ ns, the fractional high temperature area $A$ increases very slowly, but the boundary length $L$ decreases quickly. This phenomenon indicates that the
increasing of $A$ is mainly due to coalesce of previous scattered spots with high temperature. The curves
for $DT=330$K and $DT=360$K can be interpreted in a similar means.
For this shock strength, only very few material particles can obtain a temperature higher than $700$ K before the time $t=2000$ ns. Therefore, the boundary length $L$ for $DT=400$K has a meaningful increase only after the time $t=2000$ ns.

When the threshold value $DT$ is small, the condition $T > T_{th}$ is satisfied in (nearly) all of the compressed region  and $T < T_{th}$ is satisfied in the uncompressed part of the material. The condensed temperature field appears as a  highly connected structure with (nearly) equal and very small amount of high temperature and low temperature domains. So, the Euler characteristic $\chi $ keeps nearly zero in the whole shock-loading process and the mean
curvature $\kappa $ is also nearly zero. The value of $\chi $ decreases to be evidently less than zero in the unloading process, which means that the number of low temperature domains increases. (See the $\chi (t)$ curves
for cases of $DT=10$, $DT=100$ and $DT=200$ in Fig.\ref{Fig:Figu2}.) With increasing
 the threshold value $T_{th}$, more domains changes from high temperature to low temperature ones. The pattern evolution in the shock-loading process shows the following scenario: scattered high temperature domains appear gradually
with time in the background of low temperature domains. Consequently, the Euler characteristic $\chi $ is
positive and increasing with time. (See the $\chi (t)$ curves for cases of $DT=300$, $DT=330$ and $DT=360$ in Fig.\ref{Fig:Figu2}.) When the threshold value $T_{th}$ is further increased up to $700$ K, a considerable fraction of material particles could not obtain a temperature higher than $T_{th}$. The saturation phenomenon in the $\chi $ curve during the period, $550$ns $ < t < 2100$ ns, indicates that the numbers of connected domains with high and low temperatures vary with time in a similar way. The
increase of $\chi $ in the period, $2100 $ ns $ < t < 2500$ ns, is due to
that the rarefactive waves make mean-temperature decrease, some connected high temperature domains are disconnected again. For the case with $DT=500$K, the so-called low temperature domain occupies nearly all the shocked material.
 Consequently, $\chi $ is nearly zero.

\paragraph{Effects of porosity}

To study the effects of porosity, a set of snapshots for the case with a lower porosity
  are shown in Fig. \ref{Fig:Figu3}. Here $\Delta =0.286$ and the other conditions are the same as those in Fig.\ref{Fig:Figu1}. From left to right, the corresponding times are $t=500$ns, $1100$ns, $1400$ns and $1700$ns, respectively. It is easy to find that the propagation velocity of compressive wave increases with decreasing the porosity. In this case, at time $t=500$ns, the compressive wave arrives at the position with $y=1750\mu $ m; while in the case of $\Delta = 0.5$, the compressive wave arrives only at the position with $y=1000\mu $ m. In
this case, the compressive wave has arrived the top free surface and the rarefactive wave has been reflected back to the
porous material before $t=1400$ns; while in the case of $\Delta
=0.5$, the shock-loading procedure has not been finished up to the time $t=1500$ns.

The porosity effects can be more quantitatively investigated via the morphological analysis. In Fig. \ref{Fig:Figu4} we show a set of morphological analysis for the cases with various porosities,
 $\Delta =0.592$, $0.5$, $0.412$, $0.286$, $0.180$, $0.130$, $0.091$ (i.e. $\delta =2.45$, $2$, $1.7$, $%
1.4$, $1.22$, $1.15$, $1.1$  ), where $T_{th}=400$K. Via comparing the slopes of the $A(t)$ curves for the shock loading processes, it is clear that the velocity $D$ of the profile front of high temperature domains decreases with increasing porosity.
The largest value of boundary length $L_{max}$ increases as $\Delta $ decreases. In the case of $\Delta =0.091$, the $L$ obtains its maximum value at about the time $t=1250$ns. This phenomenon indicates that the highest temperature in
shocked porous material decreases when the porosity decreases. The value of $\chi $ becomes more negative when $\Delta $ decreases from $0.592$ to $0.091$, which means the
disconnected low temperature domains where $ T < 400 $ K dominate more the condensed temperature field.

Via comparing results in Figs.\ref{Fig:Figu5} and \ref{Fig:Figu6}, we can have some observation on the result dependence on the temperature threshold $T_{th}$. Figures \ref{Fig:Figu5} and \ref{Fig:Figu6} show, respectively, two sets of the Minkowski measures for the same porosities but two higher temperature thresholds, $T_{th}=500$K and $T_{th}=600$K. They present some supplementary information to that shown in Fig. \ref{Fig:Figu4}. For cases with $\Delta =0.286$, $0.180$, $0.130$ and $0.091$, only $88\%$, $55\%$, $36\%$ and $15\%$ of the material
particles obtain a temperature higher than $500$K. For cases with $\Delta =0.286$ and $0.180$, and only $16\%$ and $6\%$ obtain a temperature higher than $600$ K in the shock-loading process. When the temperature threshold $T_{th}=500$K, the case of $\Delta =0.130$ shows the maximum boundary length and the case of $\Delta =0.091$ shows the maximum Euler characteristic. When the temperature threshold $T_{th}=600$K, the case of $\Delta =0.286$ shows the maximum boundary
length and maximum Euler characteristic, which means the high temperature spots with $ T > 600$ K are
scatteredly distributed in the background of low temperature regime with $ T < 600$ K.

\paragraph{Effects of shock strength}

%%%%%%%%%%%%%%%%%%%%%%%%%%%%%%%%%%%%%%%%%%%%%%%%%%%%%%%%%%%%%%%
%%%%% file= d0.286Y120Min_V.ps
\begin{figure*}[tbp]
\centerline{\epsfig{file= 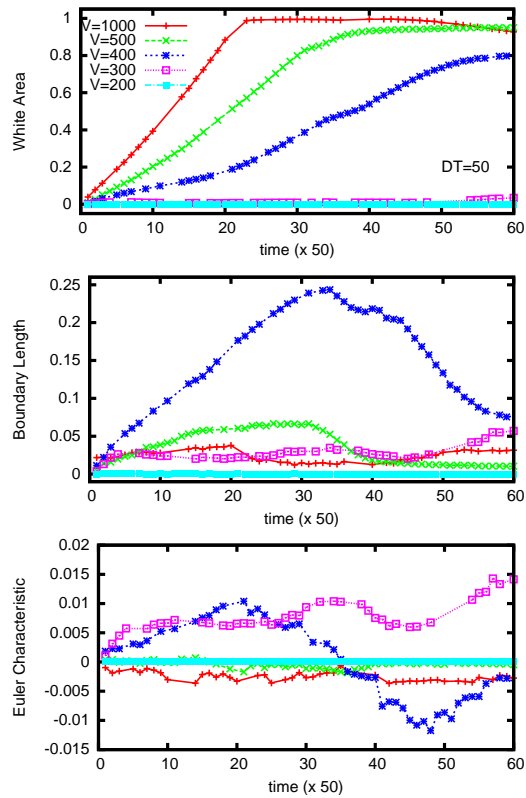, bbllx=103 pt, bblly=99
pt,bburx=509 pt,bbury=701 pt, width=0.4\textwidth,clip=}}
\caption{(Color online) Minkowski measures for cases with various shock strengths. $\protect%
\Delta=0.286$. The values of initial impacting speed $v_{init}$ are shown in
the legend.
(Adopted with permission from Ref. \cite{Xu-JPD2009}.)
}
\label{Fig:Figu11}
\end{figure*}
%%%%%%%%%%%%%%%%%%%%%%%%%%%%%%%%%%%%%%%%%%%%%%%%%%%%%%%%%%%%%%%

When study the effects of shock strength, we fix the value of porosity.
A set of snapshots for the case with $\Delta =0.286$ and $v_{init}=500$%
m/s are shown in Fig. \ref{Fig:Figu7}.
From left to right, the corresponding times are $t=500$ns, $1500$ns, $2000$ns and $2500$ns, respectively.
The first two are for the shock loading process and the latter two are for the shock unloading process.
Compared with the case shown in Fig.\ref{Fig:Figu3}, the velocity $D$ of profile front of high temperature domains and the highest temperature $T_{\max }$ decreased. The Minkowski meansures for the temperature field in this process is
shown in Fig. \ref{Fig:Figu8}. Such a shocking process could not result in high temperature domains with $T=500$K.
High-temperature area continue to increase even after some precursory compressive waves
have scanned all the material and some rarefactive waves have come back from the upper free surface. Up to the
time $t=3000$ ns, the fractional high temperature area of with $ T > 400 $K reaches $40\%$, the fractional area for $T > 380$K reaches $74\%$, that for $T > 360$ K reaches $91\%$. The temperature threshold value with $T=380K$ shows the largest boundary length at about the time $t=1500$ns when
 the high temperature spots mainly distribute scatteredly in the low temperature background. Figures \ref{Fig:Figu9} and \ref{Fig:Figu10} show the morphological analysis for cases with the same porosity but lower initial shocking strengths. Figure \ref{Fig:Figu9} is for the case with $v_{init}=400$m/s. Figure \ref{Fig:Figu10} is for the case with  $v_{init}=300$m/s. With the decrease of initial shock strength, the highest temperature $T_{\max }$ in the system becomes lower; the total fractional high temperature area $A$ for low threshold value, for example $DT=10$K, increases with time in a more linear way.

The comparison of Minkowski measures for cases with different initial shocking strengths is shown in Fig.
 \ref{Fig:Figu11}, where $\Delta =0.286$, $DT=50$K, $v_{init}=1000$ms, $500$m/s, $400$m/s, $%
300$m/s, and $200$m/s. The higher the initial shock strength, the larger the slope of $A(t)$ curve. The case with  $v_{init}=400$m/s shows the largest boundary length.
For this case, scattered high temperature spots dominate the condensed temperature field in the shock-loading process, while scattered low temperature spots dominate in the unloading process.
%%%%%%%%%%%%%%%%%%%%%%%%%%%%%%%%%%%%%%%%%%%%%%%%%%%%%%%%%%%%%%%

\paragraph{Dynamic similarities}
%%%%%%%%%%%%%%%%%%%%%%%%%%%%%%%%%%%%%%%%%%%%%%%%%%%%%%%%%%%%%%%%%%%%

\begin{figure*}[tbp]
\centerline{\epsfig{file= 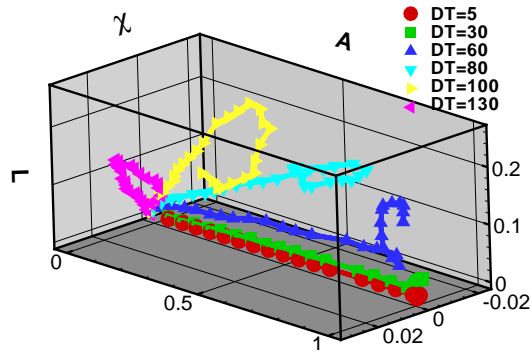, angle=0, bbllx=62 pt,bblly=253
pt,bburx=539 pt,bbury=574 pt, width=0.4\textwidth, clip=}}
\caption{(Color online) Minkowski description in the three-dimensional space
opened by $A$, $L$ and $\protect\chi$.
}
\label{Fig:DynSML_Fig2}
\end{figure*}

\begin{figure*}[tbp]
\centerline{\epsfig{file= 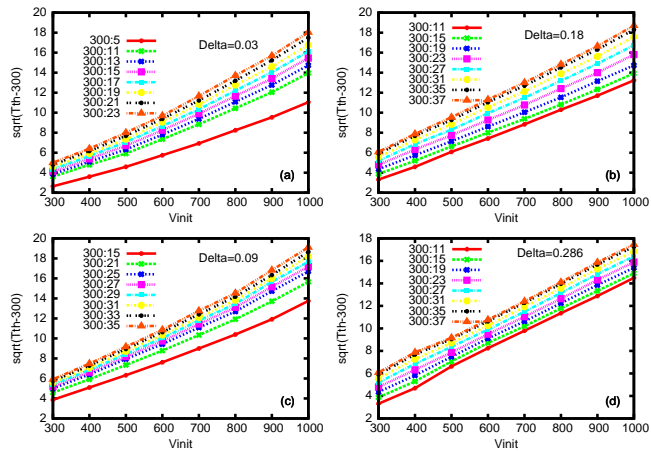, angle=0, bbllx=107 pt,bblly= 323
pt,bburx=463 pt,bbury=572 pt, width=0.49\textwidth, clip=}}
\caption{(Color online) $\protect\sqrt{T_{th}-300}$ versus $v_{init}$ for
similar processes. From (a) to (d), the porosity is 0.03, 0.18, 0.09, 0.286, respectively.
}
\label{Fig:DynSML_Fig3}
\end{figure*}

We present some results on dynamic similarities occurred in shocked porous material\cite{Xu-SciCN2010}.
Figure \ref{Fig:DynSML_Fig2} shows a set of morphological analysis for the case with $\Delta =0.09$ and  $v_{init}=800$ m/s in the three-dimensional space opened by the three Minkowski functions.
That the curves for $DT=5$K, $30$K and $60$K are closer indicates that the shape and connectivity of high temperature domains for these cases show high similarities.
We can check in which cases the condensed temperature fields show similarities. Such information is generally helpful
for classifying the shocking dynamic processes, material selection and structured material design.
For fixed material components and structures, an evolution process for a condensed temperature field is determined by the initial shock strength. We can label an evolution process of a condensed temperature field by a pair of quantities,    ($v_{init}$, $T_{th}$). We further add a subscripts, ``1'' or `` 2'', to
 index the evolution process. For cases with fixed porosity, $\Delta =0.09$, we first choose an evolution process with $v_{init1}=800$m/s and $T_{th1}=$408K, as the reference, check its process distance $d_{P}$ to the evolution
process with $v_{init2}=900$m/s and $T_{th2}$. It is easy to find that the distance $d_{P}$ takes its minimum value when $T_{th2}=442$K. In the same means, we can
know that the following evolution processes,
\textquotedblleft 300m/s,315K\textquotedblright ,
\textquotedblleft 400m/s,326K\textquotedblright ,
\textquotedblleft 500m/s,340K\textquotedblright ,
\textquotedblleft 600m/s,358K\textquotedblright ,
\textquotedblleft 700m/s,381K\textquotedblright ,
\textquotedblleft 1000m/s,489K\textquotedblright , show also high similarities to the reference process. If use the eight pairs of $v_{init}$ and $T_{th}$ to plot a curve, then we obtain the one labeled by
\textquotedblleft 300:15\textquotedblright\
in Fig.\ref{Fig:DynSML_Fig3} (c). The label ``300:15'' means processes shown in this curve show high similarities to the case where $v_{init1}=300$m/s and $DT=15$K.
If increases the temperature threshold of the reference process, we can obtain, in the same means, other curves shown in the figure. Figures \ref{Fig:DynSML_Fig3}(a),(b) and (d) are for the cases with $\Delta =0.03$, $\Delta =0.18$, $\Delta =0.286$, respectively.
From Fig. \ref{Fig:DynSML_Fig3} one can also find that, when the
porosity is high, the value of $\sqrt{T_{th}-300}$ shows nearly linearly dependence on the
impact velocity $v_{init}$. However, when the porosity is very small, the
increasing rate of $\sqrt{T_{th}-300}$ with respect to $v_{init}$ is larger. The physical reasons are as below.
The temperature increasing is mainly due to the plastic work when the porosity is high, while it is mainly from the shock compression when the porosity is very small\cite{Xu-SciCN2010}.

\begin{figure*}[tbp]
\centerline{\includegraphics*[scale=0.4,angle=270]{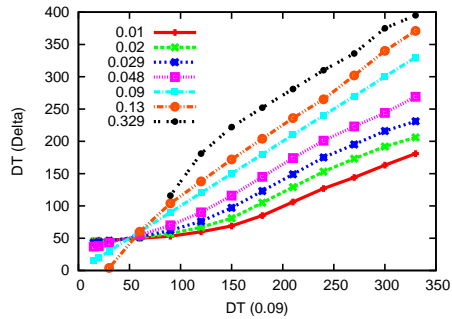}}
\caption{(Color online) $DT(\Delta_1)$ versus $DT(\Delta)$ for similar
dynamical processes. The horizontal axis shows the temperature threshold in
the reference material, $DT(\Delta_1=0.09)$. The vertical axis shows the
corresponding $DT$ values in the second material. The porosities of the
second material, 0.01,0.02,0.029,0.048,0.09,0.13, and 0.329, are shown in
the legend.
}
\label{Fig:DynSML_Fig4}
\end{figure*}

The shock strength, described by $v_{init}$, is fixed when we study the dynamical similarity $S_{P}$ versus porosity $\Delta $. In such cases, an evolution process can be labeled by the pair of values,
($\Delta $,  $T_{th}$).
We first consider the cases with $\Delta _{1}=0.09$ and $\Delta _{2}=0.01$. The initial impact velocity is fixed at $v_{init}=800$m/s.
Processes with $T_{th1} = 315$K, $320$K, $360$K, $390$K, $\cdots $, $630$K, show high similarities to
the processes with
$T_{th2} = 346$K, $347$K, $350$K, $353$K, $\cdots $, $481$K, respectively.
If use the values of $(T_{th1}-300)$ for $\Delta _{1}=0.09$ as
the $x$-coordinates and $(T_{th2}-300)$ for $\Delta _{2}=0.01$ as the $y$-coordinates, we obtain the curve labeled by ``0.01" in Fig.\ref{Fig:DynSML_Fig4}.
Figure \ref{Fig:DynSML_Fig4} shows also the cases with $\Delta _{2}= 0.02$, $0.029$, $0.048$, $0.09$, $0.13$, $0.329$.
An interesting results is that all curves meet at a point which corresponds to the temperature increment of uniform  material under the same shock strength.

\begin{figure*}[tbp]
\centerline{\includegraphics*[scale=0.4, angle=270]{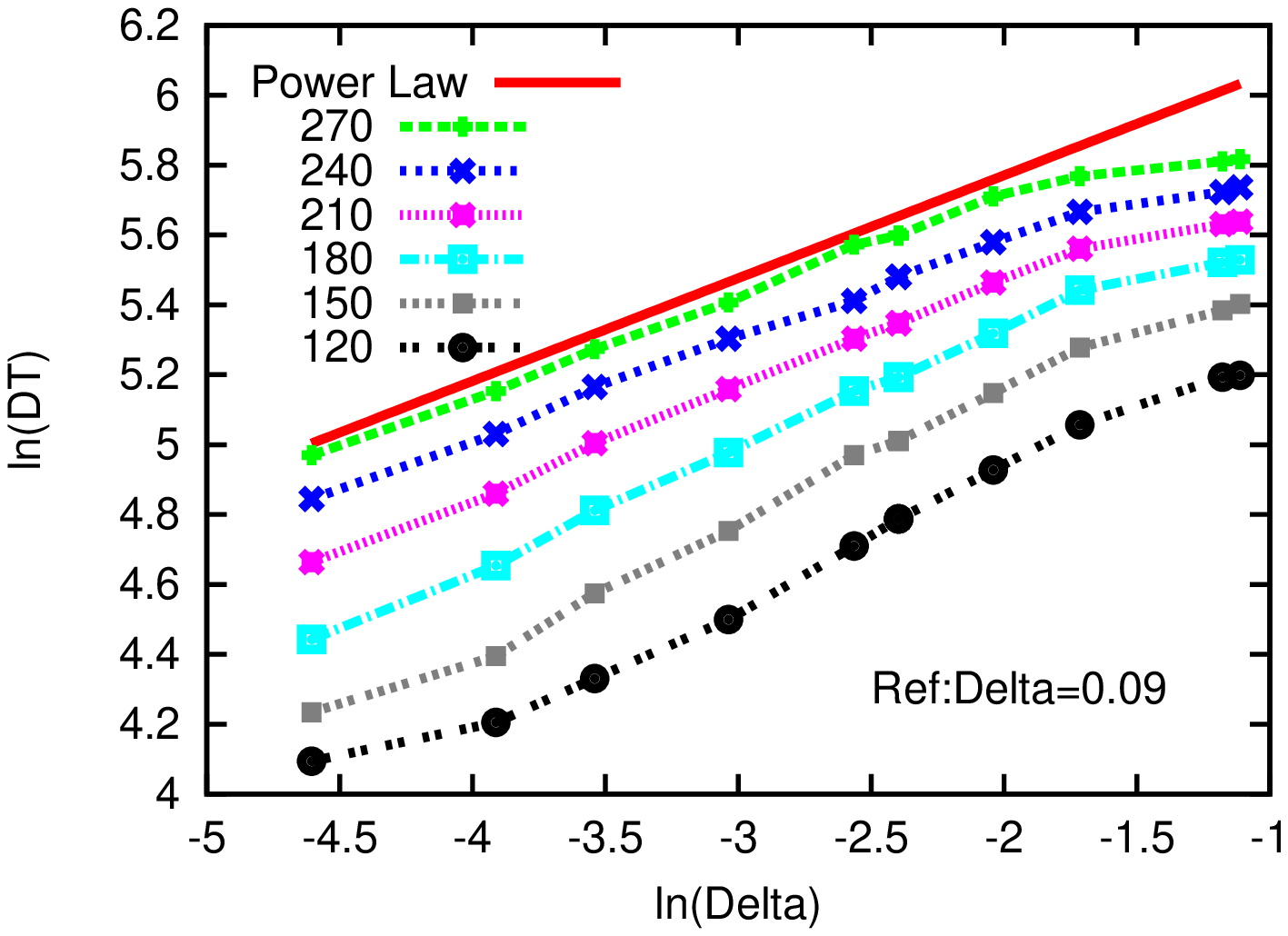}}
\caption{(Color online) Temperature threshold versus the porosity for
similar dynamical processes. The porosity of the reference material is 0.09,
the values of $T_{th}-300$, 270, 240, 210, 180, 150 and 120, are shown in
the legend. The plot is in log-log scale. A straight line is shown to guide
the eyes.}
\label{Fig:DynSML_Fig5}
\end{figure*}

We can get more observation under this line.
Under the same shock strength, the processes with
($\Delta_{2}=0.01$, $T_{th2}-300=144$K),
($\Delta _{2}=0.02$, $T_{th2}-300=173$K),
($\Delta _{2}=0.029$, $T_{th2}-300=195$K),
$\cdots $,
($\Delta _{2}=0.329$, $T_{th2}-300=336$K)
 show high similarities to the process with ($\Delta _{1}=0.09$,  $T_{th1}-300=270$K).
 If the relations between $(T_{th}-300)$ and $\Delta $ are shown in the log-log scale, we obtain the one labeled
by ``270" in Fig.\ref{Fig:DynSML_Fig5}. It is interesting to find that
the relation between $(T_{th}-300)$ and $\Delta $ follows a power-law relationship for a wide range.
There is a critical value for the porosity beyond which the power-law relation breaks.
Figure \ref{Fig:DynSML_Fig5} shows also the cases with $DT = 240$K, $210$K, $180$K, $150$K and $120$K.

In brief, the condensed temperature fields in shocked porous materials may show high similarities when the shock strength, porosity and temperature threshold are appropriately chosen.

%%%%%%%%%%%%%%%%%%%%%%%%%%%%%%%%%%%%%%%%%%%%%%%%%%%%%%%%%%%%%%%
%%%%%%%%%%%%%%%%%%%%%%            Fig.2
%%%%%%%%%%%%%%%%%%%%%%%%%%%%%%%%%%%%%%%%%%%%%%%%%%%%%%%%%%%%%%%
%%%%%%%%%%%%%%%%%%%%%%%%%%%%%%%%%%%%%%%%%%%%%%%%%%%%%%%%%%%%%%%
%%%%%%%%%%%%%%%%%%%%%%            Fig.3
%%%%%%%%%%%%%%%%%%%%%%%%%%%%%%%%%%%%%%%%%%%%%%%%%%%%%%%%%%%%%%%
%%%%%%%%%%%%%%%%%%%%%%%%%%%%%%%%%%%%%%%%%%%%%%%%%%%%%%%%%%%%%%%
%%%%%%%%%%%%%%%%%%%%%%            Fig.4
%%%%%%%%%%%%%%%%%%%%%%%%%%%%%%%%%%%%%%%%%%%%%%%%%%%%%%%%%%%%%%%
%%%%%%%%%%%%%%%%%%%%%%%%%%%%%%%%%%%%%%%%%%%%%%%%%%%%%%%%%%%%%%%
%%%%%%%%%%%%%%%%%%%%%%            Fig.5
%%%%%%%%%%%%%%%%%%%%%%%%%%%%%%%%%%%%%%%%%%%%%%%%%%%%%%%%%%%%%%%
%%%%%%%%%%%%%%%%%%%%%%%%%%%%%%%%%%%%%%%%%%%%%%%%%%%%%%%%%%%%%%%%%%%%

\subsection{ MPM investigations: Local behaviors}

In our MPM simulations, local behaviors are referred to the dynamical behaviors occurred around a single cavity in the material under shock loading or unloading process.

\subsubsection{Shock loading}

We set a single cavity in the simulated material. Due to the periodic boundary conditions applied in the horizontal conditions, what we considered corresponds to a very wide system with a row of cavities in it. We study various cases where the shock strength varies from strong to weak. In the cases with strong shock, the jet creation and the distribution of the ``hot spots'' are the main concern. When the cavity is close to the free surface, the critical condition for observing jetted material particles from the upper free surface is studied. In the cases with weak shock, we investigate the effects of cavity size, distance from the cavity center to the impacting interface, the initial yield stress of the material, etc, on the collapsing process\cite{JPCM2007}.

The main studies here include two parts: (i) the dependence of collapsing symmetry on the shock strength and other interfaces, (ii) hydrodynamic and thermodynamic behaviors ignored by the pure fluid models.
In the case with weak shock, an interesting observation is that
the cavity may not be collapsed completely and the cavity may collapse in a nearly isotropic means.
 In the case with strong shock, the jetting process is carefully studied.
 The specific collapsing process significantly affects the distribution of ``hot spots'' in the shocked material.

\begin{figure*}[tbp]
\centering
\includegraphics*[scale=0.6,angle=0,
bbllx=0pt,bblly=0pt,bburx=595pt,bbury=308pt
]{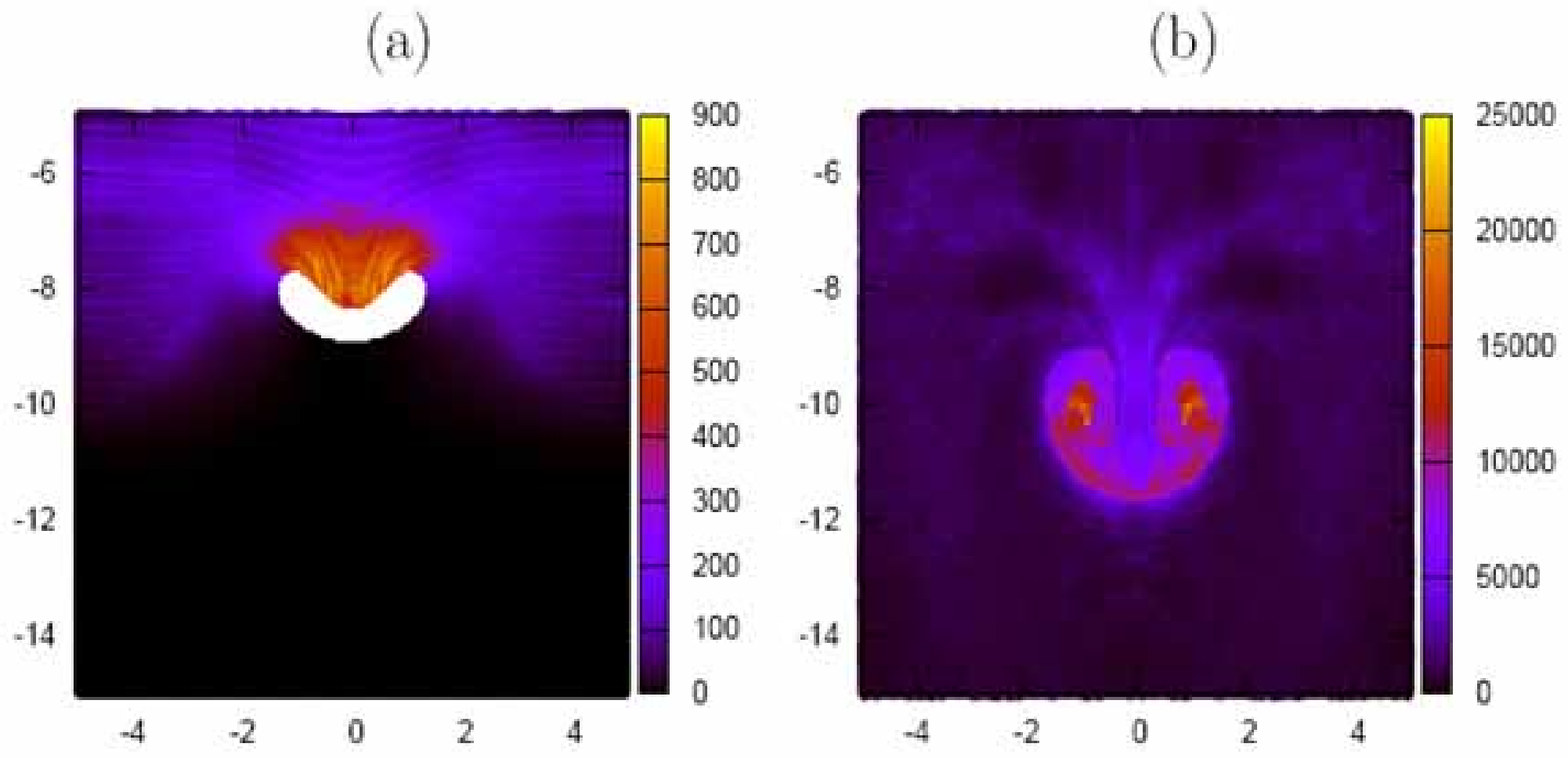}
\caption{(Color online) Snapshots of collapse of a single cavity under strong
shocks, where $v_{init}=1500$ m/s.
From black to yellow the grey-level in the figure shows the
increase of local temperature denoted by the plastic work during the
deformation procedure. The spatial unit is $\mu m$. The unit of work
is mJ. (a)t=2 ns, (b) t=5 ns.
(The gray-level version is published in Ref. \cite{JPCM2007}.)
}
\label{Fig:Figur1}
\end{figure*}
%%%%%%%%%%%%%%%
\begin{figure*}[tbp]
\centering
\includegraphics*[scale=0.6,angle=0,
bbllx=0pt,bblly=0pt,bburx=575pt,bbury=408pt
]{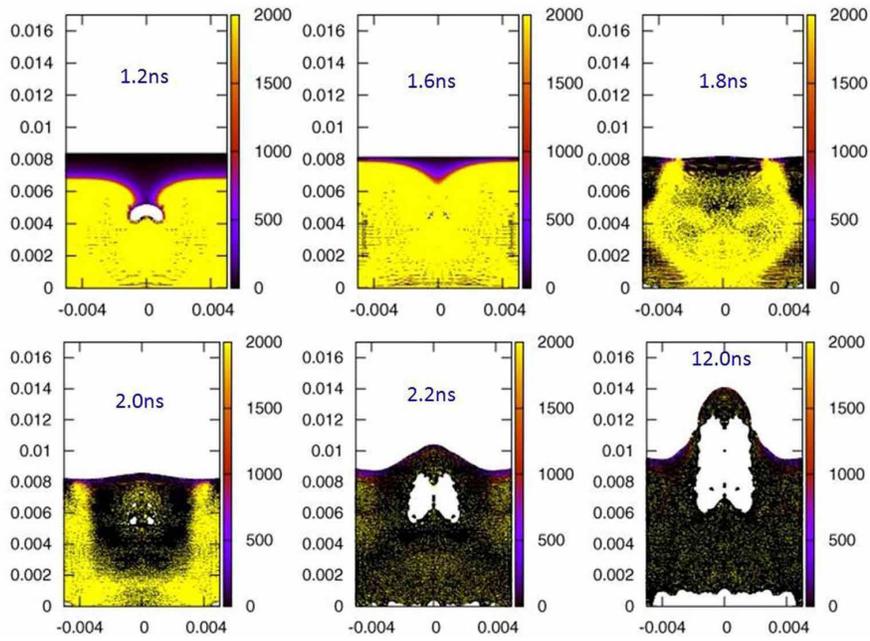} \caption{(Color online) Snapshots
of collapse of a single cavity under a strong shock. For the
pressure contour, from black to yellow the value increases. The
corresponding times at Figs.(a)-(f) are 1.2, 1.6, 1.8, 2.0, 2.2, 12
ns, respectively. The spatial unit is mm. The unit of pressure is
MPa.
(The gray-level version is published in Ref. \cite{JPCM2007}.)
}
\label{Fig:Figur2}
\end{figure*}

\begin{figure*}[tbp]
\centering
\includegraphics*[scale=0.6,angle=0,
bbllx=0pt,bblly=0pt,bburx=571pt,bbury=414pt
]{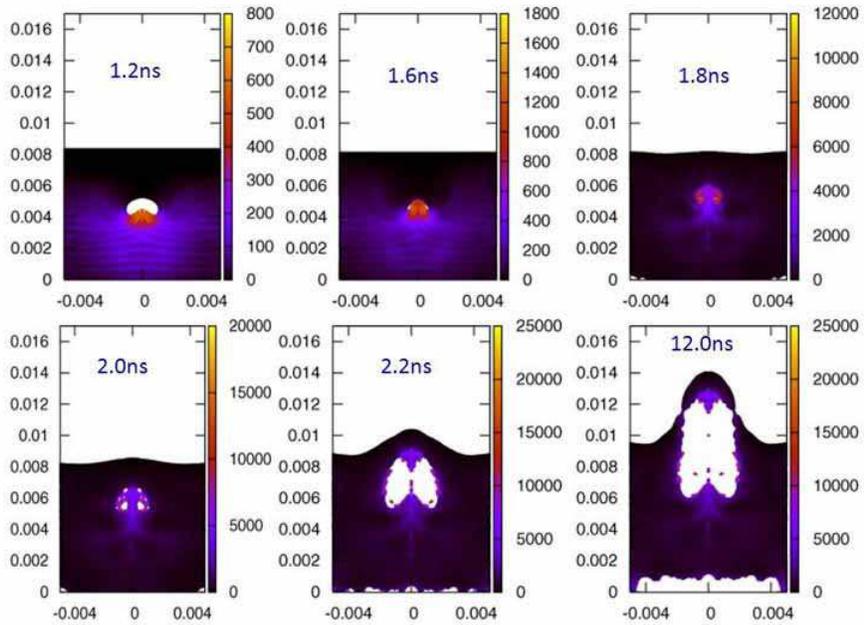}
\caption{(Color online) Configurations with local temperature denoted by the
plastic work during the deformation. The unit of work is mJ.
Figs.(a)-(f) here correspond to Figs.(a)-(f) in Fig.2, respectively.
(The gray-level version is published in Ref. \cite{JPCM2007}.)
}
\label{Fig:Figur3}
\end{figure*}

\begin{figure*}[tbp]
\centering
\includegraphics*[scale=0.7,angle=0,
bbllx=0pt,bblly=0pt,bburx=210pt,bbury=110pt
]{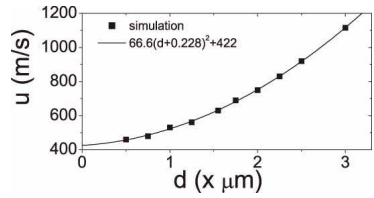}
\caption{Critical impact velocity versus the thickness of the downstream
wall of the cavity. The symbols are simulation results and the solid line is
the fitting result.
(Adopted with permission from Ref. \cite{JPCM2007}.)
}
\label{Fig:Figur4}
\end{figure*}
%%%%%%%%%%%%%%%%%%%%%%%%%%%%%%%%%%%%%%%%
\begin{figure*}[tbp]
\centering
\includegraphics*[scale=0.7,angle=0,
bbllx=0pt,bblly=0pt,bburx=415pt,bbury=265pt
]{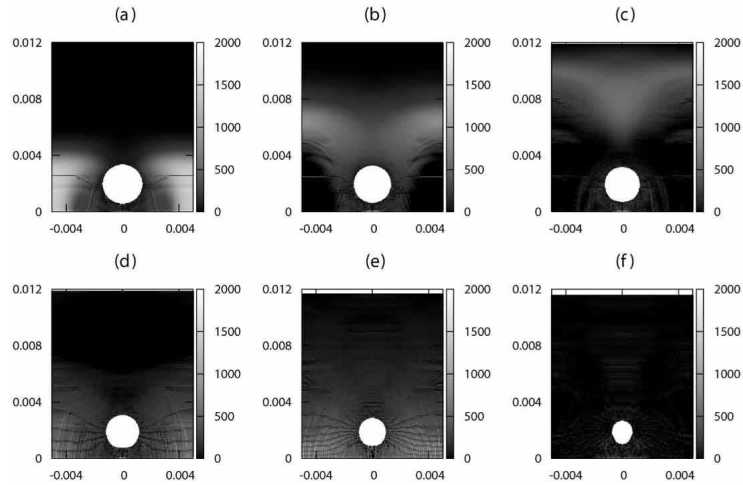}
\caption{ Snapshots of collapse of a single cavity under a weak
shock. From black to white the local pressure increases. The spatial unit is millimetre. The unit of
pressure is megapascal. (a) t=1.0 ns, (b) t=1.6 ns, (c) t=2.2 ns, (d) t=3.0
ns, (e) t=5.4 ns, (f) t=16.0 ns.
(Adopted with permission from Ref. \cite{JPCM2007}.)
}
\label{Fig:Figur5}
\end{figure*}

\begin{figure*}[tbp]
\centering
\includegraphics*[scale=0.7,angle=0,
bbllx=0pt,bblly=0pt,bburx=411pt,bbury=265pt
]{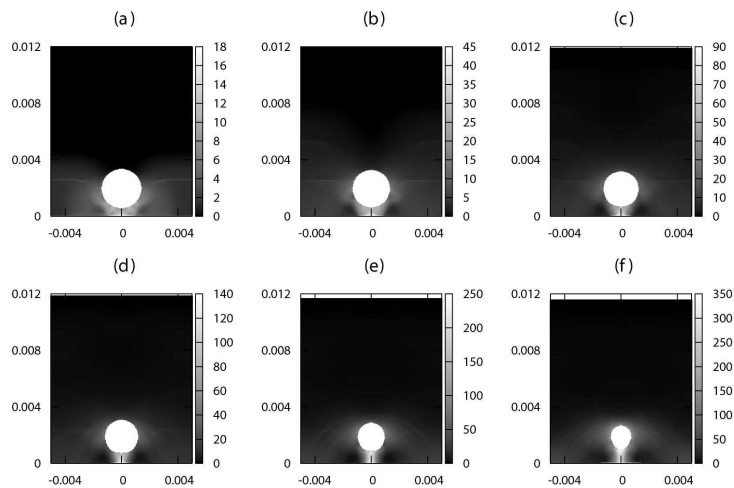}
\caption{ Configurations with local temperature denoted by the
plastic work during the deformation.
From black to white the
grey level in the figure shows the increase of local temperature. The spatial unit is millimetre. The unit of work is millijoule.(a) t = 1.0 ns, (b) t = 1.6 ns, (c) t = 2.2 ns, (d) t = 3.0 ns,
(e) t = 5.4 ns, (f) t = 16.0 ns.
(Adopted with permission from Ref. \cite{JPCM2007}.)
}
\label{Fig:Figur6}
\end{figure*}
%%%%%%%%%%%%%%%%%%%%%%%%%%%%%%%%%%%%%%%%

\begin{figure*}[tbp]
\centering
\includegraphics*[scale=0.6,angle=0,
bbllx=0pt,bblly=0pt,bburx=513pt,bbury=231pt
]{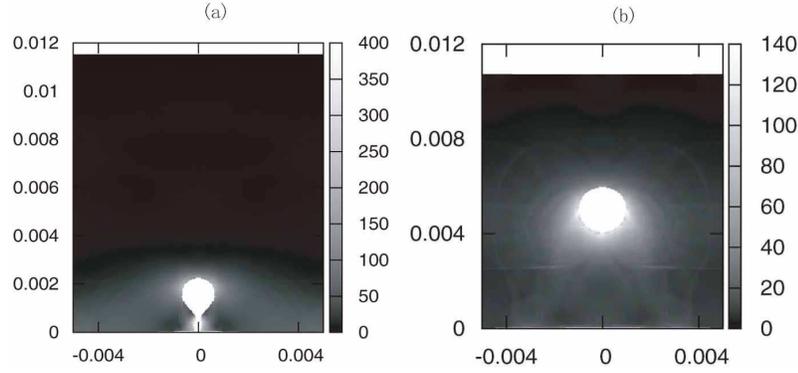}
\caption{ Transition of symmetry of collapsing. (a) Asymmetric in
vertical direction near the impacting face; (b) Nearly symmetric
collapse. The gray level in the figure corresponds to the plastic
work. The spatial unit in the figure is millimetre. The unit of energy is
millijoule.
(Adopted with permission from Ref. \cite{JPCM2007}.)
}
\label{Fig:Figur7}
\end{figure*}

\begin{figure*}[tbp]
\centering
\includegraphics*[scale=1.3,angle=0,
bbllx=0pt,bblly=0pt,bburx=251pt,bbury=178pt
]{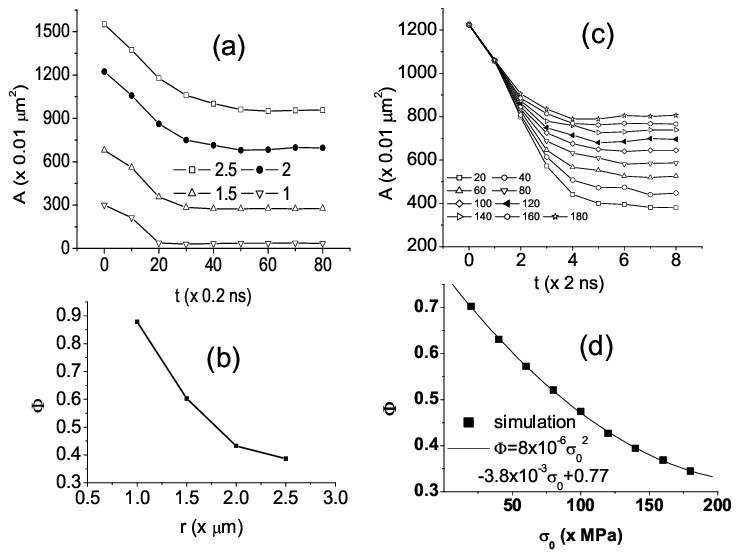}
\caption{ Effects of cavity size [(a) and (b)] and initial yield [(c)
and (d)] to the collapsing procedure. (a) Area of cavity versus time
for different initial radius. What shown in the legend are the
initial radii, the unit is $\mu$ m; (b) Collapsibility versus
initial radius of cavity; (c) Area of cavity versus time for
different initial yield stresses. What shown in the legend are the
initial yield stresses, the unit is megapascal. (d) Collapsibility versus
initial yield stress.
(Adopted with permission from Ref. \cite{JPCM2007}.)
}
\label{Fig:Figur8}
\end{figure*}
%%%%%%%%%%%%%%%%%%%%%%%%%%%%%%%%%%%%%%%%
%%%%%%%%%%%%%%%%%

 A snapshot for a case with high shock strength is shown in Fig. \ref{Fig:Figur1}.
The initial pressure loaded by the shock is about 30 GPa which is less than the critical value, 270 GPa. The global scenario is follows: (i) When the shock wave arrives at the upper-stream wall of the cavity, plastic deformation begins to occur. The shock waves at the two sides of the cavity propagate forwards to the free surface. The propagation speed of compressive waves at the two sides is larger than the deforming speed of the upper-stream wall of the cavity. (ii) The cavity continues to collapse, a configuration with a turned ``C" occurs. A jet phenomenon occurs.
``Hot spots'' occurs at the tip regime of the jet. (See Fig.\ref{Fig:Figur1}(a).) (iii) The propagation speed of jetted material increases with time. Along the initial shocking direction, the tip of the jet catches up, then exceeds the compressive waves at its two sides. (iv) The jetted material particles hit the down-stream wall of the cavity, leads to a pair of vortices rolling in opposite directions. The ``hot spots" appear at the vortex centers. (See Fig. \ref{Fig:Figur1} (b).)

 When the cavity is close to the upper free surface, if the shocking is strong enough, the material particles jetted into the cavity will break the down-stream wall of the cavity, and consequently, some material particles will be jetted out of the upper free surface. Such a behavior has been observed in experiments and has long been concerned.
A dynamical process is shown in Fig. \ref{Fig:Figur2}, where the initial cavity is under the free surface by a distance $d = 4.5 \mu$ m and the initial radius is $r = 1.5 \mu $m, $v_{init}= 1120$ m/s. The corresponding times for Figs. (a)-(f) are 1.2 ns, 1.4 ns, 2.0 ns, 2.4 ns, 4.4 ns, and 12.0 ns, respectively. Figure \ref{Fig:Figur2}(a) shows a snapshot where the shock wave has passed most part of the cavity. At this time the cavity has been substantially deformed and some material particles have been projected into the cavity. When $t=1.6$ ns,
the cavity has been nearly filled with the jetted material (see Fig. \ref{Fig:Figur2}(b)). The compressive wave first arrives at the upper free surface and rarefactive waves are reflected back from the two sides. The rarefactive waves reflected back into the shocked material decrease the pressure and new cavitation may occur around the region of the original cavity. (See Figs. \ref{Fig:Figur2} (c) - \ref{Fig:Figur2}(d).) Compared with those at the two sides, material in the middle has a much higher pressure and has much more kinetic energy. The newly created cavities coalescence and become a larger one with time. (See Fig. \ref{Fig:Figur2}(e).) If the upper wall of
the newly created cavity possesses enough kinetic energy, it will break. (See Fig. \ref{Fig:Figur2}(f).) The corresponding configurations with temperature is shown in Fig. \ref{Fig:Figur3}.
In Fig.(a) the ``hot spot'' appears at the tip of the material tongue. In Figs. (b)-(c) the ``hot spot'' occurs at regime hit by the material tongue. When new cavities are created, ``hot spots'' occur at the inner wall of the cavity, especially the upper and bottom walls. (See Figs. (d)-(f).) Whether or not there are material particles jetted from the upper free surface depends on the initial shocking strength and the width $d$. The critical initial velocity $u$ increases parabolically with $d$. (See Fig. \ref{Fig:Figur4}.)

With decreasing the shock strength, the collapsing process becomes slower. When the initial impact velocity decreases to about 200 m/s, the dynamical scenario has been significantly different. It is interesting to find that the cavity can not be fully collapsed, and the final configuration around the cavity varies from (nearly) symmetric to asymmetric, or vice verse, with changing the distance from the cavity center to the impacting surface. Figure \ref{Fig:Figur5} shows an asymmetric dynamical scenario where the cavity collapsed less in shocking direction.
The behavior can be understood via
``recovering" or ``magnifying" the system in such a way: The
rigid walls at the two sides and at the bottom of the system can be regarded as ``mirrors". The extended system is
symmetric about the ``mirrors". The distance between two neighboring cavities in the horizontal direction is $d_H$; and  the distance between the two cavities in the computed and the fictitious material bodies, respectively,  is $d_V$. The
cavities reflect rarefactive waves when compressive waves come, which makes lower the pressure in between two cavities. If the distance $d_V$ in vertical direction is much less than $d_H$ in horizontal direction, the rarefaction effects
in between the two cavities in vertical direction are more pronounced.
Therefore, the pressure in this region is lower than those in the surrounding regions.
At the same time, when the compressive waves arrive at the upper free surface, rarefactive waves will be reflected
back and propagate towards the cavity. This is a second reason why the cavity collapses less in the vertical direction. Compared with the rarefactive waves reflected from the fictitious cavity below the bottom, the ones from the upper free surface make effects in a much wider range, which is responsible for the phenomenon that
the collapsing of the lower part of the cavity is more pronounced.
Fig. \ref{Fig:Figur6} shows such an case. The highest temperature appears in the region below the cavity.
Figures \ref{Fig:Figur5}(a) and \ref{Fig:Figur6}(a) show that, although the rarefactive waves lower  the pressure in the influenced region, but may increase the temperature.
Because the rarefactive waves may make plastic work.

The temperature in the ``hot spot" continues to increase in the process of collapsing. The lower part of the cavity will collapse more pronouncedly if the distance between the cavity and the lower impacting face is further decreased. Figure  \ref{Fig:Figur7}(a) shows the final steady state for such a case, where the lower boundary of the cavity just locates at the impacting face. In contrast, if the distance between the cavity and the lower impacting face is increased, the
collapsed cavity will be more symmetric. Figure \ref{Fig:Figur7}(b) shows such a case, where the collapsing is nearly isotropic.

When study the effects of cavity size on the collapsing behaviors, the other conditions are kept unchanged. For the checked cases, the collapsibility increases with decreasing the initial cavity. The evolution of area of the
cavity is shown in Fig. \ref{Fig:Figur8}(a), where four different sizes are used.
If define the collapsibility as $\Phi=(A_0-A)/A_0$, then $\Phi$ decreases with increasing the
cavity radius, where $A_0$ and $A$ are the areas of the
cavity in the initial and final states. (See Fig. \ref{Fig:Figur8}(b).)
The effects of material strength is also investigated via changing the initial yield stress of
the material and keeping unchanged all other parameters. The corresponding collapsing processes are shown in
Fig. \ref{Fig:Figur8}(c). The collapsibility decreases nearly parabolically with increasing the initial yield. (See Fig. \ref{Fig:Figur8}(d).)

\subsubsection{Shock unloading}

Response and failure of ductile materials under dynamic loading are important and fundamental issues in the fields of science and engineering. The process of material failure is very complicated because various physical and mechanical mechanisms, in a wide spatial-temporal scales, are involved and coupled. The spallation and fragmentation of metal materials consist mainly of the following typical stages: nucleation, growth and coalescence of microscopic voids and/or larger scale cavities. The quasi-static growths of voids and cavities have been extensively studied. However, the  dynamical growth is much more complex and far from being well understood.

Early in 1972 Carroll and Holtz\cite{42} showed that the compression effect on the cavity growth is not pronounced if the material is not sensitive to the loading rate. The study was extended to the case of visco-plastic materials by Johnson\cite{43} in 1981. In 1987 Becker\cite{Becker1987} numerically analyzed the influences of a nonuniform distribution of porosity on the flow localization and failure in a porous material.
In this study, the void distribution and properties were obtained from measurements on partially consolidated and sintered iron powder. An elastic visco-plastic constitutive relation for porous plastic solids was used. The model considers local material failure via considering the dependence of flow potential on void volume fraction. The region modeled is a small portion of a larger body under various stress conditions. Under imposed periodic boundary conditions, both the plane strain and axisymmetric deformations were investigated.
It was found that interactions between regimes with higher fractions of void promote the plastic flow localization into a band. Local failure occurs through void growth and coalescence within the band. This study suggested a failure criterion based on a critical void volume fraction. The latter is only weakly dependent on the stress history. The critical void fraction is dependent on the initial void distribution and material hardening characteristics.
In 1992 Ortiz and Molinari\cite{44} investigated the influences of strain hardening and rate sensitivity on the dynamic growth of a void in a plastic material. They pointed out that the effects of inertia, hardening and loading rate can significantly influence the void growth.  The results of Benson\cite{45} in 1993 and those of Ramesh and Wright\cite{46} in 2003 indicate that the inertia effect is responsible for the stable growth of cavity.
In 1998, within the scope of the local approach methodology, Pardoen, et al.\cite{Pardoen1998} investigated the ductile fracture of round copper bars. They analyzed two damage models and comparatively studied four coalescence criteria.
The two damage models are the Rice-Tracey model and the Gurson-Leblond-Perrin model. The four coalescence criteria are as follows: (i) a critical value of the damage parameter, (ii) the Brown and Embury criterion, (iii) the Thomason criterion and (iv) a criterion based on the reaching of the maximum von Mises equivalent stress in a Gurson type simulation. Both the ellipsoidal void growth and void interaction were accounted for. In this study, as far as possible, all the parameters of the models were identified from experiments and physical observations.  Via using specimens which presents a wide range of notch radii, the effect of stress triaxiality was studied. Via comparing the behaviour of the material in the cold drawn state and in the annealed state, the effect of strain-hardening was analyzed.
In 2000 an extended model for void growth and coalescence was proposed by Pardoen et al.\cite{Pardoen2000}. This model integrated two existing contributions. The first is the Gologanu-Leblond-Devaux model extending the Gurson model to void shape effects. The second is the Thomason scheme for the onset of void coalescence. Each of these was extended heuristically to describe the strain hardening. To supplement the criterion for the onset of coalescence, a simple micro-mechanically based constitutive model for the void coalescence stage was proposed. The fully enhanced Gurson model depends on both the flow properties of the material and the dimensional ratios of the void-cell representative volume element. The effect of void shape, relative void spacing, strain hardening, and porosity are incorporated into it.
In 2001 an inelastic rate-dependent crystalline constitutive formulation and specialized computational schemes
 were developed by Orsini, et al.\cite{Orsini2001}. They aim
 to obtain a detailed understanding of the interrelated physical mechanisms which can result in ductile material failure in rate-dependent porous crystalline materials subjected to finite inelastic deformations.
It was shown that ductile failure can occur either due to void growth parallel to the stress axis or void interaction along bands. The former results in void coalescence normal to the stress axis. The latter are characterized by intense shear-strain localization and intersect the free surface at regions of extensive specimen necking.
In 2002, two mechanisms of ductile fracture, void by void growth and multiple void interaction, were discussed
by Tvergaard and Hutchinson\cite{2002a}; the plastic flow in porous material was discussed by
 Zohdi, et al.\cite{2002b}.

In most of current studies on cavity/void growth, the main concern is focused on its relevance on macroscopic behaviors\cite{42,43,Becker1987,44,45,46,Pardoen1998,Pardoen2000,Orsini2001,2002a,2002b}. The quantitative relations are determined via fitting experimental results. So, those studies were not capable of revealing or indicating the underlying idiographic physical and mechanical mechanisms of cavity/void growth. The cavity coalescence is the final stage of spallation developed from mesoscopic scale to macroscale\cite{27}. But this stage is also the least-known stage\cite{47,48,Zurek1998,49,50,51,52,53}. Continuous damage mechanic theory adopts fluid or solid model supplemented by some damage modeling. The damage is generally accounted for by an internal variable. The internal variable is generally defined by the variation of some mechanical behavior. It is not dynamically relevant to the particular structures.

\paragraph{Global scenario}
In our numerical experiments, the simulated aluminum material body is initially located within the volume, $[-20,20] \times [-20,20] \times [0,50]$. The length unit is $\mu$m. A spherical cavity with radius $r=5$ $\mu$m is located at the position $(0,0,z)$ within the material body. A rigid wall with $z=0$ is connected with the aluminum material body. The mesh size is $1 \mu$m. The diameter of the material particle is $0.5 \mu$m. Periodic boundary conditions are used in the horizontal directions. Free boundary condition is used in the upper surface. The rigid wall is assumed to be the same kind of material with the material body\cite{Xu-FoP2013}.
The material body starts to move upwards at the velocity $v_{z0}$ at the time $t=0$. Thus, the tensional wave or  rarefactive wave occurs at the interface with $z=0$. The rarefaction wave propagates upwards within the material body.

\begin{figure*}[tbp]
\center\includegraphics*%
[bbllx=0pt,bblly=0pt,bburx=533pt,bbury=476pt,angle=0,width=0.6\textwidth]
{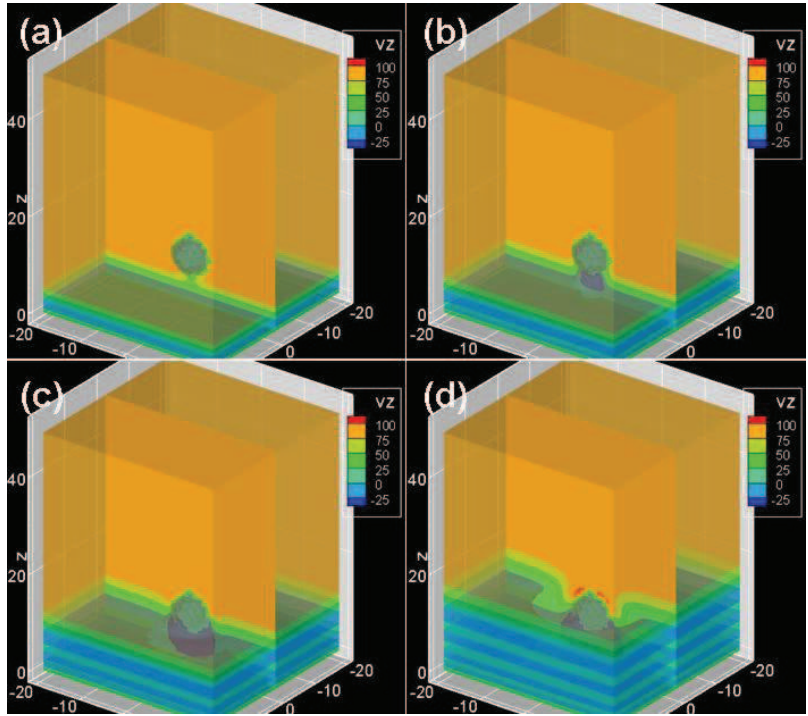}
\caption{(Color online) Configurations with $v_z$ field at four different times for the case with $z=10 \mu$m and initial $v_{z0}=100$m/s. The contours for $v_z=0$ are shown in the plots. (a)$t=0.8$ns, (b)$t=1.2$ns, (c)$t=2.0$ns and (d)$t=3.0$ns.
(Adopted with permission from Ref. \cite{Xu-FoP2013}.)
}
\label{Fig:Figure1}
\end{figure*}

A set of snapshots of configurations with $v_z$ field are shown in Fig. \ref{Fig:Figure1}, where $z=10 \mu$m, $v_{z0}=100$m/s. The contours for $v_z=0$ are shown. The velocities at the nodes within the cavity are equal zero
because no material particles are located within the cavity. Before the arrival of the global tensional wave, the  upper contour with $v_z=0$ within the body presents the initial morphology of the cavity. The moving upwards of the lower contour with $v_z=0$ shows the propagation of tensional wave. The lower contour for $v_z=0$ is approaching the lower boundary of the cavity at about $t=0.8$ns. Before that time the velocities of particles below the cavity had begun to decrease. It can also be found that, below the lower $v_z=0$ contour, some material particles show negative velocities. With propagating of the tensional wave, the lower contour for $v_z=0$ begins to get connection with that corresponding to the cavity. When the tensional wave arrives at the cavity, compression wave is reflected back. Under the compression wave, more material particles obtain negative velocities, and their amplitudes continue to increase.[See Figs.(b) and (c).] Compared with the propagation of tensional wave surrounding the cavity, the cavity deformation is a slower behavior. The surrounding tensional waves begin to converge after passing the cavity. Thus, higher negative pressure occurs on the top of the cavity. Consequently, material particles on the top of the cavity are accelerated by the upward stresses. At about $t=3$ns some material particles obtain velocities larger than $100$m/s. (See Fig.(d)).

\begin{figure*}[tbp]
\center\includegraphics*%
[bbllx=0pt,bblly=0pt,bburx=583pt,bbury=497pt,angle=0,width=0.6\textwidth]
{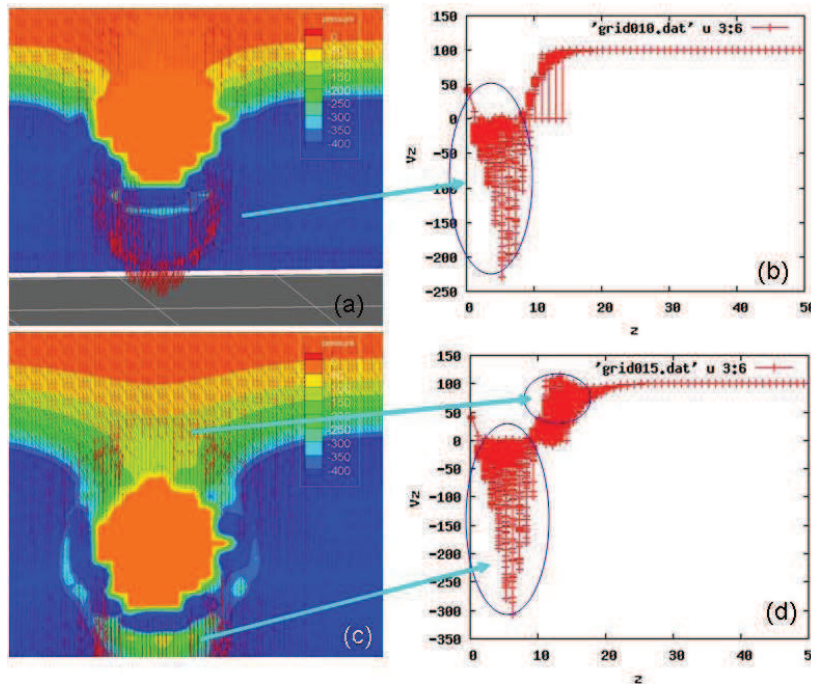}
\caption{(Color online) Configurations with pressure and velocity fields in the plane with $x=0$ [see (a) and (c)] and $v_z$ distribution in the tensile direction [see (b) and (d)]. $t=2$ns in (a) and (c). $t=3$ns in (b) and (d).
(Adopted with permission from Ref. \cite{Xu-FoP2013}.)
}
\label{Fig:Figure2}
\end{figure*}

Figure \ref{Fig:Figure2}(a) shows the configuration with pressure and velocity fields within the $x=0$ plane, where  $t=2$ns. Figure \ref{Fig:Figure2}(c) shows that for $t=3$ns.  From blue to red, the color corresponds to the increase of pressure. One can observe the deformation of the cavity under tensile loading.
To study the behaviors in amplitudes of particle velocities, the distributions of $v_z$ along the $z$-direction are shown in Figs. \ref{Fig:Figure2}(b) and \ref{Fig:Figure2}(d), respectively. When $t=2$ns, the maximum downward particle velocity is about $230$m/s. When $t=3$ns, it is about $300$m/s. Figure \ref{Fig:Figure2}(d) shows also that some material particles have upward velocities larger than $100$m/s.
The morphology irregularities of the cavities shown in the figures result from the following three reasons. (i) Due to the discreteness of the material particles in the MPM, the initial cavity is not strictly spherical.
(ii) The mesh size is not negligible compared with the dimension of the cavity.
(iii) The compressive waves reflected back from the cavity result in the Richtmyer-Meshkov(RM) instability, which is the main physical reason for the initial irregularities of the deformed cavity.
For the first two points, it should be pointed out that, in practical simulations,
we have to make compromise between the size of simulated material body, cavity size and the mesh unit. It should also be commented that the cavities in practical materials are generally not strictly spherical, which is partly simulated by the finite size effects of mesh and particles.

With the increasing of upward stress acting on particles on the top region of the cavity, the accelerations and velocities of particles increase. The global tensional wave arrives at the upper free surface at bout $t=7.2$ns. The maximum value of particle velocities on the top region of the cavity is about $430$m/s. There exists a region below the cavity within which the particles have large downward velocities at this moment. The largest value of downward velocity is about $325$m/s. There exists a valley between the peak and the tensional wave front in the curve of $v_z$ versus $z$. The smallest value of particle velocities is about $6$m/s. When the tensional wave arrives at the upper free surface,  compression wave is reflected back into the material. The material particles, in the region scanned by the reflected compression wave, obtain downward accelerations. Several characteristics are typical for the compression wave.
(i) The velocities of material particles, representing the upper free surface, decrease.
(ii) The valley continues to move toward the upper free surface.
(iii) The maximum velocity between the valley and the cavity continues to increase.
At the same time, the region with maximum downward particle velocity moves toward the bottom.
The simulation results for the case with  single cavity are also indicative for interaction of neighboring cavities
because periodic boundary conditions are used in the horizontal directions.
From the pressure field, the negative pressures within regions among the neighboring cavities are weaker. The contours for negative pressure with small amplitudes are connected. The compression wave reflected back from the cavity becomes stronger with the increasing of the strength of tensional wave. Consequently, regions with local positive pressures occur among the neighboring cavities. The occurrence of positive pressures within the background regime, where material particles are scanned by the global tensional wave, is a typical cavity effect.

\begin{figure*}[tbp]
\center\includegraphics%
[bbllx=0pt,bblly=0pt,bburx=545pt,bbury=323pt,angle=0,width=0.5\textwidth]
{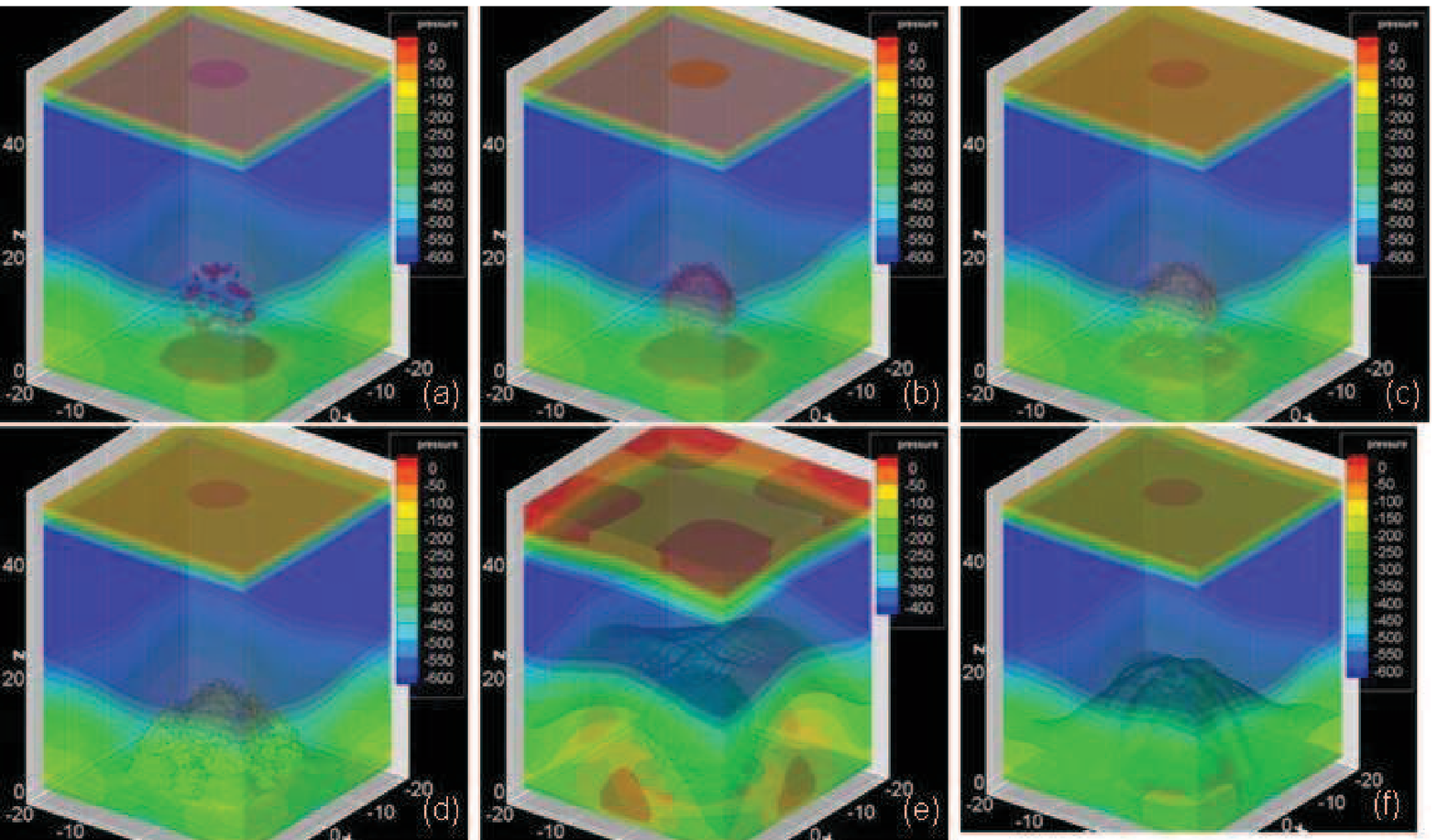}
\caption{(Color online) Configurations with pressure field at the time $9$ns. The pressure contours in (a)-(f) correspond to $-300$MPa, $-350$MPa, $-400$MPs, $-450$MPa, $-500$MPa, and $-550$MPa, respectively.
(Adopted with permission from Ref. \cite{Xu-FoP2013}.)
}
\label{Fig:Figure3}
\end{figure*}

\begin{figure*}[tbp]
\center\includegraphics*%
[bbllx=0pt,bblly=0pt,bburx=547pt,bbury=316pt,angle=0,width=0.5\textwidth]
{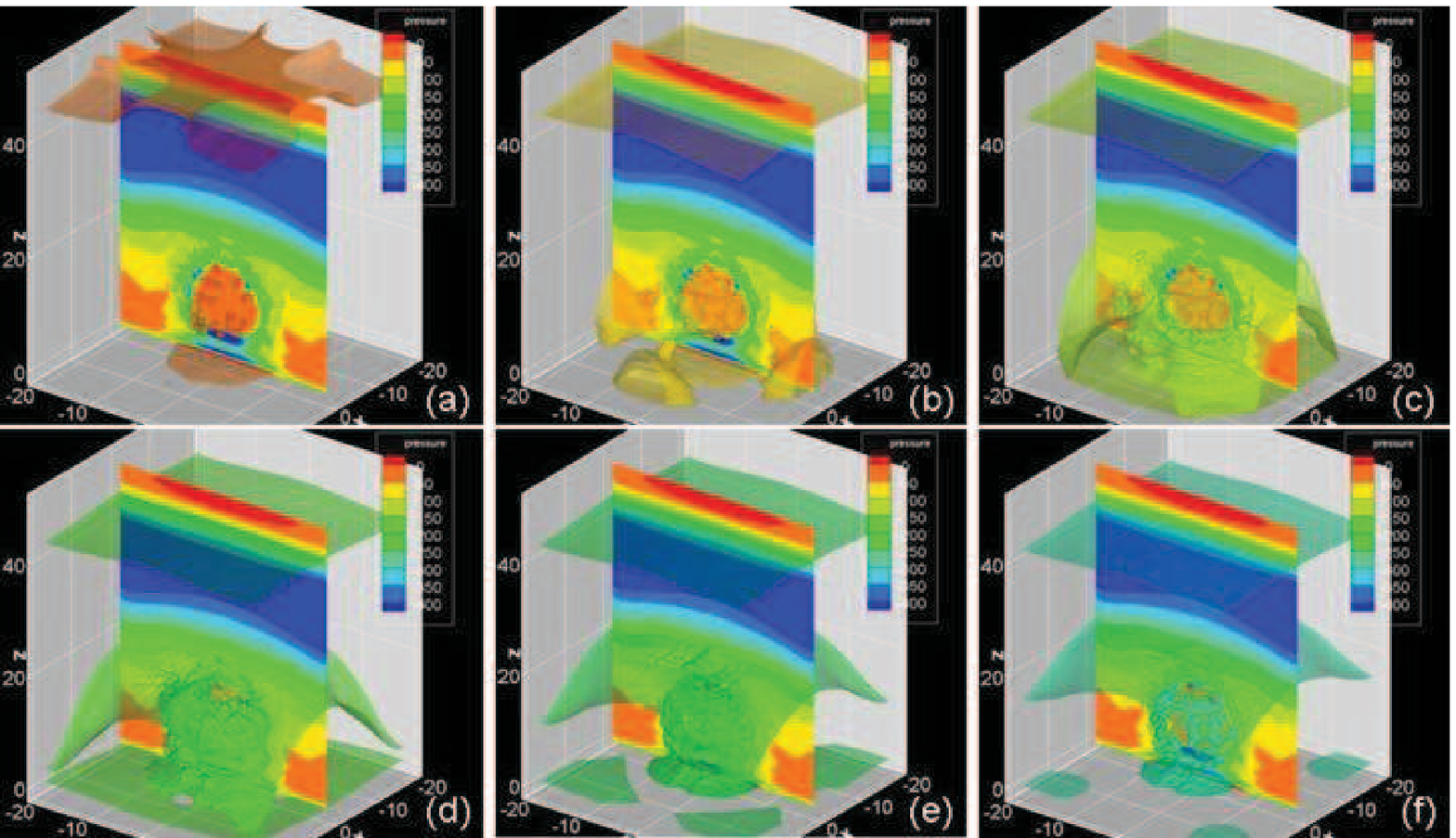}
\caption{(Color online) Configurations with pressure field at the time $11$ns. The pressure contours in (a)-(f) correspond to $0$MPa, $-50$MPa, $-100$MPs, $-150$MPa, $-200$MPa, and $-250$MPa, respectively.
(Adopted with permission from Ref. \cite{Xu-FoP2013}.)
}
\label{Fig:Figure4}
\end{figure*}

The deformation of the cavity is still controlled by the tensile loading
before the reflected compression wave arrives. As an example,
we analyze the pressure distributions at two times, 9 ns and 11 ns. The configurations with pressure field at the time 9ns is shown in Fig. \ref{Fig:Figure3}, where the contour pressures are $-300$Mpa, $-350$Mpa, $-400$Mpa, $-450$Mpa, $-500$Mpa and $-550$Mpa, respectively. Around the cavities,
the contours for pressure lower than $-300$ Mpa are connected. The neighboring cavities interact via the connection of pressure contours. There is still no positive pressure occur among the neighboring cavities when $t = 9$ns. Figure \ref{Fig:Figure4} shows various pressure contours, where  $t = 11$ ns, and the contour pressures are $0$Mpa, $-50$Mpa, $-100$Mpa, $-150$Mpa, $-200$MPa and $-250$Mpa, respectively. Observations on pressure distribution around the cavity are as below. (i) The pressure surrounding the cavity is zero. (ii) The corresponding contour moves away from the cavity and its surface area becomes larger
 with the increasing of pressure.
(iii) Among cases shown in the figure, the contour for $-150$Mpa has the maximum area.
The contour area becomes smaller if further increase the pressure value. Observations on pressure distribution between the cavity and rigid wall are as below. There are four regimes, around the cavity, where the pressures are positive. The pressure contours for -100Mpa, between the nearest cavities, are connected. The contours for $-150$Mpa have a higher connectivity. All contours for $-200$Mpa, $-250$Mpa, etc. are connected. The pressure distribution on the top of the cavity shows the following behaviors. The highest pressure regime does not locate above the cavity but above the middle of neighboring cavities.  Because the tensional wave propagates more quickly within the solid region, the wave firstly arrives at the upper free surface. The weaker the negative pressure, the closer to the upper free surface and the planar the corresponding pressure contour.

%%%%%%%%%%%%%%%%%%%%%%%%%%%%%%%%%%%%%%%%
%%%%%%%%%%%%%%%%%%%%%%%%%%%%%%%%%%%%%%%%%

%%%%%%%%%%%%%%%%%%%%%%%%%%%%%%%%%%%%%%%%
%%%%%%%%%%%%%%%%%%%%%%%%%%%%%%%%%%%%%%%%%
%%%%%%%%%%%%%%%%%%%%%%%%%%%%%%%%%%%%%%%%
%%%%%%%%%%%%%%%%%%%%%%%%%%%%%%%%%%%%%%%%%

%%%%%%%%%%%%%%%%%%%%%%%%%%%%%%%%%%%%%%%%
%%%%%%%%%%%%%%%%%%%%%%%%%%%%%%%%%%%%%%%%%
%%%%%%%%%%%%%%%%%%%%%%%%%%%%%%%%%%%%%%%%
\begin{figure*}[tbp]
\center\includegraphics*%
[bbllx=0pt,bblly=0pt,bburx=536pt,bbury=476pt,angle=0,width=0.6\textwidth]
{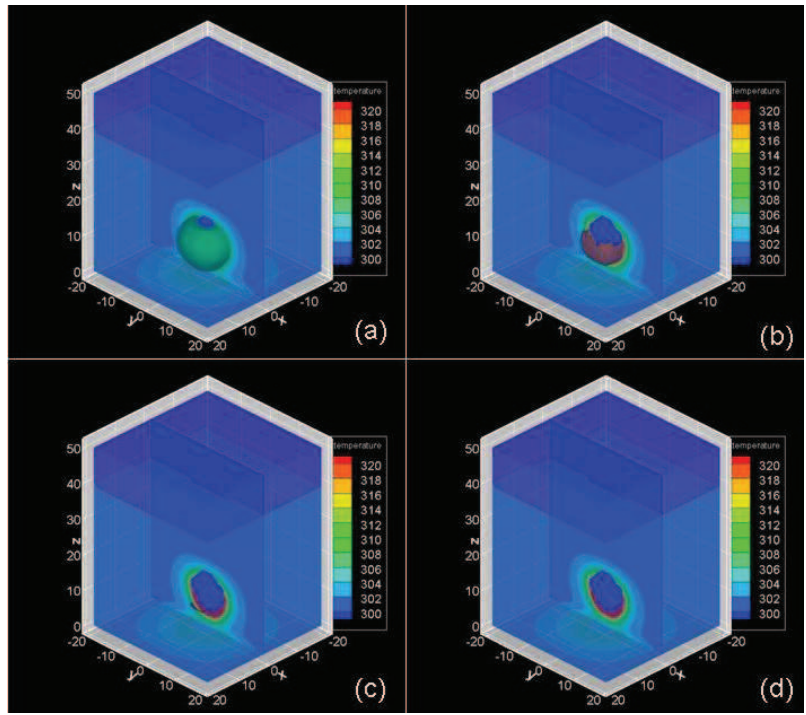}
\caption{(Color online) Configurations with temperature field at the time $6$ns. The contours in (a)-(d) correspond to 310K, 320K, 330K and 340K, respectively.
(Adopted with permission from Ref. \cite{Xu-FoP2013}.)
}
\label{Fig:Figure5}
\end{figure*}
%%%%%%%%%%%%%%%%%%%%%%%%%%%%%%%%%%%%%%%%%

%%%%%%%%%%%%%%%%%%%%%%%%%%%%%%%%%%%%%%%%
\begin{figure*}[tbp]
\center\includegraphics*%
[bbllx=0pt,bblly=0pt,bburx=550pt,bbury=749pt,angle=0,width=0.40\textwidth]
{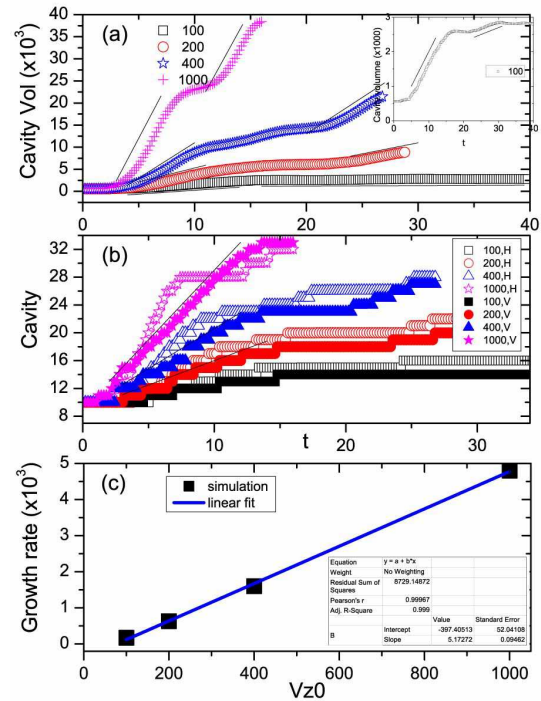}
\caption{(Color online) Evolution of the cavity morphology. (a) Cavity volume versus time. (b) Cavity dimensions in the Horizontal(H) and Vertical(V) directions versus time. (c) The linear growth rate versus initial tensile velocity. The sizes of the initial tensile velocity $v_{z0}$, 100, 200, 400 and 1000, are shown in the legend of Fig.(a). The unit is m/s.  In Figs.(a) and (b) the points are simulation results and the lines are plotted to guide the eyes. An enlarge portion of the curve for $v_{z0}=100$ is shown in the inset of Fig.(a). In Fig.(c) the points are for the slopes of fitting lines in Fig. (a) for the first linear growth stage, and the line are linear fitting result for the points.
(Adopted with permission from Ref. \cite{Xu-FoP2013}.)
}
\label{Fig:Figure6}
\end{figure*}
%%%%%%%%%%%%%%%%%%%%%%%%%%%%%%%%%%%%%%%%%

%%%%%%%%%%%%%%%%%%%%%%%%%%%%%%%%%%%%%%%%
\begin{figure*}[tbp]
\center\includegraphics*%
[bbllx=0pt,bblly=0pt,bburx=554pt,bbury=609pt,angle=0,width=0.5\textwidth]
{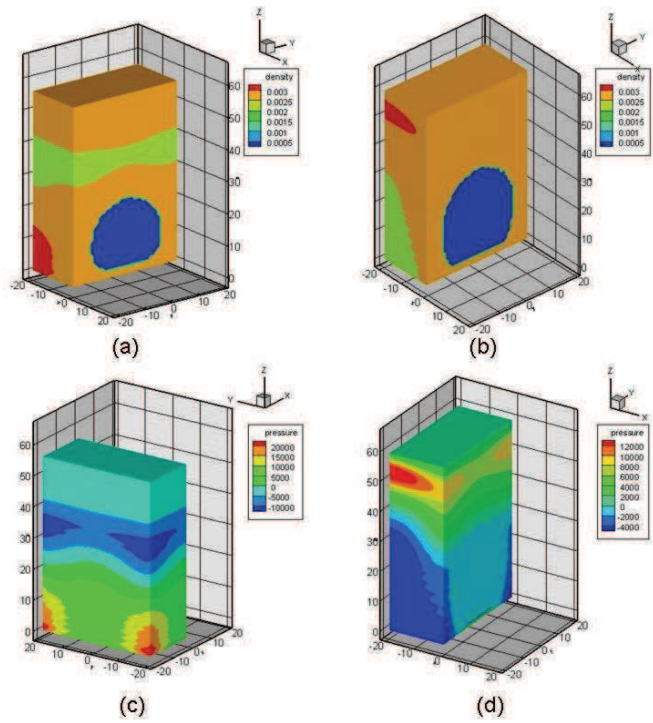}
\caption{(Color online) Configurations with density field [(a) and (b)] and configurations with pressure field [(c) and (d)] at two times, 7.2ns and 12ns. Only the portion with $-20 \leq x \leq 0$ is shown in each plot.
(Adopted with permission from Ref. \cite{Xu-FoP2013}.)
}
\label{Fig:Figure7}
\end{figure*}
%%%%%%%%%%%%%%%%%%%%%%%%%%%%%%%%%%%%%%%%%
%%%%%%%%%%%%%%%%%%%%%%%%%%%%%%%%%%%%%%%%
\begin{figure*}[tbp]
\center\includegraphics*%
[bbllx=0pt,bblly=0pt,bburx=552pt,bbury=600pt,angle=0,width=0.4\textwidth]
{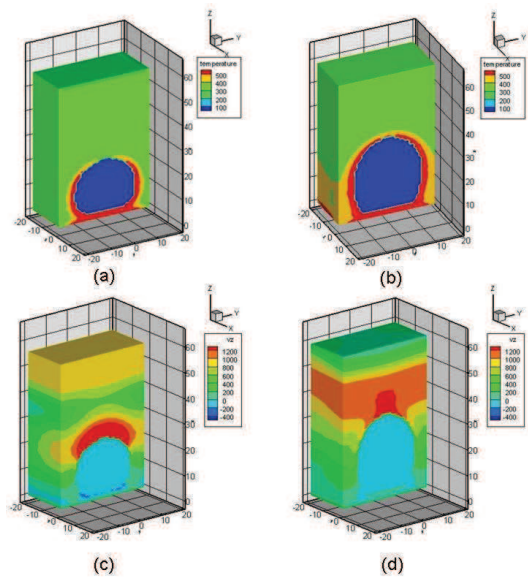}
\caption{(Color online) Configurations with temperature field [(a) and (b)] and configurations with $v_z$ field [(c) and (d)] at two times, 7.2ns and 12ns. Only the portion with $-20 \leq x \leq 0$ is shown in each plot.
(Adopted with permission from Ref. \cite{Xu-FoP2013}.)
}
\label{Fig:Figure8}
\end{figure*}
%%%%%%%%%%%%%%%%%%%%%%%%%%%%%%%%%%%%%%%%%

%%%%%%%%%%%%%%%%%%%%%%%%%%%%%%%%%%%%%%%%
\begin{figure*}[tbp]
\center\includegraphics*%
[bbllx=0pt,bblly=0pt,bburx=579pt,bbury=604pt,angle=0,width=0.4\textwidth]
{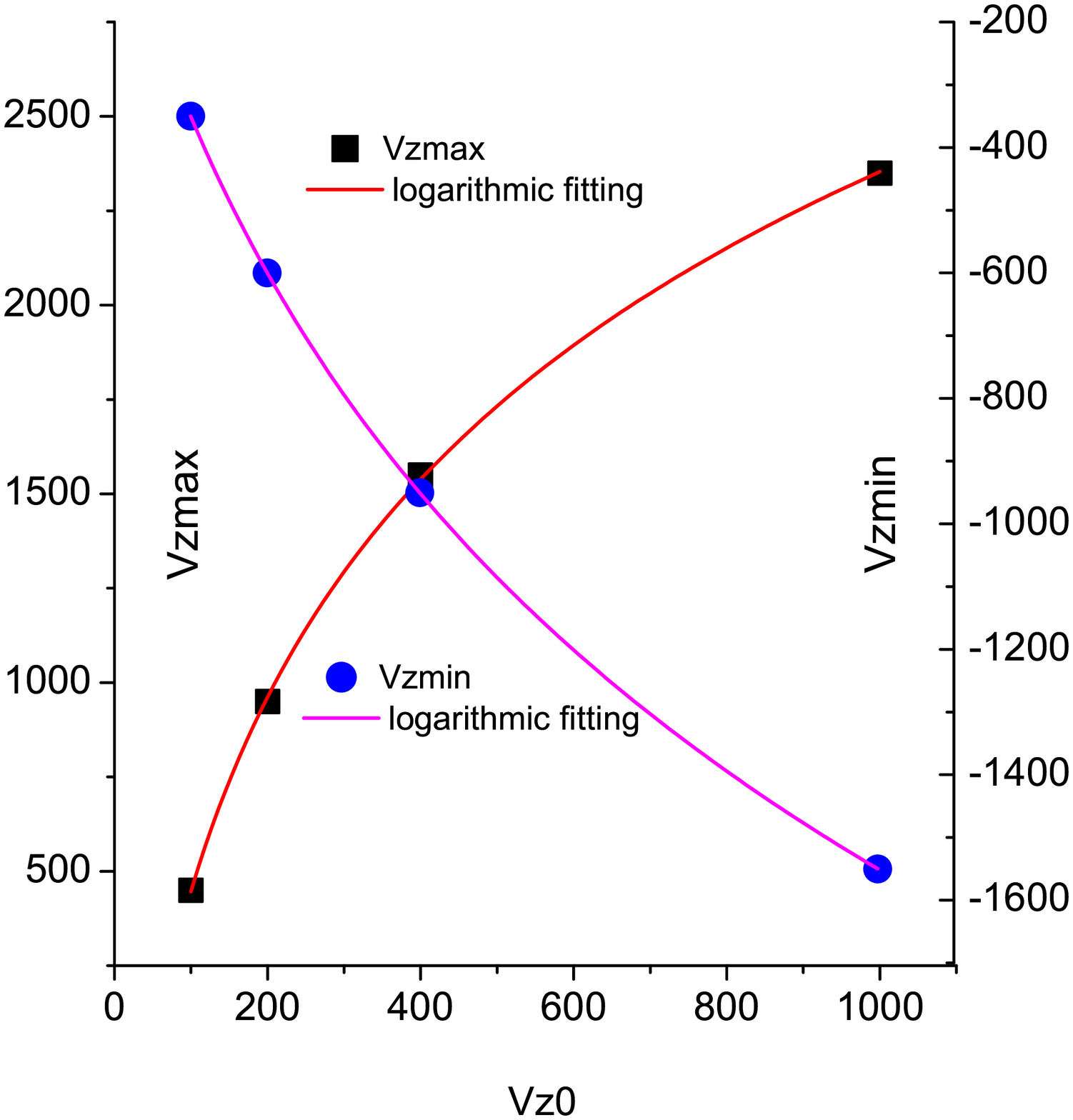}
\caption{(Color online) Maximum upward particle velocity and maximum downward particle velocity versus initial tensile velocity $v_{z0}$. The points are simulation results and the lines are logarithmic fitting results.
(Adopted with permission from Ref. \cite{Xu-FoP2013}.)
}
\label{Fig:Figure9}
\end{figure*}
%%%%%%%%%%%%%%%%%%%%%%%%%%%%%%%%%%%%%%%%%

\paragraph{Morphology versus tensile strength}

The configurations with temperature field are shown in
Fig. \ref{Fig:Figure5}, where $t = 6$ ns and the contour temperatures are $310$K, $320$K, $330$K and $340$K, respectively.  The thermal process is much slower compared with the dynamical process. It is the plastic work that determines the temperature and distribution of hot-spots.

All tensional waves arrive at the upper free surface at the same time because they propagates in the same speed, sound speed. The growth rate of the cavity gets larger with the increase of tensile strength.
The evolution of the cavity morphology is shown in Fig.\ref{Fig:Figure6}. Figure \ref{Fig:Figure6}(a) is for the cavity volume versus time. The points show simulation results and the lines are plotted to guide the eyes.
The initial tensile velocities, 100, 200, 400 and 1000, are shown in the legend, where the unit is m/s. The inset
shows an enlarged portion of the curve for $v_{z0}=100$m/s.
The growth of cavity can be described by the following stages. (i) initial slow growth stage, (ii) linear growth stage,  which ends when the global rarefaction wave arrives at the upper free surface, (iii) slower growth stage, which ends when the reflected compression wave arrives at the cavity, (iv) quicker growth stage, and (v) linear growth stage.
The evolutions of the cavity dimensions in Horizontal(H) and Vertical(V) directions are shown in Fig.\ref{Fig:Figure6}(b).
The points show simulation results and the lines are plotted to guide the eyes.
An interesting phenomenon is that the growth rate in horizontal direction is larger than that in vertical direction. Such a phenomenon corresponds to the ``necking effect" in macroscopic scale. The growths of cavity dimensions show also a linear stage. With the increase of the strength of tensile loading, the growth rates becomes larger.
The initial linear growth rate of cavity volume versus initial strength of tensile loading $v_{z0}$ is shown in
Fig.\ref{Fig:Figure6}(c). The points show the slopes of fitting lines in Fig. (a) for the first linear growth stage, and the line are linear fitting results for the points.
Within the checked range, with increasing the initial tensile velocity $v_{z0}$, the volume growth rate increases linearly.
Figure \ref{Fig:Figure7}(a)   shows the density fields of the material at the time 7.2ns.
Figure \ref{Fig:Figure7}(b) is for the time, $t=12$ns.
The corresponding pressure fields are shown in Figs. \ref{Fig:Figure7}(c) and \ref{Fig:Figure7}(d).

\paragraph{Energy transformation versus tensile strength}

During the process of tensile loading, kinetic energy of the material transforms gradually to elastic potential energy and plastic work. For the case with uniform material, those energies distribute uniformly in planes parallel to the rigid wall. The dynamical and thermodynamical process is in fact one-dimensional,
even though the material is three-dimensional.
The situation becomes much more complex when cavities exist.
The configurations with temperature field are shown in Figs. \ref{Fig:Figure8}(a) and \ref{Fig:Figure8}(b). The times are the same as in Fig.\ref{Fig:Figure7}. Besides the cavity morphology, one can understand better the energy transformation from kinetic to thermal. A high temperature layer, surrounding the deformed cavity, exists. Because the plastic work by the stresses is pronounced in that region. The configurations with $v_z$ field at the same two times are shown in Figs. \ref{Fig:Figure8}(c) and \ref{Fig:Figure8}(d), respectively. With the reflecting back of compression wave from the upper free surface, the range with high particle velocity decreases.

Figure \ref{Fig:Figure9} shows
 both the dependence of maximum upward particle velocity, $v_{z\max}$, above the cavity on the initial tensile velocity $v_{z0}$ and the dependence of maximum downward particle velocity, $v_{z\min}$, below the cavity versus $v_{z0}$.
 The points show simulation results and the lines show fitting results. Both $v_{z\max}$ and $|v_{z\min}|$ increase logarithmically with $v_{z0}$.

\subsection{ DBM investigations}

As mentioned above, many behaviors in heterogenous materials under strong shock can be  described and investigated by fluid models. At the same time, the non-equilibrium phase transition kinetics studied by the liquid-vapor model can help to understand the solid-solid phase transition in the metal under shock from two sides, the morphological and non-equilibrium behaviors.
When such a flow system is in a unstable state, the free energy of the system is too high compared with that in its
ground or metastable state. The inter-particle force will drive changes, and the gradient force induced by gradients of macroscopic quantities will opposes them.
From the view of phase transition kinetics, for a system instantaneously quenched from a disordered state into a two-phase coexistence one, the flows are generally known to undergo two TNE stages,
the early spinodal decomposition stage and the later domain growth stage,
before approaching the finial totally separated equilibrium state.
Previous studies were focused mainly on the domain growth law in the second stage, which is partly due to the uncertainty in quantitative determining of transition time from the former to the latter stages.
This uncertainty is due to the existence of large variety of complex spatial patterns, especially during the spinodal decomposition stage. In a recent work\cite{GXZ-EPL2012}, with the help of morphological analysis, our group presented a geometrical criterion for separating the two stages and some new insights into the first stage have been obtained. But roughly speaking, the spinodal decomposition stage has, far from, being well-understood.
At the same time, as a typical case of non-equilibrium phase transition, the TNE behaviors during the whole process are barely investigated.

The DBM is a kinetic method making it possible to investigate effectively the complex interplay between various non-equilibrium behaviors. Based on the measured TNE quantities, one can further define new quantities, like the TNE strength, to roughly estimate the deviation from the corresponding thermodynamic equilibrium.
In a recent study\cite{SoftMatter2015}, it is found that the TNE strength attains its maximum at the end of the spinodal decomposition stage. Consequently, the TNE intensity presents a convenient physical criterion to separate the two stages of phase separation. The effects of latent heat and surface tension on phase separation
were also studied via the DBM simulation. Some specific results are as below.

%%%%%%%%%%%%%%%%%%%%%%%%%%%%%%%%%%%%%%%%%%%%%%%%%%%%%%%%%%%%%%%%%%%%%%%%%%%%%%%%
\begin{figure*}[tbp]
\centering
\includegraphics*[ scale=0.7,angle=0,bbllx=106pt,bblly=270pt,bburx=576pt,bbury=635pt]{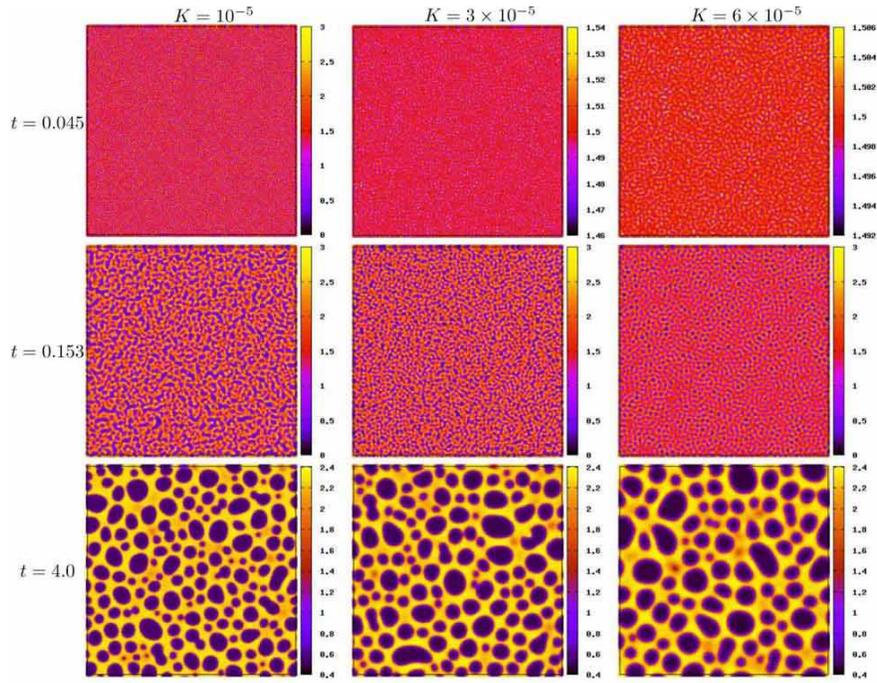}
\caption{(Color online) Density patterns at three representative times during thermal phase separation processes, $t=0.045$ (the first row), $t=0.153$
(the second row) and $t=4.0$ (the third row). From left to right, the three columns
correspond to cases with $K=10^{-5}$, $3\times 10^{-5}$ and $6\times10^{-5}$, respectively.
(Adopted with permission from Ref. \cite{SoftMatter2015}.)
}
\label{DBMFig9}
\end{figure*}
%%%%%%%%%%%%%%%%%%%%%%%%%%%%%%%%%%%%%%%%%%%%%%%%%%%%%%%%%%%%%%%%%%%%%%%%%%%%%%%%

%%%%%%%%%%%%%%%%%%%%%%%%%%%%%%%%%%%%%%%%%%%%%%%%%%%%%%%%%%%%%%%%%%%%%%%%%%%%%%%%
\begin{figure*}[tbp]
\centering
\includegraphics*[ scale=0.7,angle=0,bbllx=0pt,bblly=0pt,bburx=574pt,bbury=348pt]{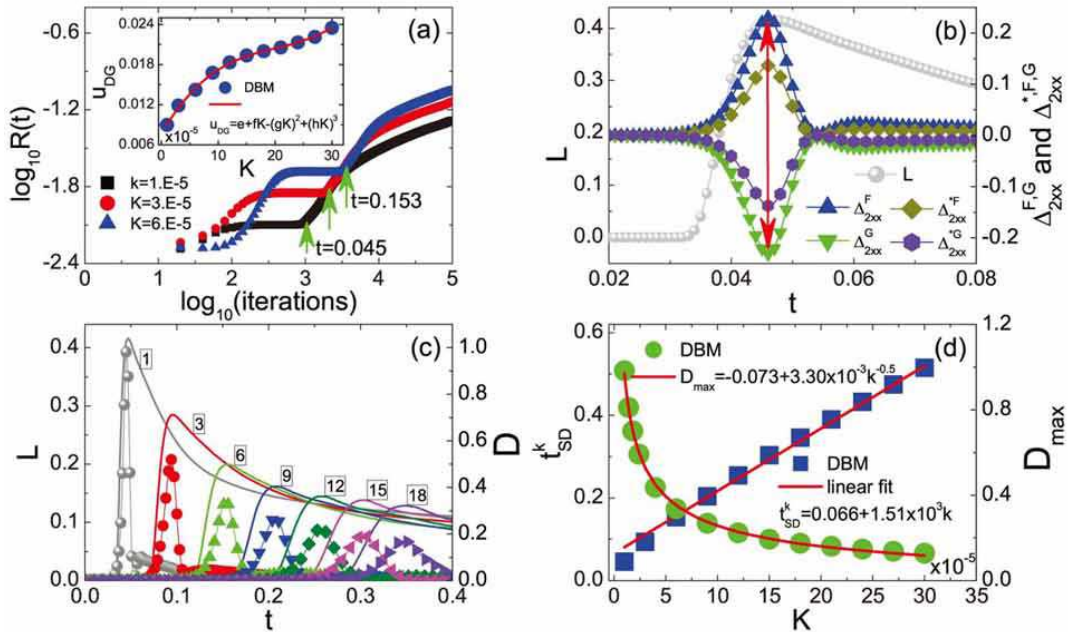}
\caption{(Color online) (a) Evolutions of the characteristic domain sizes $R$ for the procedures shown in Fig. \ref{DBMFig9}.
(b) Evolutions of the boundary length $L$ and the $xx$ component of some TNE manifestations for the phase separation process with $K=10^{-5}$. (c) Evolutions of the boundary lengths $L$ (solid curves) and the corresponding TNE intensities $D$ (curves with solid symbols) for phase separation processes with various $K$. Here $1,3,6,...,18$ labeled on the $L$-curves indicate cases with $K=10^{-5}$, $3\times10^{-5}$, $6\times 10^{-5}$,...,$1.8 \times 10^{-4}$, respectively.
(d) Duration of the spinodal decomposition stage $t_{SD}$ and the maximum TNE intensity $D_{\text{max}}$ as functions of $K$.
(Adopted with permission from Ref. \cite{SoftMatter2015}.)
}
\label{DBMFig10}
\end{figure*}
%%%%%%%%%%%%%%%%%%%%%%%%%%%%%%%%%%%%%%%%%%%%%%%%%%%%%%%%%%%%%%%%%%%%%%%%%%%%%%%%

Figure \ref{DBMFig9} shows three sets of snapshots for the density field in the phase separation process. The difference for the three columns is the value of $K$.  $K$ is a coefficient describing the strength of interfacial energy.
 The interfacial tension strongly influences the pattern morphology, the evolution speed and the depth of phase separation. At the time $t=0.045$,  numerous mini domains with large density ratio occur for the case with small $K=10^{-5}$, which indicates that the evolution has already entered the spinodal decomposition stage. While for cases of larger $K$, at the time time, the density variance is small, and it decreases with the increasing of coefficient  $K$. Nevertheless at the time $t=0.153$, the mean domain size and the phase separation depth are nearly the same for the three cases; all of them proceed to the domain growth stage. In the later times, it is observed that the larger the coefficient $K$, the faster the phase separation, the bigger the mean size, the fewer the domain number and the wider the interface.

These observations can be quantitatively studied via the time evolution of the characteristic domain size $R(t)$.
 See Fig. \ref{DBMFig10}(a). As a rough estimation of the phase separation process, the $R(t)$ curves behave similarly and distinguish approximately the phase separation process into two stages. During the first stage, the characteristic domain size $R(t)$ increases and arrives at a platform which is marked by the green arrow. It should be pointed out that, the marked point corresponds to the end of the spinodal decomposition stage. The plateau is dependent on the initial state described by the intensity of random noise, the depth of temperature quench,  and also the interfacial tension. The larger the interfacial tension, the longer the duration $t_{SD}$ of the spinodal decomposition stage, and the larger the domain size $R_{SD}$ for the spinodal decomposition stage.

During the phase separation process,
the potential energy transforms into the thermal energy and the interfacial energy. The physical scenario is as below. Under the action of inter-particle force, a liquid (vapor) embryo continuously gains (loses) molecules due to the condensation (evaporation), then the interface emerges and part of the potential energy transforms into the interfacial energy. Since the interfacial energy is proportional to $K$, an increasing $K$ means an increasing interfacial energy.
Since the interfacial tension always resists the appearance of new interface to minimize the interfacial energy,
  an increasing $K$ means an increasing $t_{SD}$ required for completing such an energy conversion process. The larger the interfacial tension, the longer it takes for sharp interfaces to form.

In the second stage, under the action of interfacial tension, small domains merge to minimize the interfacial energy. The domain size $R(t)$ continuously grows with time. The slope of the $R(t)$ curve presents a phase separation speed
$u_{DG}$ for the domain growth stage. It can be found that $u_{DG}$ increases with increasing $K$.
Thus, during the second stage, the phase separation process is significantly accelerated by the interfacial tension. Specifically here, the curve of $R(t)$ for $K=6\times10^{-5}$ crosses with the other two at the time $t=0.153$, then rises quickly and exceeds the former two.
When the coefficient $K$ varies from $10^{-5}$ to $3\times10^{-4}$, the dependence of separation speed $u_{SD}$ on $K$ can be fitted by
\begin{equation}
u_{DG}=e+fK-(gK)^{2}+(hK)^{3},
\end{equation}
with $e=0.00764$, $f=1.51\times 10^{2}$, $g=8.06\times 10^{2}$, $h=1.02\times 10^{3}$, as shown in the legend of Fig. \ref{DBMFig10}(a).

To numerically determine the duration $t_{SD}$,  we check the evolution of the second Minkowski measure: boundary length $L(t)$ for the density field. The density threshold is chosen as $\rho_{th}=1.70$
in Fig. \ref{DBMFig10}(b). Because the density pattern has the largest boundary length for this case. Some TNE manifestations are exhibited in the same panel. One can find that the peak of the $L(t)$ curve exactly coincides with the peaks or troughs of the TNE curves.

Since each component of $\boldsymbol{\Delta}$ or $\boldsymbol{\Delta}^{*}$ describes the TNE from its own aspect, a  ``TNE strength" can be defined as
\begin{equation}
D=\sqrt{\boldsymbol{\Delta_{2}^{*2}}+\boldsymbol{\Delta_{3}^{*2}}+\boldsymbol{\Delta_{3,1}^{*2}}+\boldsymbol{\Delta_{4,2}^{*2}}}.
\end{equation}
to  roughly and averagely estimate the deviation amplitude from the thermodynamic equilibrium.
 One can also use $\sqrt{\boldsymbol{\Delta_{2}^{2}}+\boldsymbol{\Delta_{3}^{2}}+\boldsymbol{\Delta_{3,1}^{2}}+\boldsymbol{\Delta_{4,2}^{2}}}$ (or its $F$, or $G$ component).
 Before using the concept, one needs first nondimensionalize the discrete Boltzmann equation.
 Here, the discrete Boltzmann equation is dimensionless, so do $\boldsymbol{\Delta}$ and $D$.
Thus,  $D=0$ indicates that the system is in thermodynamic equilibrium, and $D>0$ indicates that the system is out of the thermodynamic equilibrium.  Figure \ref{DBMFig10}(c) shows the evolutions of $L(t)$ (solid curves) and $D(t)$ (curves with solid symbols), where $D(t)$ is
 calculated from $\boldsymbol{\Delta}^{*F}$) for various $K$.
   The numbers, $1$,$3$,$6$,...,$18$, labeled on the $L(t)$-curves indicate cases with $K=10^{-5}$, $3\times10^{-5}$, $6\times 10^{-5}$,...,$1.8 \times 10^{-4}$, respectively. One can find a a perfect coincidence between the peaks of $L(t)$ and $D(t)$ in pairs. Thus, the behavior of $D(t)$ curve provides a convenient and efficient physical means to discriminate the stages of spinodal decomposition and domain growth. The left side of the peak corresponds to the spinodal decomposition stage, and the right side corresponds to the domain growth stage. Compared to the morphological method, the extension of the current approach to three dimensions is straightforward. Because the calculation of the interface area in three dimensional case is much more complex in coding.

From Fig. \ref{DBMFig10}(c) one can observe that,
when $K$ varies in the range $[10^{-5}, 3\times10^{-4}]$, the dependence of $t_{SD}$ on $K$ can be fitted by  the following equation,
\begin{equation}
t_{SD}=a+bK\text{,}  \label{t_SD-k}
\end{equation}
with $a=0.066$ and $b=1.51\times 10^{3}$. The specific result is shown in Fig. \ref{DBMFig10}(d).
Go to a further step, because of the length of interface, the depth of phase separation, as well as the gradient force and inter-particle force obtain their peak values at the end of the stage of spinodal decomposition, the TNE effect is the strongest at this moment.
The interfacial tension effects are found to decrease the maximum of the TNE strength $D_{\text{max}}$. The relation follows roughly
\begin{equation}
D_{\text{max}}=c+dK^{-0.5}\text{,}  \label{dmax-k}
\end{equation}
with $c=-0.073$ and $d=3.30\times 10^{-3}$. See Fig. \ref{DBMFig10}(d).
From the physical side, the Knudsen number is usually used to measure the TNE level.
 It is defined as the ratio between the molecular mean-free-path $\lambda$ and a macroscopic character length $L$.
 For a phase separation process, $L$ can be roughly taken the domain size $R_{SD}$ at the end of the spinodal decomposition stage. Thus, the mean Knudsen number $\text{Kn} = \lambda/2R_{SD}$.
 Since $R_{SD}$ increases with $K$, $\text{Kn}$ and the TNE strength decrease with $K$. This can also be understood as follows. A larger coefficient $K$  broadens the interfacial width, reduce the gradient force and refrain the TNE intensity.
The details of the DBM and more results are referred to Ref.\cite{SoftMatter2015}.

As for system under detonation by DBM, we just briefly mention a few results. As an initial application, various non-equilibrium behaviors around the detonation wave in one-dimensional detonation process were preliminarily probed.
It is found that, at the von Neumann pressure peak, the system is in a state being close the its local thermodynamic equilibrium; in front of and behind the von Neumann pressure peak, the system deviates from the thermodynamic equilibrium state in opposite directions.
The following TNE behaviors,
exchanges of internal kinetic energy between different displacement degrees of freedom and between displacement and internal degrees of freedom of molecules, have been observed.
It was found that the system viscosity (or heat conductivity)
decreases the local TNE, but increases the global TNE around the detonation
wave. Even locally, the system viscosity (or heat conductivity) results
in two competing trends, i.e. to increase and to decrease the TNE
effects. The physical reason is that the viscosity (or heat conductivity)
takes part in both the thermodynamic and hydrodynamic responses to the
corresponding driving forces. The ideas to formulate DBM with the smallest
number of discrete velocities and DBM with flexible discrete velocity model
are presented. A double-distribution-function DBM for combustion is referred to Ref.\cite{XuLin-DDF-DBM}.

\subsection{MD investigations}
MPM simulation is based on continuum assumption. The material properties are described by constitutive equation. Consequently, the minimum element of the material used in the simulation should be larger than 1 $\mu m$. When some critical phenomena,
such as initiating of phase transition,
localization of plasticity,
creation of damage and fracture,
have to be considered, the original continuum modeling and constitutive equation do not work any more and need to be improved based on more fundamental mechanisms in smaller scales. The MD simulation can study the dynamic behaviors of microscopic structures in the scales of nanometer and sub-nanoseconds. The bridge connecting the microscopic MD simulation and the mesoscopic MPM simulation is still a grand challenge in nowadays. We hope to incorporate the microscopic evolution mechanisms of these critical phenomena into the larger-scale model via some coarse-grainning techniques.

Our MD simulations can be classified into two groups.
  The first group aims to study the creation mechanisms of micro-structures like dislocation, void, cavity, and new phase grain in metal materials\cite{Lu-ActaPhysSin2012,Pang-CJCP2011,Pang-SciCN2012,Pang-SciRep2014C}. The second group aims to study the evolution behaviors of these micro-structures\cite{Yang-Acta2008,Pang-AMR2013,Pang-SciRep2014A,Pang-SciRep2014B}.

\subsubsection{Creation mechanisms of micro-structures}

\begin{figure*}[tbp]
\centering
\includegraphics*[scale=0.65]{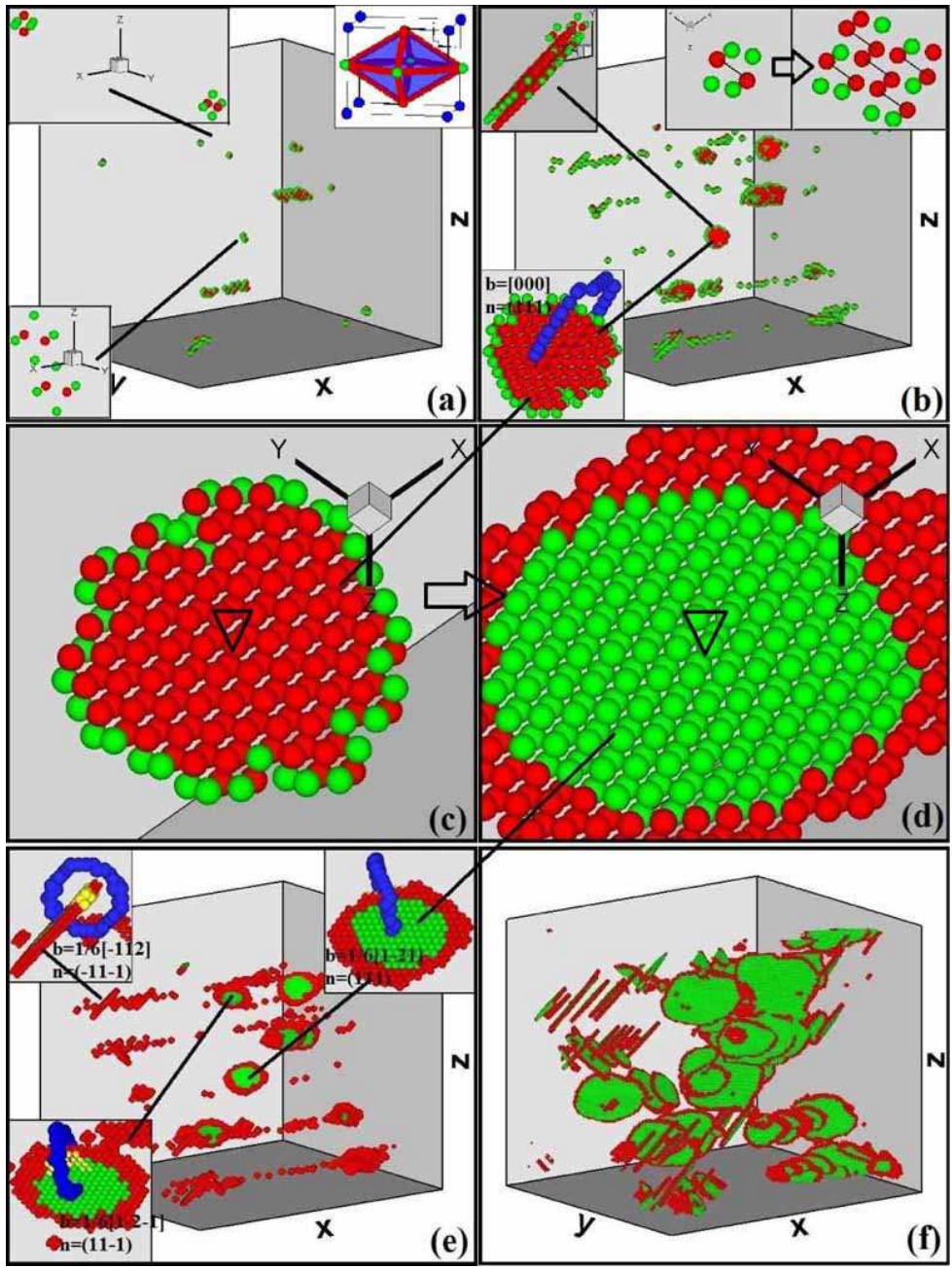}
\caption{(Color online)
Molecular-dynamics simulation snapshots that provide a general
three-stage physical picture for the generation of dislocations and the
corresponding non-zero Burgers vectors in FCC ductile metals under
high-strain-rate uniaxial stretch. Panel (a) shows that FOSs (for a detailed
view, see the top right inset) are firstly activated in the metals by thermal
fluctuations. Panel (b) shows that FOSs begin to stack on the close-packed
plane to form double-layer defect clusters (see the top left inset for closer
view). This stacking process is shown in the top right inset. The Burgers
vector for the double-layer defect cluster structure is calculated to be zero,
as shown in the bottom left inset. Panel (c) and panel (d) shows the
transformation of the double-layer defect clusters into stacking faults. Panel
(e) gives a few non-zero Burgers vectors of the nucleated dislocations that
surround the stacking faults. Panel (f) shows the growth of stacking faults
and dislocations. In panels (a)-(c)the coordination numbers of red and green
atoms are 13 and 12, respectively, while in panels (d)-(f) the CNA values of
red and green atoms are 5 (dislocation atoms) and 2 (HCP atoms), respectively.
(Adopted with permission from Ref. \cite{Pang-SciRep2014C}.)
}
\label{Fig:Fg1}
\end{figure*}

Due to the fundamental importance for dynamic fracture modeling, the dislocation and void nucleation processes have attracted extensive theoretical and experimental studies.
But due to the complexities, the
physical picture on the self-organized atomic collective motions during dislocation creation,
 and
 the mechanisms for the void nucleation, obscured by the extreme diversity, in structural configurations surrounding the void nucleation core,
 are still open problems and far from being well clarified.
 Via MD simulations we investigated the origin of dislocation creation and void nucleation  in face-centered-cubic (FCC) ductile metals under uniaxial high strain rate tensile loading.
The dislocation creation process can be described by three distinguished stages (See Fig.\ref{Fig:Fg1}): In the first stage, thermal fluctuations randomly activate Flattened octahedral structures (FOSs) in the material. In the second stage, double-layer defect clusters occur via self-organized stacking of FOSs on the
close-packed plane. In the third stage, stacking faults appear and the Shockley partial dislocations are created from the double-layer defect clusters.
The dislocation nucleation and slip dot not release bulk stress (negative pressure), even though they can release part of the shear stress. Plenty of energies accumulate in the system with the increase of the tensile strain. At the weak points in material, some voids or cracks occur to release the accumulated energy.

\begin{figure*}[tbp]
\centering
\includegraphics*[scale=0.65]{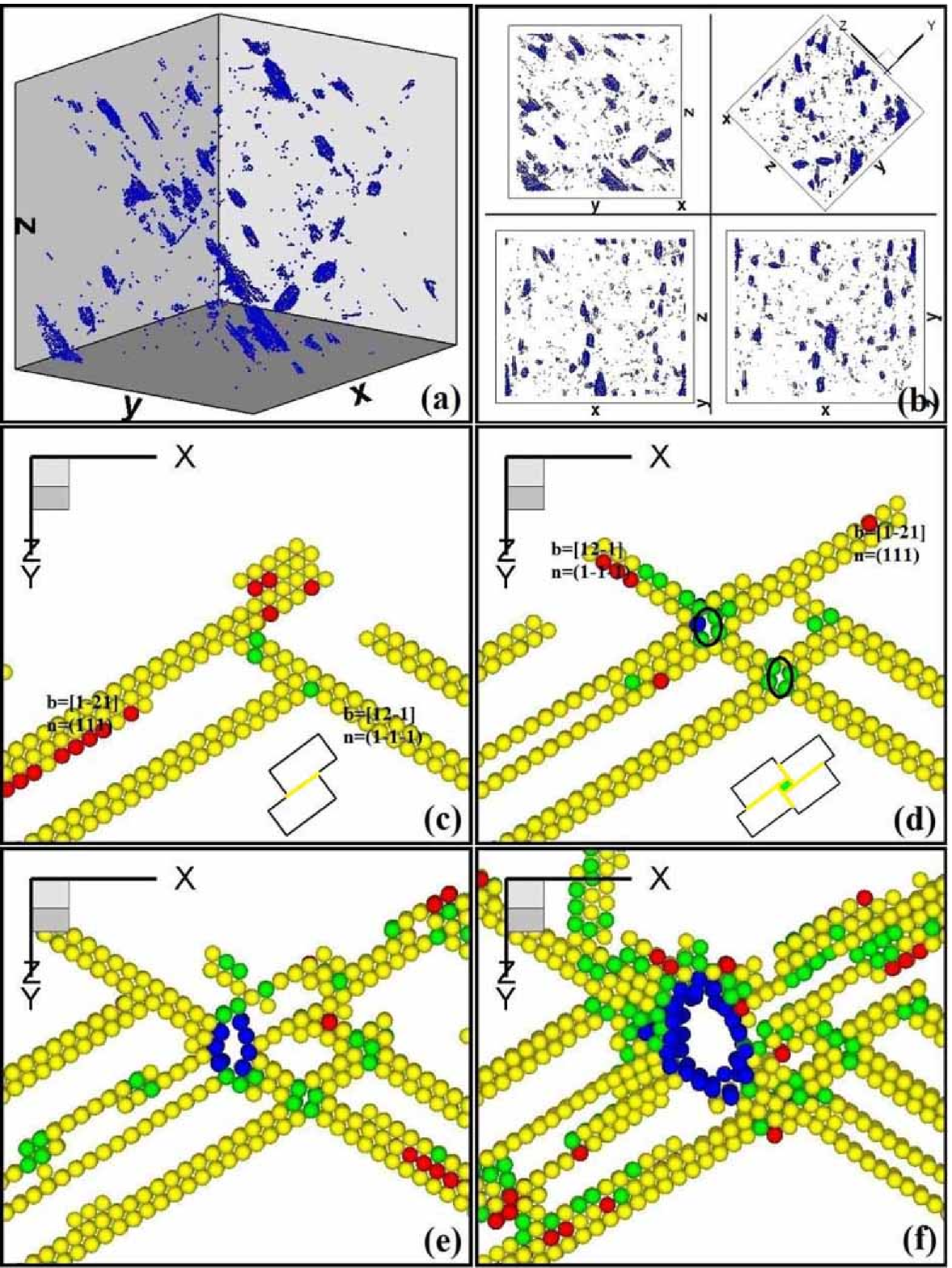}
\caption{(Color online)
Incipient void nucleation phenomenon and its two-stage mechanism.
Panel (a) and (b) shows the void nucleation phenomenon.
Panels (c)-(f) show the evolutionary process of these atoms in a picked slice,
and therein the insets are the schematic diagrams of stacking faults creation and vacancy string formation.
Panel (c) shows that
four stacking faults appearing as lines nucleate from double-layer defect
clusters. Panel (d) shows that two pillar-like vacancy strings are generated
from the the intersections of stacking faults, see the upper and lower black
circles. Panel (e) shows that the upper vacancy string transforms into a void
via emitting dislocations, while the lower one retains its size. Panel (f)
shows that nucleated voids grow gradually and neighboring vacancy strings
disappear.
(Adopted with permission from Ref. \cite{Pang-SciRep2014C}.)
)
}
\label{Fig:Fg3}
\end{figure*}

The cavity creation process can be described by the following two stages. In the first stage, stacking faults, with
different normal directions, evolve to intersect with each other. They generate pillar-like vacancy strings located at the intercrossing lines. (See Fig. \ref{Fig:Fg3} and the inset inside.)
In the second stage, these vacancy strings grow into voids via emitting dislocations. The growth of the nucleated voids releases stress, which suppresses the growth of neighboring vacancy strings.

The above process of vacancy string creation could be regarded as two successive plastic deformations. The first deformation brings a stacking fault into the system(see the inset in Fig. \ref{Fig:Fg3}(c)), and the atoms have a displacement of the corresponding Burgers vector along the plane. During the second deformation process, the atoms further have a corresponding displacement along the other plane, which results in a volume variation(see the inset in Fig. \ref{Fig:Fg3}(d)). The plastic deformation resulting from stacking faults can be described by a distortion tensor
$\mbox{\boldmath$\beta$} =\mbox{\boldmath$\delta$}(\Sigma)\mathbf{b}
= \int\int_{\sum}d\mathbf{s^{\prime}\delta(r^{\prime}-r)b}$, where $\mbox{\boldmath$\delta$}(\Sigma)$ is the surface Dirac function, $\Sigma$ is stacking fault plane, and $\mathbf{b}$ is the Burgers vector. The relative volume variation is $\delta V/V = \mbox{Tr (\boldmath $\beta$)}$. For the case of single stacking fault,
$\mbox{Tr ({\boldmath$\beta$})} = \mbox{\boldmath$\delta$}(\Sigma)\mathbf{\cdot b} = 0$, herefore, there is no density variation in the system. For the case of two stacking faults intersecting with each other, the distortion tensor is
$\mbox{\boldmath$\beta$} = \mbox{{\boldmath$\beta$}$_1$+{\boldmath$\beta$}$_2$}\cdot(\mathbf{I%
}+\mbox{{\boldmath$\beta$}$_1$})$, where \mbox{{\boldmath$\beta$}$_{1}$} and %
\mbox{{\boldmath$\beta$}$_{2}$} are the distortion tensors of the two stacking faults, respectively. Therefore, the volume variation is $\delta V = \mathrm{\int} dV
\mbox{Tr ({\boldmath$\beta$}$_1 \cdot$ {\boldmath$\beta$}$_2$)} =(\mathbf{b}_{2}\cdot\mathbf{n}_{1})(\mathbf{b}%
_{1}\cdot\mathbf{n}_{2})L/|\mathbf{n}_{1}\times\mathbf{n}_{2}|$, where $%
\mathbf{n}_{1}$ and $\mathbf{n}_{2}$ are normal directions of the stacking faults and $L$ is the length of the vacancy string. Here, we arrive at that the cross-section area of vacancy string resulting from the intersection of two different stacking faults is $(\mathbf{b}_{2}\cdot\mathbf{n}_{1})(%
\mathbf{b}_{1}\cdot\mathbf{n}_{2})/|\mathbf{n}_{1}\times\mathbf{n}_{2}|$, and the direction of vacancy string is $\mathbf{n}_{1}\times\mathbf{n}_{2}$.

The dislocations generated from the double-layer defect clusters are all Shockley partial dislocations. Since the corresponding Burgers vectors have been obtained,  the initial vacancy strings have a
typical cross-section area, $\sqrt{2}a^{2}/36$, according to the above expression,
where $a$ is the lattice constant. They have six possible types of distribution directions. But
only two types of voids are observed in Fig. \ref{Fig:Fg3}(a). The corresponding directions are
 $[011]$ and $[0\bar{1}1]$ which are perpendicular to the loading direction. The physical reasons
 are as follows:
 The energy released via growth of a vacancy string is
 $\int\int d \mathbf{s} \cdot\mbox{\boldmath$\sigma$} \cdot\delta\mathbf{r}$,
 where $\delta\mathbf{r}$ is the growth displacement, $\mathbf{s}$ the surface area of the vacancy string, and $\mbox{\boldmath$\sigma$}$  the applied stress.
 In the numerical simulation, the applied stress is $\sigma_{xx}$. Only when the vacancy string is perpendicular to $[100]$, does its growth release more energy and evolve into voids. The numerical results shown in Fig. \ref{Fig:Fg3}(a) confirm this analysis.

%%%%%%%%%%%%%%%%%%%%%%%%%%%%%%%%%%%%%%%%%%%%%%%%%%%%%
\begin{figure*}[ptb]
\begin{center}
\includegraphics[width=0.8\linewidth]{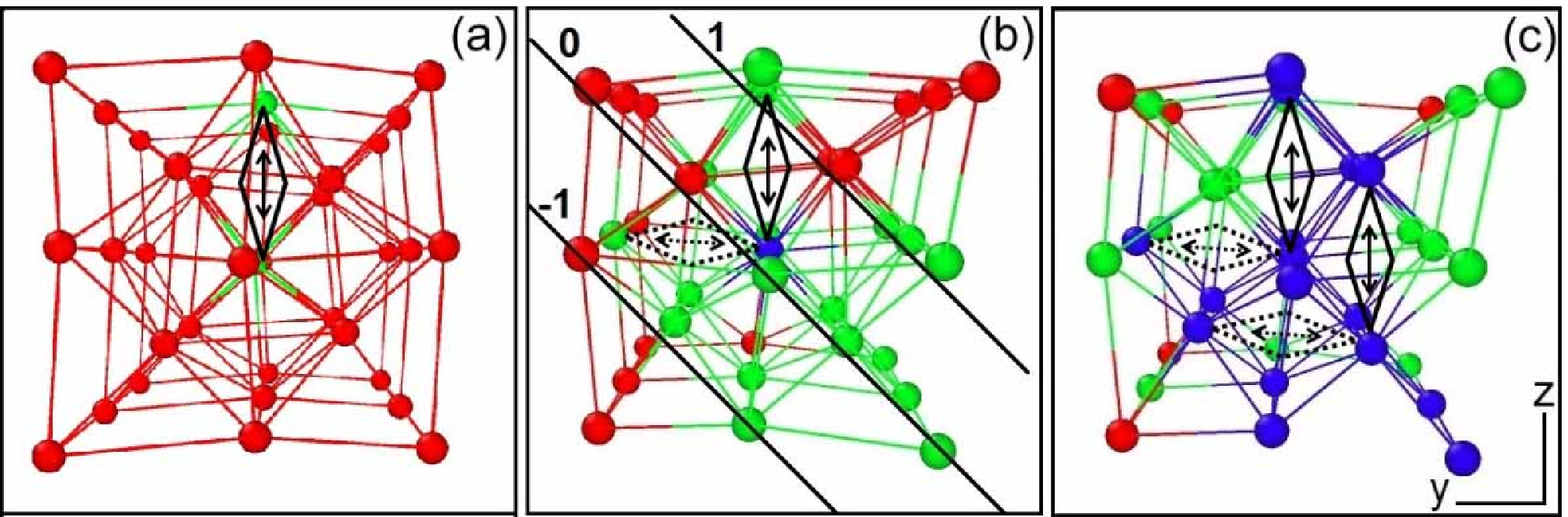}
\end{center}
\caption{(Color online)
The nucleation and growth mechanisms of hcp domain by means of the formation of FOSs. Panel (a) shows the structure of formed FOS. Panel (b) shows the nucleation of hcp domain. Panel (c) shows the growth of hcp domain.
(Adopted with permission from Ref. \cite{Pang-SciRep2014A}.)
}%
\label{Fig:Figggg2}%
\end{figure*}
%%%%%%%%%%%%%%%%%%%%%%%%%%%%%%%%%%%%%%%%%%%%%%%%%%%%%
\begin{figure*}[ptb]
\begin{center}
\includegraphics[width=0.8\linewidth]{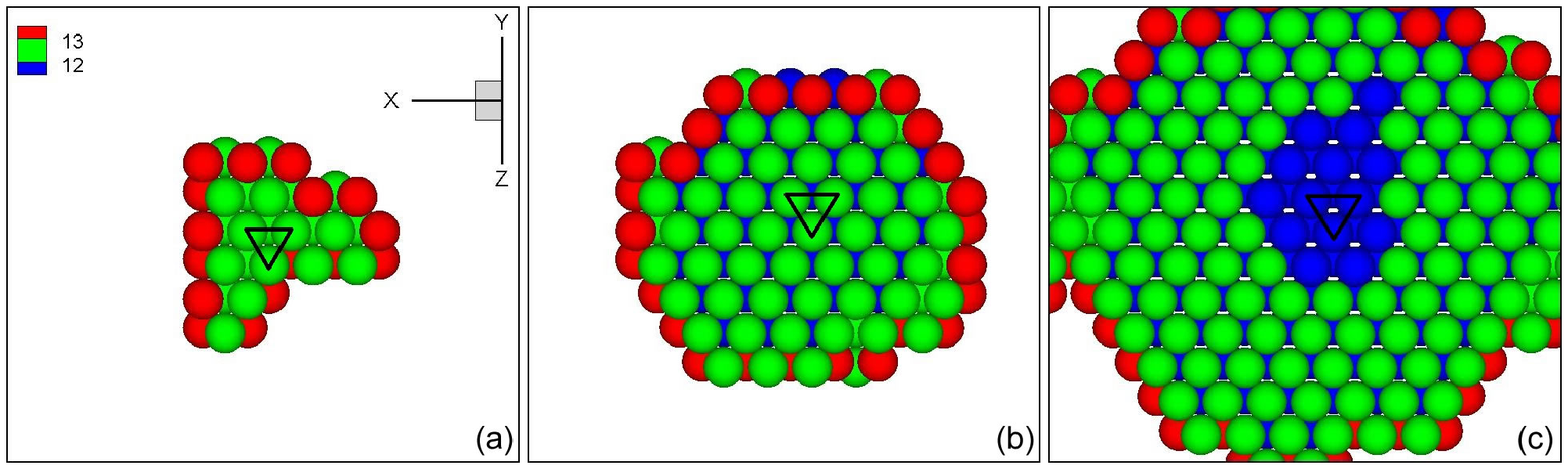}
\end{center}
\caption{(Color online)
The relatively slip process between the atomic layers observed from the normal direction of phase plane. Panel (a) shows that some FOSs simply aggregate into a thin stratified structure. Panel (b) shows that the atoms in the central region of the middle layer are transformed into hcp structure. Panel (c) shows that the atoms in the central region of three layers are transformed into hcp structure.
(Adopted with permission from Ref. \cite{Pang-SciRep2014A}.)
}%
\label{Fig:Figggg3}%
\end{figure*}
%%%%%%%%%%%%%%%%%%%%%%%%%%%%%%%%%%%%%%%%%%%%%%%%%%%%%

With the development of the high-pressure technology, structure phase transition in metal has attracted wide notice and
been one of the focuses in material physics. As for the structure transition from bcc to hcp occurred in iron under a pressure higher than 13 GPa, before our works, even though the transition process has been roughly understood, how the phase transition initiate and how the new phase grain nucleate are still far from clear.
The MD simulation shows that, similarly to case of dislocation nucleation, the nucleation process of new phase grain can also be described by three stages. In the first stage, as shown in Fig. \ref{Fig:Figggg2}, with the aid of thermal fluctuations,  some atoms deviate from their equilibrium positions to form FOSs with two different deformation directions in the local region. In the second stage, as shown in Figs. \ref{Fig:Figggg3} (a)-(b), the FOSs with different deformation directions aggregate to form a thin stratified structure like twin-crystal configuration. In the third stage, as shown in Figs. \ref{Fig:Figggg3} (b)-(c), the thin stratified structure undergoes a relative slip to form the new hcp phase.

\subsubsection{Evolution mechanisms of micro-structures}

\begin{figure*}[tbp]
\centering
\includegraphics*[ scale=0.6,angle=0,
bbllx=0pt,bblly=0pt,bburx=591pt,bbury=497pt
]{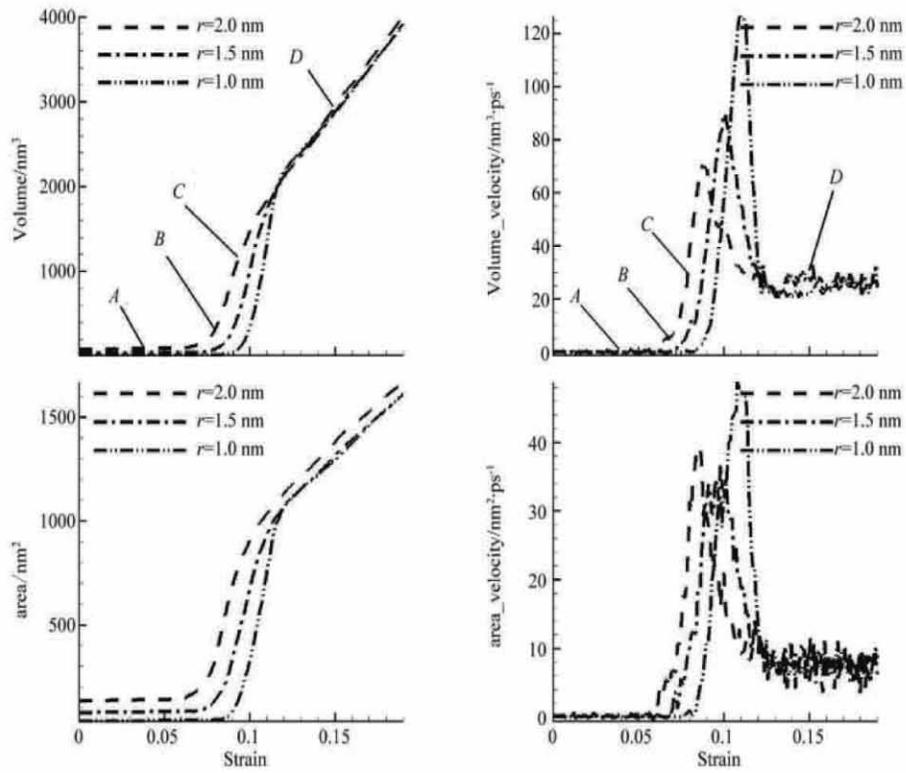}
\caption{ Volumes and areas of voids with different size.
(Adopted with permission from Ref. \cite{Pang-CJCP2011}.)
}%
\label{Fig:CJCPFig1}%
\end{figure*}

\begin{figure*}[tbp]
\centering
\includegraphics*[ scale=0.6,angle=0,
bbllx=0pt,bblly=0pt,bburx=591pt,bbury=428pt
]{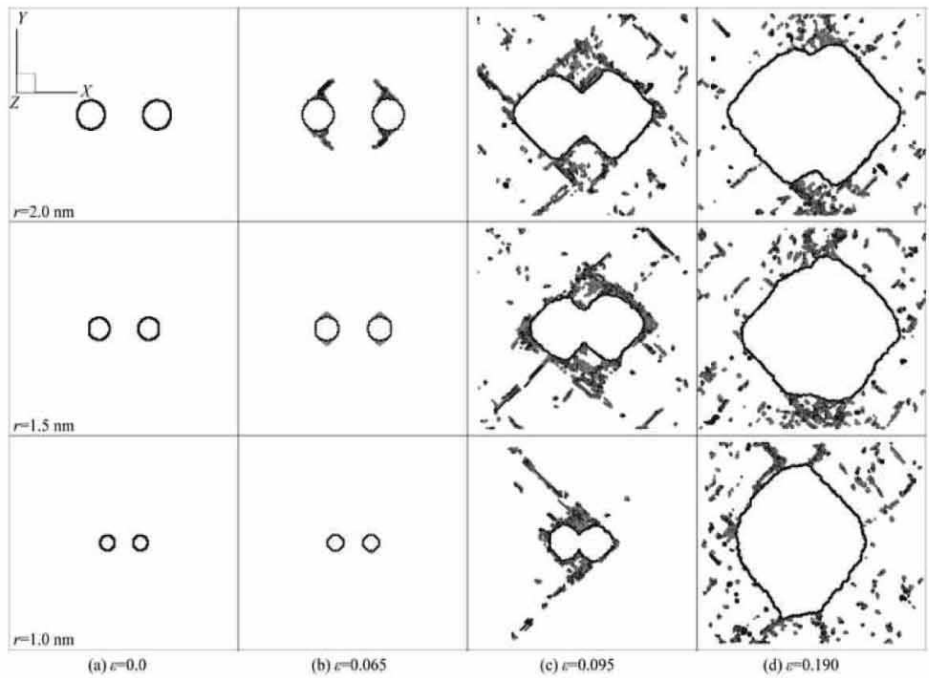}
\caption{ Growth and coalescence of voids with different size.
(Adopted with permission from Ref. \cite{Pang-CJCP2011}.)
}%
\label{Fig:CJCPFig2}%
\end{figure*}

When model the damage and micro-fracture, the growth and coalescence of voids have to be considered. We investigated the dynamics of a pair of voids located along loading direction in crystal copper under uniaxial tension. Voids with  different sizes grow and coalesce through dislocation nucleating on void surfaces. In early elastic stage, voids grow along the loading direction, then the vertical direction and finally form octahedral-like structures in plastic stage. Critical yield stress increases with decreasing of void size.
%%%%%%%%%%%%%%%%%%%%%%%%%
(See Fig.\ref{Fig:CJCPFig1}).
%%%%%%%%%%%%%%%%%%%%%%%%%
If radius is large, dislocations nucleate and migrate symmetrically. Voids elongate in loading direction, and evolution process is similar. If radius is small,dislocations nucleate asymmetrically and voids elongate along vertical direction. The process of void growth may be characterized by elastic deformation, independent growth,coalescence and steady growth.
%%%%%%%%%%%%%%%%%%%%%%%%%
(See Fig.\ref{Fig:CJCPFig2}).
%%%%%%%%%%%%%%%%%%%%%%%%%
 Independent growth stage diminishes gradually as void size becomes smaller.

\begin{figure*}[ptb]
\begin{center}
\includegraphics[width=0.8\linewidth]{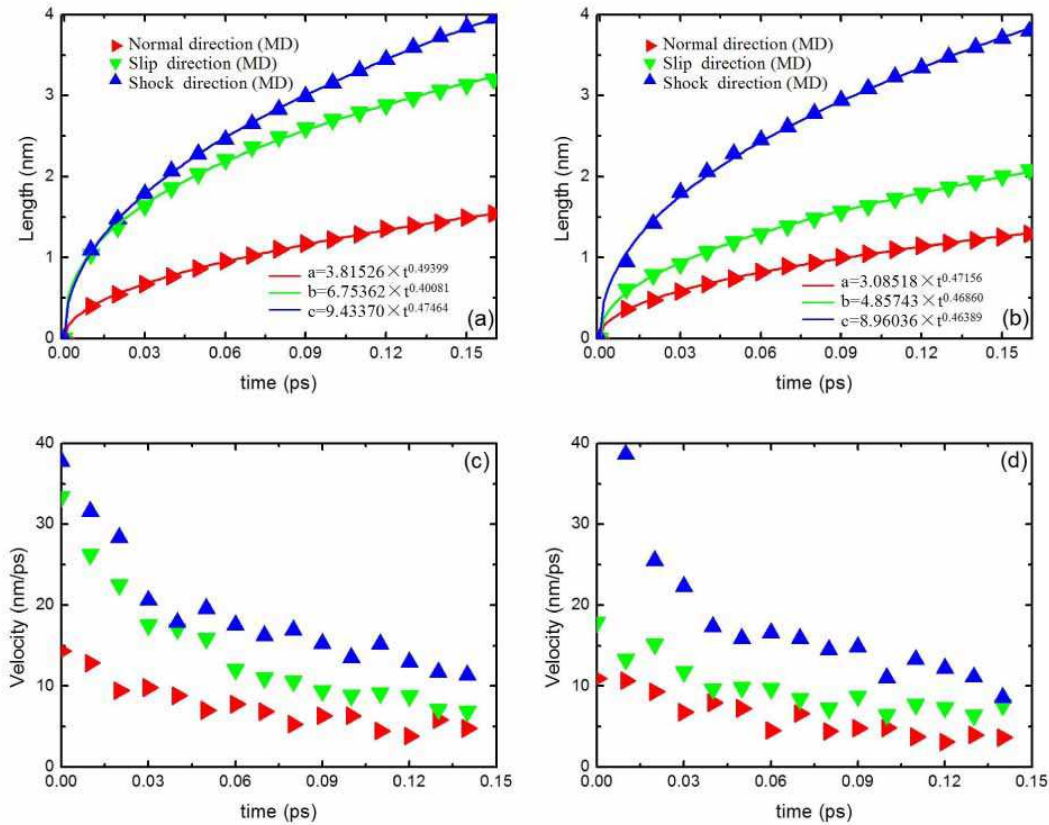}
\end{center}
\caption{(Color online) Principal-axis lengths and growth speeds of two HCP phase domains which form at different times under the same shock velocity.
Panels (a) and (b) are the principal-axis lengths, while panels (c) and (d) are the growth speeds. The symbols are the MD simulated results and the lines are the fitting results.
(Adopted with permission from Ref. \cite{Pang-SciRep2014B}.)
}%
\label{Fig:Figgg3}%
\end{figure*}

\begin{figure*}[ptb]
\begin{center}
\includegraphics[width=0.8\linewidth]{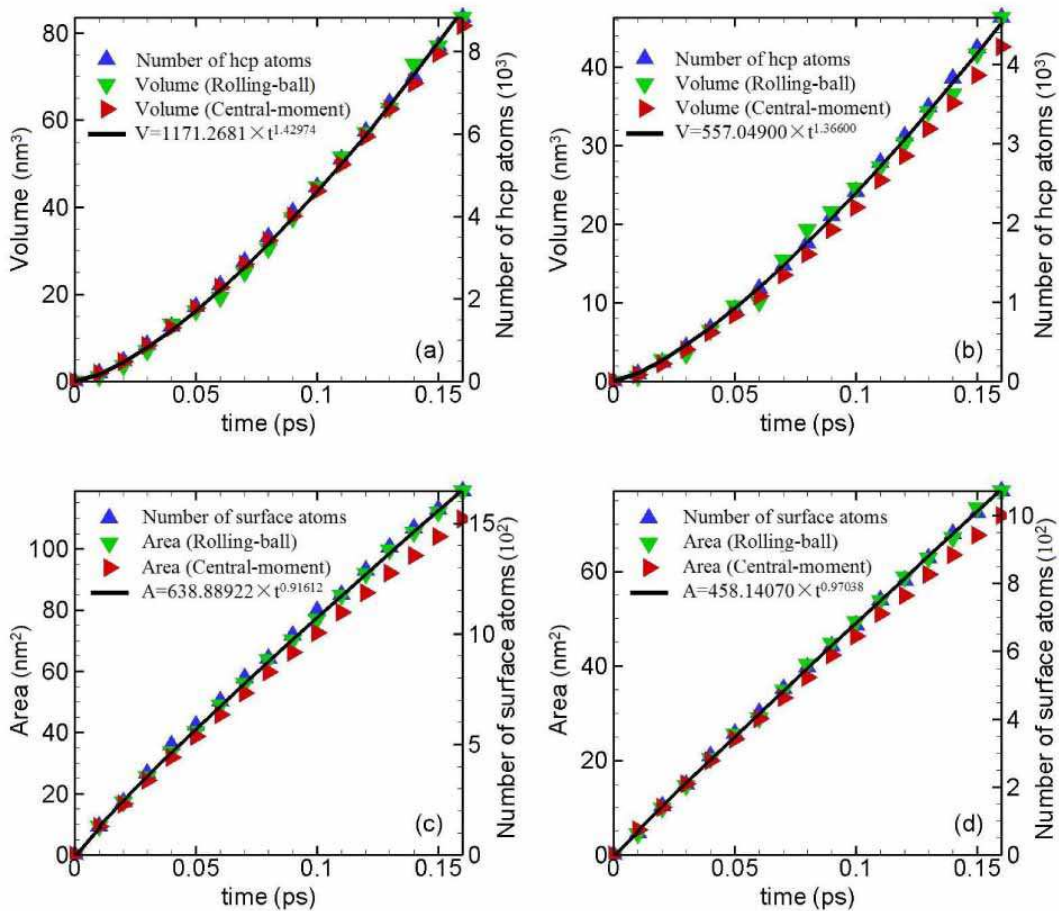}
\end{center}
\caption{(Color online) Time evolutions of surface areas, volumes, and the corresponding atom numbers of the same two phase domains as those used in Fig. \ref{Fig:Figgg3}. The symbols are the MD simulated results and the lines are the fitting results.
(Adopted with permission from Ref. \cite{Pang-SciRep2014B}.)
}%
\label{Fig:Figgg4}%
\end{figure*}

The growth mechanism of new phase domain has not obtain quantitative characterization. The main reason is due to lacking technology to precisely determine the new phase domain. To calculate and analyze the morphology and growth speed of the hcp phase domains, a central-moment method and a rolling-ball algorithm were designed. To clarify our derived growth law of the phase domains, a phase transition model was proposed. Studies show that the new-phase evolution process undergoes three distinguished stages with different time scales of the hcp phase fraction in the system.
In the initial independent growth stage, the morphology of domain is ellipsoid-like where the three principal axes are approximately along [100], [011], and [01$\bar{1}$] directions. The growth in the three directions shows different velocities. See Figs.\ref{Fig:Figgg3} (a)-(b). The growth velocity depends also on the morphology of phase domain.
The growth speed of a single phase domain is supersonic. It is in a range $4 \times 10^{4}$ to $5 \times 10^{3}$ m/s. See Figs.\ref{Fig:Figgg3} (c)-(d). The time evolutions of the size, surface area, and volume of the single hcp domain follow roughly $L \sim t^{0.5}$, $A\sim t^{1.0}$, $V \sim t^{1.5}$, respectively.  See Fig.\ref{Fig:Figgg4}.
The FOS is the primary structural unit of the embryo nucleus and phase interface of hcp domains. The interfacial energy is reduced via formation of FOSs. The hcp phase domain grows up via forming new FOSs along the phase boundary. The growth rate of single phase domain depends on the loading means \emph{and its occurrence time.}

%%%%%%%%%%%%%%%%%%%%%%%%%%%%%%%%%%%%%%%%%%%%%%%%%%%%%
%%%%%%%%%%%%%%%%%%%%%%%%%%%%%%%%%%%%%%%%%%%%%%%%%%%%%
%%%%%%%%%%%%%%%%%%%%%%%%%%%%%%%%%%%%%%%%%%%%%%%%%%%%%

\subsubsection{From micro to meso scales: coarse-grain modeling}

\begin{figure*}[ptb]
\begin{center}
\includegraphics[width=0.8\linewidth]{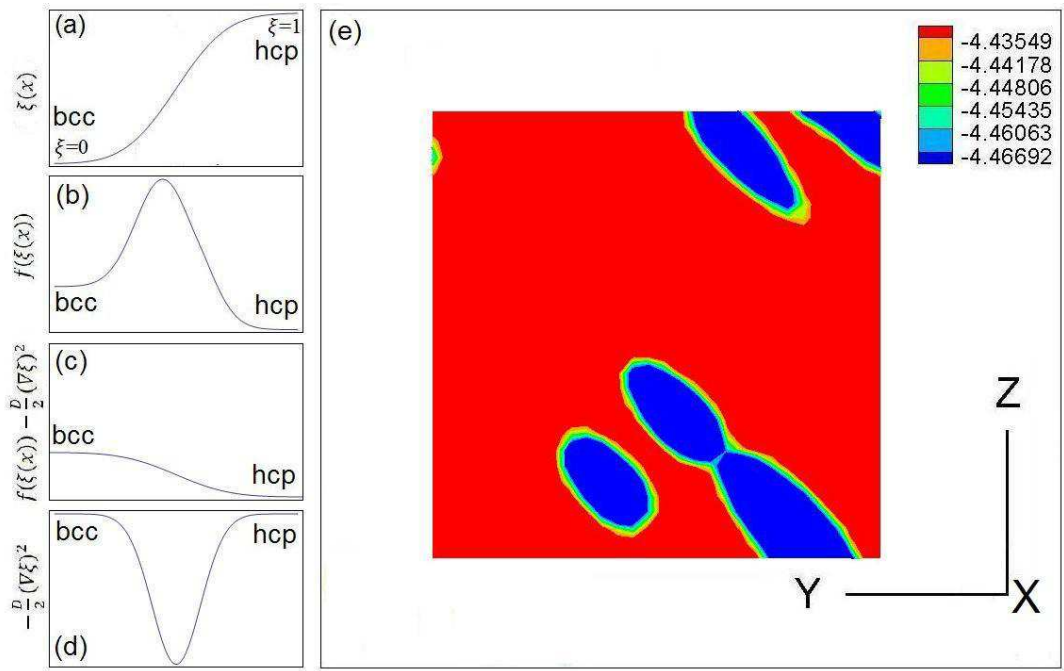}
\end{center}
\caption{(Color online) The potential energy distribution and schematic diagram of phase transition. Panels (a)-(d) are the schematic diagram of phase transition, where the horizontal axis represents the distance away from the phase interface. Panel (e) shows the potential energy distribution of a slice in shocked iron.
(Adopted with permission from Ref. \cite{Pang-SciRep2014B}.)
}%
\label{Fig:Figgg5}%
\end{figure*}

As an specific example of coarse-grain modeling from the micro to mesoscales, based on the order parameter theory of Ginzburg-Landau, we propose a phase transition model to describe the shocking kinetic process in iron. To this aim, we choose the slippage $\xi$ of the lattice as the order parameter. It varies from $\xi=0$ in bcc structure to $\xi=1$ in hcp structure, as schematically shown in Fig. \ref{Fig:Figgg5}(a), where the horizontal axis represents the distance away from the phase interface.
For the uniform bulk phase transition in metal iron, to undergo transformation from bcc to hcp structure via the lattice slippage, the system needs to overcome a potential barrier \cite{Liu}, as schematically shown in Fig. \ref{Fig:Figgg5}(b).
However, surely, what observed in experiments and simulations are not uniform bulk structure transition, but complex processes with creation and growth of new phase domains.
Therefore, a solely uniform description is insufficient. The phase domain effects must be reasonably included in a phase transition model. In the stage of new phase domain nucleation, within the local region with stress concentration,  the atoms overcome the potential barrier with the aid of collective thermal fluctuations.
 Simulation results have shown that, in the growth stage of a phase domain, the growth speed is supersonic, and the stress wave has no enough time to propagate in the domain of hcp phase. It is mainly the interfacial energy that drives the growth of a new phase domain.
The potential energy field of a slice in the simulated system is shown in Fig. \ref{Fig:Figgg5}(e), where the regions with red, blue, and other colors represent the bcc, hcp, and interface structures, respectively. The potential energy in the interfacial region is in between those for hcp and bcc phases, from which we obtain
 the schematic shown in Fig. \ref{Fig:Figgg5}(c).
 The interface energy is negative, from which we obtain the schematic shown in Fig. \ref{Fig:Figgg5}(d). It
 is the interfacial energy that reduces prominently the potential barrier between two phases.
 Consequently, the transition process becomes easier.

Based on this physical scenario, the relation between system energy and order parameter can be roughly described by
\begin{equation}
F=\int[f(\xi)-\frac{D}{2}(\bigtriangledown\xi)^{2}]d^{3}\mathbf{r},
\label{energy}%
\end{equation}
where the first and second terms in the integrand are the bulk free energy and interface energy of the system, respectively. Specifically,
\[
f(\xi)=\frac{a}{2}\xi^{2}-\frac{a+1}{3}\xi^{3}+\frac{1}{4}\xi^{4}
\]
is a bi-stable function with two stable points, $\xi=0$ and $\xi=1$. Here, the dependence of bulk free energy on temperature and pressure is described by the parameter $a$. Under low pressure, $a>1/2$, the bcc phase is stable, while under high pressure, $a<1/2$, the hcp phase is stable. The possible anisotropy in the interface energy has been ignored for simplicity. The evolution equation of the order parameter reads
\begin{equation}
\partial_{t}\xi=\frac{\delta F}{\delta\xi}=f^{\prime}(\xi)-D\nabla^{2}\xi.
\label{evolution-equation}%
\end{equation}
For the steady growth of one-dimensional phase domain,
\[
\xi=\xi(\eta)\equiv \xi(x-c_{0}t).
\]
It satisfies the following eigenvalue equation
\begin{equation}
\left\{
\begin{array}
[c]{l}%
f^{\prime}(\xi)-D\partial_{\eta}^{2}\xi+c_{0}\partial_{\eta}\xi=0,\\
\xi\mid_{\eta\rightarrow-\infty}=0,\\
\xi\mid_{\eta\rightarrow+\infty}=1,
\end{array}
\right.  \label{one-dimensional}%
\end{equation}
where $c_{0}$ is the growth speed of the hcp phase domain, instead of the sound speed. For the convenience of describing  the growth of a three-dimensional phase domain,
we adopt the local coordinate system, instead of the Cartesian coordinates,
\[
\mathbf{r}=\mathbf{r}_{0}+\lambda \mathbf{n}+\mu\mathbf{t}_{1}+\nu\mathbf{t}_{2},
\]
where $\mathbf{r}_{0}$ represents a point at the interface, and
$\mathbf{n},\mathbf{t}_{1},\mathbf{t}_{2}$ represent the normal and two principal tangential unit
vectors of the interface at position $\mathbf{r}_{0}$, respectively. The evolution equation of order parameter reads
\begin{equation}
\partial_{t}\xi=f^{\prime}(\xi)-D(\partial_{\lambda}^{2}+\partial_{\mu}%
^{2}+\partial_{\nu}^{2}+(k_{1}+k_{2})\partial_{\lambda}-k_{1}\partial_{\mu
}-k_{2}\partial_{\nu})\xi, \label{curvilinear-coordinates}%
\end{equation}
where $k_{1}$ and $k_{2}$ are the curvatures along the two principal tangential directions, respectively. According to the relation
\[\xi=\xi (\eta)\equiv\xi(\lambda-vt),
\]
 the above expression can be reduced to
\begin{equation}
f^{\prime}(\xi)-D\partial_{\eta}^{2}\xi+(-Dk+v)\partial_{\eta}\xi=0
\label{three-dimensional}%
\end{equation}
with the boundary conditions
\[\xi\mid_{\eta\rightarrow-\infty}=0\text{,} \xi \mid_{\eta\rightarrow+\infty}=1,
\]
 and $k=k_{1}+k_{2}$. Comparing Eqs. $(\ref{one-dimensional})$ and $(\ref{three-dimensional})$ giving the growth speed of new phase domain evaluated by
\begin{equation}
v=c_{0}+Dk. \label{growth speed}%
\end{equation}

For structure phase transition induced by shock wave, the interface energy is related to the pressure surrounding the phase domain. So, the parameter $D$ is a function of pressure. Simulated results and theoretical analysis show that the growth speed is supersonic and the value of $D$ is almost a constant in the phase domain growth process. Therefore, equation (\ref{growth speed}) indicates that the growth speed of the phase domain is a function of the local curvature. In the initial stage, the volume of phase domain is small, the local curvature is large, and consequently the energy triggering phase transition is relatively more concentrated, which is responsible for the relatively quick growth.
With the growth of the phase domain, the interfacial area becomes larger, and the local curvature decreases, which is responsible for the relatively slow growth.

\begin{figure*}[ptb]
\begin{center}
\includegraphics[width=0.8\linewidth]{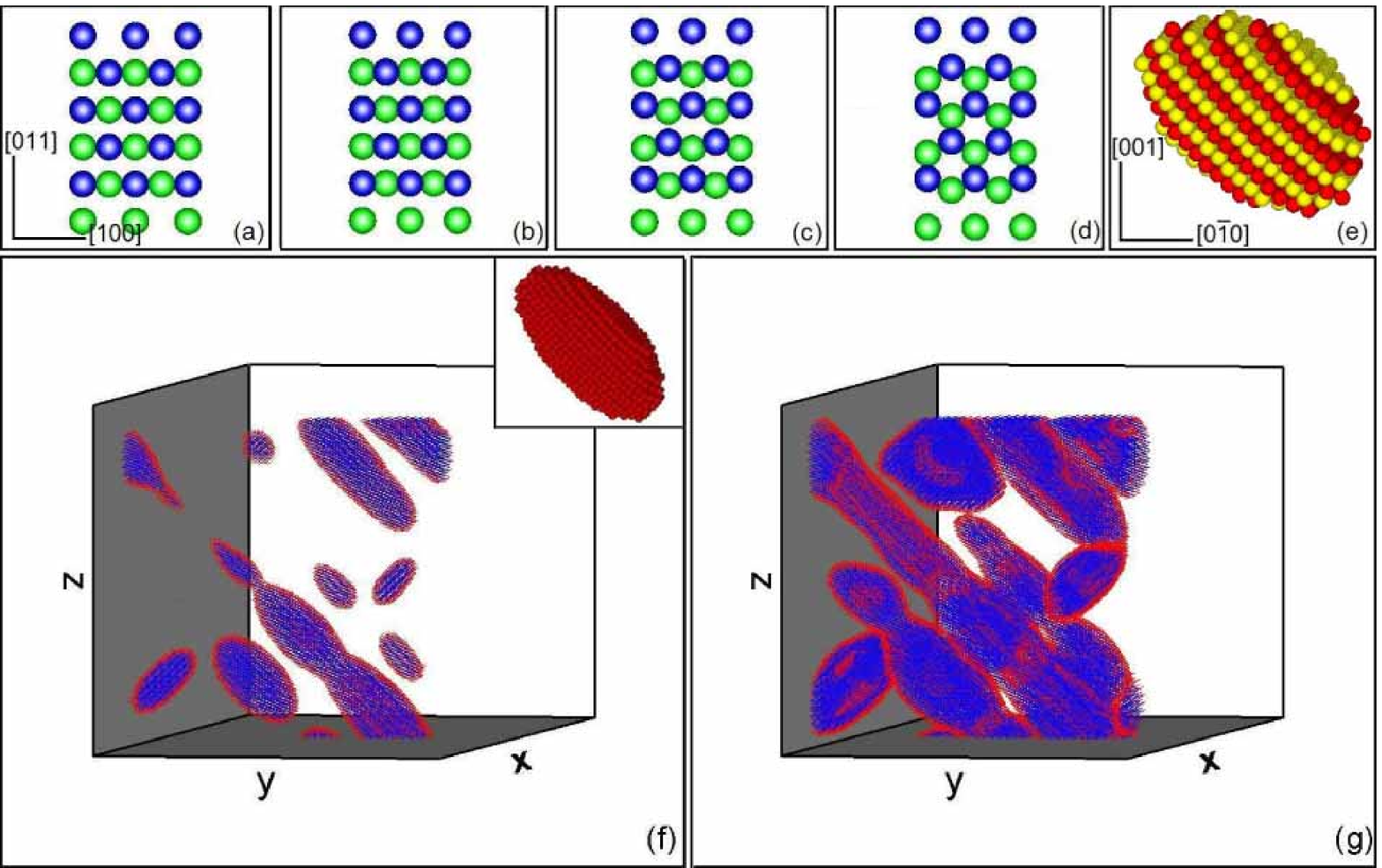}
\end{center}
\caption{(Color online) Microstructure evolution in iron from BCC to HCP structure during the shock process.
Panels (a)-(e) show the phase transition mechanism, while panels (f)-(g) show the formed new phase domains in the shocked region.
(Adopted with permission from Ref. \cite{Pang-SciRep2014B}.)
}%
\label{Fig:Figgg1}%
\end{figure*}

For an ellipsoidal phase domain, the local curvature is non-uniform on the surface of the phase domain. The portion with a larger curvature grow more quickly. As a result, the phase domain becomes more and more flat with time. Actually, simulation results in Fig. \ref{Fig:Figgg1} confirm that various phase domains evolve to be disc-like, spherical, columnar, elongated, etc., in the later stage. Now, we can go a further step to give a very simple but illustrative estimation on the dynamics of domain growth. For this purpose, we assume the phase domain is spherical with radius $R$, the growth speed in Eq. \ref{growth speed} gives
\[
\dot{R} = c_{0}+2 D / R.
 \]
Further assuming $c_{0} \rightarrow 0 $ gives
\[
  R(t)=\sqrt{4 D t} .
\]
Our MD simulation results show that the linear length of an ellipsoidal phase domain roughly follows $L\sim t^{0.465}$, which is close to what this analytical expression indicates.
The difference bettwen the curvatures of the sphere and the ellipsoid is responsible for deviation of exponent 0.465
from 0.5.

%%%%%%%%%%%%%%%%%%%%%%%%%%%%%%%%%%%%%%%%%%%%%%%%%%%%%

%%%%%%%%%%%%%%%%%%%%%%%%%%%%%%%%%%%%%%%%%%%%%%%%%%%%%%%%%

A second example of mesoscopic modeling is used to study the impulse propagation in granular chain.
Solitary signal has been observed and well understood in a horizontal granular chain without considering dissipation. But in vertical chain under gravity, more complex and interesting phenomena occur.
Via MD-like simulations,  signal is found to change its speed and form with propagating depth. The grain velocity and the propagation speed, amplitude and width of the signal follow power laws in depth. Physically, this is due to the power-law type changing of elastic properties in a medium under gravity. The impulse propagation can be classified into two types in terms of the behavior of the power-law exponents. The latter are further dependent on the strength of the nonlinearity. The power-law exponents are invariant with the strength of the impulse in the weakly nonlinear regime, while they vary with the strength of the impulse in the strongly nonlinear regime\cite{Hong-PRE2001,Xu-CTP2001B}.
For the case with impurities inside the vertical granular chain under gravity, different characteristics in the backscattered signal are found. These characteristics are dependent on the presence of light and heavy impurities inside  the chain. Physically, that the soliton can be confined in the region with light impurity but can not in the region with heavy impurity is responsible for this difference. This difference can be used, in nondestructive inspection of impurities in granular medium, to discriminate between light and heavy impurities. It may also be used in design of grain layer protector from shocks\cite{Hong-APL2002,Xu-CTP2001C}.

%\include{Introduction}
%\include{Models_and_SimulationTools}
%\include{ComplexFiledAnalysis}

%%%%%%%%%%%%%%%%%%%%%%%%%%%%%%%%%%%%%%
%\include{Observations_MPM}
%\include{Observations_DBM}
%\include{Observations_MD}
%%%\include{Observations_MD2}

\section{Conclusion and discussions}

The macroscopic mechanical properties of a heterogeneous material under shock rely not only on the loading/unloading process, but also on the stable or slowly-varying structures within the material, such as dislocations in plastic deformation, new phase domains in phase transition, cavities and shearing bands in the damaging or fracture process, non-equilibrium vortices and jets behind the shock front, etc. Those structures generally cover a wide range of spatiotemporal scales. To understand and model the macroscopic behaviors of a heterogeneous material, those mesoscopic
structures and their evolution mechanisms, their energies and means of dissipation, must be taken into account.

To access the complex structures and physical fields, various models and simulation schemes are needed. For such typical multi-scale problems, there are mainly two complementary categories of studies. In the first category, the bridging of schemes in neighboring scales, for example, the connection of molecular dynamics and finite element methods, is the main concern. In the second category, the complicated problem is first decomposed into different scales. One constructs/chooses a model according to the scale and main mechanisms working at that scale. Perform numerical simulations using the relatively mature schemes. The physical information is transferred between neighboring scales in such a way: The statistical information of results in smaller scale contributes to establishing the constitutive equation in larger one. Except for the microscopic molecular dynamics model, both the mesoscopic and macroscopic models can be further classified into two categories, solidic and fluidic models, respectively.
A cellular automata model for elastic solid material was attempted in 2013\cite{Dong-CTP2013}.
No matter which model and simulation tool are used, what obtained are large quantity of data. How to analyze the data and pick out reliable structural information is the first issue to resolve. The solution to this issue is the footstone
of further studies.

The dynamical responses of heterogeneous materials are studied from the sides of strength, inertia and dissipation. To access the strength behaviors, the molecular dynamics is used to simulate the evolutions of microscopic structures, such as dislocations, phase domains, microscopic voids under uniform deformation and shock loading; the material point method is used to simulate the mesoscopic behaviors of cavities under shock loading and unloading. To access the inertia and dissipation behaviors, the discrete Boltzmann method is used to simulate the non-equilibrium phase transition kinetic processes, and the material point method is used to simulate porous media under shock.

It is found that, the creation and evolution of microscopic structures in crystal are mainly determined by the properties of local active regimes. For example, dislocations tend to occur in regimes with dense FOS structures; Microscopic voids tend to occur in regimes where fault stacks cross; The FOS structures near a phase domain make easier the phase transition. The morphology of mesoscopic cavities in the collapsing process is determined by the complex interactions between various wave series. For example, when the compression is weak, the cavity may keep its spherical structure; when the compression is strong, the regime around the cavity, where the waves focus,
may show a tip even jet. During the phase transition process, the growth of the new phase domain is determined by the competition of transition energy, surface energy, and thermal energy. For example, during the spinnodal decomposition, more surfaces occur and consequently the non-equilibrium effects resulted from the mass flux and energy flux become more pronounced with time. This observation can be used as a new physical criterion to discriminate the two stages, spinnodal decomposition and domain growth, in the phase transition process. In porous materials under shock, the vortices and jets resulted from all cavities behind the shock front show some kinds of similarities for various cavity size, shock strength, etc. Therefore, the global scenarios of porous materials under shock show some structural and dynamical similarities. For example, the temperature fields occurred in two separate shocking processes with two sets of shocking strengths and porosities may show similarities. The DBM results for phase separation are helpful for understanding the structural phase transition occurred in metal under shock.

After all, physics is an experimental science. The simulation results are to be checked, directly or indirectly, with experiments, and are used to anticipate possible new results. Due to the complexity of the problem, some phenomenological or semi-phenomenological models have to be taken into account when necessary.

Up to now, all the above observations are based on intuitive understanding on relevant structures. How to quantitatively bridge the structures and macroscopic dynamical properties of the material is still a challenging and open problem. This involves two important and complimentary issues, coarse-grained modeling of mesoscopic structures and homogenization of complex heterogeneous systems. The former is relevant to parametrization of mesoscopic structures, energy relations and dissipation mechanisms. The latter is relevant to the averaging and transition between neighboring scales. The two issues are currently demanding much more corporations and efforts of scientists in related fields. The field of heterogeneous materials under shock will see significant development in the following years.

\vspace*{2mm} \Acknowledgements{The authors thank Prof. Hua Li
for fruitful discussions. This work is supported by the Science Foundation of LCP, National Natural Science Foundation of China
[under Grant Nos. 11475028 and  11325209], the opening project of State Key
Laboratory of Explosion Science and Technology (Beijing Institute of
Technology) [under Grant No. KFJJ14-1M] and the Open Project Program of
State Key Laboratory of Theoretical Physics, Institute of Theoretical
Physics, Chinese Academy of Sciences, China [under Grant No. Y4KF151CJ1].}

\end{multicols}


\begin{thebibliography}{999}

\bibitem{Heterogeneous-book1}
 S. Nemat-Nasser and M. Hori. \emph{Micromechanics: overall properties of heterogeneous materials} (Elsevier, Oxford, 1999).

\bibitem{Heterogeneous-book2}
 V. F. Nesterenko, \emph{Dynamics of Heterogeneous Materials}(Springer-Verlag, New York, 2001).

\bibitem{Zhu-Review2010}
J. S. Zhu, X. M. Hu, P. Wang, J. Chen, and A. G. Xu. Adv. Mech, \textbf{40}, 400 (2010).

\bibitem{Review-FoP2012}
A. G. Xu, G. C. Zhang, Y. B. Gan, F. Chen, and Y. J. Li. Front. Phys, \textbf{7}, 582 (2012).

\bibitem{Review-PiP2014}
A. G. Xu, G. C. Zhang, Y. J. Li, and H. Li. Prog. Phys, \textbf{34}, 136 (2014).

\bibitem{Review-APC2015}
A. G. Xu, G. C. Zhang, and Y. J. Ying. Acta. Phys. Sini, \textbf{64}, 184701 (2015).

\bibitem{Succi-DBM2015}
M. L. Rocca, A. Montessori, P. Prestininzi, and S. Succi. J. Comput. Phys, \textbf{284}, 117 (2015).

\bibitem{SoftMatter2015}
Y. B. Gan, A. G. Xu, G. C. Zhang, and S. Succi. Soft Matter, \textbf{11}, 5336 (2015).


\bibitem{MD1} B. J. Alder and T. E. Wainwright. J. Chem. Phys, \textbf{31}, 459 (1959).

\bibitem{MD2} A. Rahman. Phys. Rev, \textbf{136}, A405 (1964).

\bibitem {Daw} M. S. Daw and M. T. Baskes. Phys. Rev. B, \textbf{29}, 6443 (1984).

\bibitem {Harrison} R. Harrison, A. F. Voter, and S. P. Chen. \emph{Atomistic Simulation Of Materials}, edited by V. Vitek and D. J. Srolovitz (New York: Plenum Press, 1989)


\bibitem{CModel} F. Auricchio and L. B. da Veiga. Int. J. Numer. Meth. Engng, \textbf{56}, 1375 (2003).

\bibitem{ZhangX2006} X. Zhang, K. Y. Sze, and S. Ma. Int. J. Numer. Meth. Engng, \textbf{56}, 689 (2006).

\bibitem{MaS2009} S. Ma, X. Zhang, Y. Lian, and X. Zhou. CMES-Computer Modeling In Engineering and Sciences, \textbf{39}, 101 (2009).


\bibitem{ZhangXiong2009} S. Ma, X. Zhang, and X. M. Qiu. Int. J. Impact. Engineering, \textbf{36}, 272 (2009).

\bibitem{MaPhDThesis} S. Ma. Material Point Meshfree Methods for Impact and Explosion Problems (in Chinese). Ph. D thesis. Supervisor: Xiong Zhang. Beijing: Tsinghua University, 2006.

\bibitem{H1964} F. H. Harlow. Methods for Computational Physics, \textbf{3}, 319--343, Adler B, Fernbach S, Rotenberg M (eds). New York: Academic Press, 1964.

\bibitem{MPM1} D. Burgess, D. Sulsky, and J. U. Brackbill. J. Comput. Phys, \textbf{103}, 1 (1992).

\bibitem{MPM2} S. Bardenhagen, J. Brackbill, and D. Sulsky. Comput. Methods. Appl. Mech. Eng, \textbf{87}, 529 (2000).

\bibitem{MPM3} Y. J. Guo and J. A. Nairn. Computer Modeling in Engineering \& Sciences, \textbf{1}, 11 (2006).

\bibitem{MPM4} N. P. Daphalapurkar, H. Lu, D. Coker, and R. Komanduri. Int. J. Fract, \textbf{143}, 79 (2007).

\bibitem{MPM5} S. Ma, X. Zhang, X. M. Qiu. Int. J. Impact. Eng, \textbf{36}, 272 (2009).

\bibitem{JPCM2007} A. G. Xu, X. F. Pan, G. C. Zhang, and J. S. Zhu. J. Phys: Condens Matter, \textbf{19}, 326212 (2007).

\bibitem{CTP2008} X. F. Pan , A. G. Xu, G. C. Zhang, P. Zhang, J. S. Zhu, S. Ma, and X. Zhang. Commun. Theor. Phys, \textbf{49}, 1129 (2008).

\bibitem{JPD2008} X. F. Pan, A. G. Xu, G. C. Zhang, and J. S. Zhu. J. Phys. D: Appl. Phys, \textbf{41}, 015401 (2008).

\bibitem{explosion} B. P. Zhang, Q. M. Zhang, and F. L. Huang. \emph{Explosion physics}(Beijing: Ordance Industry Press of China, 1997).

\bibitem{Succi-Book} S. Succi. \emph{The Lattice Boltzmann Equation for Fluid Dynamics and Beyond} (New York: Oxford University Press, 2001).

\bibitem{XuLin-PRE2014} C. D. Lin, A. G. Xu, G. C. Zhang G, Y. J. Li, and S. Sauro. Phys. Rev. E, \textbf{89}, 013307 (2014).

\bibitem{XuLin-PRE2015} A. G. Xu, C. D. Lin, G. C. Zhang G, and Y. J. Li. Phys. Rev. E,  \textbf{91}, 043306 (2015).

\bibitem{XuLai2015} H. L. Lai, A. G. Xu, G. C. Zhang, Y. B. Gan, and Y. J. Ying.
Non-equilibrium thermo-hydrodynamic effects on the Rayleigh-Taylor instability in compressible flows. arXiv: 1507.01107.

\bibitem{XuZhang2016} Y. D. Zhang. Modeling and research of detonation based on discrete Boltzmann method. Master degree thesis. Supervisors: Chengmin Zhu and Aiguo Xu. Beijing: Beijing University of Aeronautics and Astronautics, 2015.

\bibitem{GLS} G. Gonnella, A. Lamura, and V. Sofonea. Phys. Rev. E, \textbf{76}, 036703 (2007).

\bibitem{Xu-PRE2011} Y. B. Gan, A. G. Xu, G. C. Zhang, Y. J. Li, and H. Li. Phys. Rev. E, \textbf{84}, 046715 (2011).

\bibitem{Xu-EPL2012} Y. Gan, A. G. Xu, G. C. Zhang, P. Zhang, and Y. J. Li. EPL, \textbf{97}, 44002, (2012).

\bibitem{XZG-arXiv2014} A. G. Xu, G. C. Zhang, and Y. B. Gan. Discrete Boltzmann modeling of liquid-vapor system. arXiv:1403.3744.

\bibitem{Onuki-PRL} A. Onuki A. Phys. Rev. Lett, \textbf{94}, 054501 (2005).

\bibitem{CS} N. F. Carnahan and K. E. Starling. J. Chem. Phys, \textbf{51}, 635 (1969).

\bibitem{XuYan-FoP2013} B. Yan, A. G. Xu, G. C. Zhang, Y. J. Ying, and H. Li. Front. Phys, \textbf{8}, 94 (2013).

\bibitem{XuLin-CTP2014} C. D. Lin, A. G. Xu, G. C. Zhang, and Y. J. Li. Commun. Theor. Phys, \textbf{62}, 737 (2014).

\bibitem{XuLin-DDF-DBM}
C. D. Lin, A. G. Xu, G. C. Zhang, and Y. J. Li. Double-distribution-function discrete Boltzmann model for combustion. Combustion and Flame (2015), arXiv: 1405.5500.

\bibitem{Pang-SciRep2014C} W. W. Pang, P. Zhang, G. C. Zhang, A. G. Xu, and X. G. Zhao. Sci. Rep, \textbf{4}, 6981 (2014).

\bibitem{Pang-SciRep2014A} W. W. Pang, P. Zhang, G. C. Zhang, A. G. Xu, and X. G. Zhao. Sci. Rep, \textbf{4},  5273, (2014).

\bibitem{Pang-SciRep2014B} W. W. Pang, P. Zhang, G. C. Zhang, A. G. Xu, and X. G. Zhao. Sci. Rep, \textbf{4}, 3628, (2014).

\bibitem{XGL-PRE2003}
A. G. Xu, G. Gonnella, and A. Lamura. Phys. Rev. E, \textbf{67}, 056105 (2003).

\bibitem{XGL-EPL2005}
A. G. Xu, G. Gonnella, A. Lamura, G. AmatI, and F. Massaioli. Europhys Lett, \textbf{71}, 651 (2005).

\bibitem{XGL-PRE2006}
A. G. Xu, G. Gonnella, and A. Lamura. Phys. Rev. E, \textbf{74}, 011505 (2006).

\bibitem{Xu-JPD2009} A. G. Xu, G. C. Zhang, X. F. Pan, P. Zhang, and J. S. Zhu. J. Phys. D: Appl. Phys. \textbf{42}, 075409 (2009).

\bibitem{Xu-CTP2009B} A. G. Xu, G. C. Zhang, P. Zhang, X. F. Pan, and J. S. Zhu. Commun. Theor. Phys, \textbf{52}, 901 (2009).

\bibitem{Xu-CTP2009A} A. G. Xu, G. C. Zhang, X. F. Pan, and J. S. Zhu. Commun. Theor. Phys, \textbf{51}, 691 (2009).

\bibitem{MD-SHT-Book} G. C. Zhang, A. G. Xu, and G. Lu. \emph{General Index and Its Application in MD Simulations}. in MOLECULAR INTERACTIONS edited by Aurelia Meghea.Published by InTech, Janeza Trdine 9, 51000 Rijeka, Croatia.

\bibitem{Zhang-SciCN2010} G. C. Zhang, A. G. Xu, G. Lu, and Z. Y. Mo, Sci. China. Phys. Mech. Astron, \textbf{53}, 1610 (2010).




%%%%%%%%%%%%%% Minkowsi

\bibitem{Serra1982}
J. Serra, \emph{Image Analysis and Mathematical Morphology}. (New York: Academic, Vols. 1 and 2, 1982).

\bibitem{PRE1996} K. R. Mecke. Phys. Rev. E, \textbf{53}, 4794 (1996).

\bibitem{JChemP2000} A. Aksimentiev, K. Moorthi, R. Holyst. J. Chem. Phys, \textbf{112}, 1 (2000).

\bibitem{Sofonea} K. R. Mecke and V. Sofonea. Phys. Rev. E, \textbf{56}, R3761 (1997).

\bibitem{PR}  W. T. G\'{o}\'{z}d\'{z} and R. Holyst. Phys. Rev. Lett, \textbf{76}, 2726 (1996).

%%%%%%%%%%%%%% End of Minkowsi

\bibitem{Kelchner}
C. L. Kelchner, S. J. Plimpton, J. C. Hamilton. Phys. Rev. B, \textbf{58}, 11085 (1998).

\bibitem{Faken} D. Faken and H. Jonsson. Computational Materials Science, \textbf{2}, 279 (1994).

\bibitem{Mencl} E. Mencl and H. Muller. Scientific Visualization Conference, Dagstuhl, Germany, June 1997

\bibitem{Boissonant} J. D. Boissonant.ACM Transactions on Graphics, \textbf{3}, 266 (1984).


\bibitem{Hoppe} H. Hoppe, T. DeRose, T. Duchanp, J. Mcdonald, and W Stuetzle. ACM Computer Graphics, \textbf{26}, 71 (1992).

\bibitem{Zhao}
H. K. Zhao, S. Osher, B. Merriman, and M. Kang. Computer Vision and Image Processing, \textbf{80}, 295 (2002).

\bibitem{Bernardini}
F. Bernardini, J. Mittlelman, H. Rushmeir, C. Silva, and G. Taubin. IEEE. Trans. Visual. Comput. Graphics, \textbf{5}, 349 (1999).


\bibitem{WangZhang2013} S. Wang, G. Lu, and G. Zhang. Computational Materials Science \textbf{68}, 396 (2013).

%%%%%%%%%%%% MPM simulations

\bibitem{Xu-SciCN2010} A. G. Xu, G. C. Zhang, H. Li, Y. J. Ying, X. J. Yu, and J. S. Zhu.  Sci. China. Phys. Mech. Astron, \textbf{53}, 1466 (2010).

\bibitem{Xu-CAMWA2011} A. G. Xu, G. C. Zhang, Li H, Y. J. Ying, and J. S. Zhu. Comput. Math. Appl, \textbf{61}, 3618 (2011).


\bibitem{Xu-PhysScr2010} A. G. Xu, G. C. Zhang, Y. J. Ying, P. Zhang, and J. S. Zhu. Phys. Scr, \textbf{81}, 055805 (2010).


%%%%%%%%%%%%%
\bibitem{42} M. M. Carroll, A. C. Holt. J. Appl. Phys, \textbf{27}, 1626 (1972).

\bibitem{43} J. N. Johnson.Journal of Applied Physics, \textbf{52}, 2812 (1981).

\bibitem{Becker1987} R. Becker. Journal of the Mechanics \& Physics of Solids, \textbf{35}, 577 (1987).

\bibitem{44} M. Ortiz M and A. Molinari. Journal of Applied Mechanics, \textbf{59}, 48 (1992).

\bibitem{45} D. J. Benson. \emph{The Numerical Simulation of the Dynamic Compaction of Powders}.Chapter:High-Pressure Shock Compression of Solids, IV.
Part of the series High-Pressure Shock Compression of Condensed Matter, pp 233--255

\bibitem{46} X. Y. Wu, K. T. Ramesh, T. W. Wright. Journal of Mechanics Physics of Solids, \textbf{51}, 1 (2003).

\bibitem{Pardoen1998} T. Pardoen, I. Doghri, F. Delannay. Acta. Materialia, \textbf{46}, 541 (1998).

\bibitem{Pardoen2000} T. Pardoen and J. W. Hutchinson. J. Mech. Phys. Solids, \textbf{48}, 2467 (2000).

\bibitem{Orsini2001} V. C. Orsini and M. A. Zikry. Int. J. Plast, \textbf{17}, 1393 (2001).

\bibitem{2002a} V. Tvergaard and J. W. Hutchinson. International Journal of Solids \& Structures, \textbf{39}, 3581 (2002).
.
\bibitem{2002b} T. I. Zohdi, M. Kachanov, I. Sevostianov. International Journal of Plasticity, \textbf{18}, 1649 (2002).


\bibitem{27} D. R. Curran, L. Seaman, D. A. Shockey. Phys. Rep, \textbf{147}, 253 (1987).

\bibitem{47} E. T. Seppala, J. Belak. Phys. Rev. Lett, \textbf{93}, 245503 (2004).

\bibitem{48} A. K. Zurek, W. R. Thissell, J. N. Johnson, D. L. Tonks, and R Hixson. Journal of Materials Processing Technology, \textbf{60}, 261 (1996).

\bibitem{Zurek1998} A. K. Zurek, J. D. Embury, A. Kelly, W. R. Thissell, R. L. Gustavsen1, J. E. Vorthman and R. S. Hixson. AIP Conf. Proc. \textbf{429}, 423 (1998). doi:http://dx.doi.org/10.1063/1.55658 (4 pages)

\bibitem{49} D. L. Tonks, A. K. Zurek, and W. R. Thissell. AIP Conf. Proc, \textbf{620}, 611 (2002). doi:http://dx.doi.org/10.1063/1.1483613 (4 pages)

\bibitem{50} J. P. Bandstra, D. M. Goto, and D. A. Koss. Materials Science and Engineering: A, \textbf{249}, 46 (1998).

\bibitem{51} J. P. Bandstra and D. A. Koss. Materials Science and Engineering: A, \textbf{319}, 490 (2001).

\bibitem{52} D. A. Koss, A. Geltmacher, P. Matic, and R. K. Everett. Materials Science and Engineering: A, \textbf{366}, 269 (2004).

\bibitem{53} M. F. Horstemeyer, M. M. Matalanis, A. M. Sieber, and M. L. Botos. Int. J. Plasticity, \textbf{16}, 979 (2000).

\bibitem{Xu-FoP2013} A. G. Xu, G. C. Zhang, Y. J. Ying, and X. J. Yu. Front. Phys, \textbf{8}, 394 (2013).

\bibitem{GXZ-EPL2012} Y. B. Gan, A. G. Xu, G. C. Zhang, and Y. J. Li. EPL, \textbf{97}, 44002 (2012).


%%%%%%%%%%%%%%%%%%%%%%%%%%%%%%%%%%%%%%%%%%%%%%%%%%%%%%%%%%%%%%%%%%%%%%%%%%%%%%%
%%%%%%% 2015-9-29
%%%%%%%%%%%%%%%%%%%%%%%%%%%%%%%%%%%%%%%%%%%%%%%%%%%%%%%%%%%%%%%%%%%%%%%%%%%%%%%
%%%%%%%%%%%%%%%%%%% MD

\bibitem{Lu-ActaPhysSin2012} G. Lu, S. C. Wang, G. C. Zhang, and A. G. Xu. Acta. Phys. Sin, \textbf{61}, 073102 (2012).

\bibitem{Pang-CJCP2011} W. Pang, G. C. Zhang, A. G. Xu, and G. Lu. Chin. J. Comp. Phys, \textbf{28}, 540 (2011).


\bibitem{Pang-SciCN2012} W. W. Pang, P. Zhang, G. C. Zhang, and X. G. Zhao. Sci. Sin-Phys. Mech. Astron, \textbf{42}, 464 (2012).

\bibitem{Yang-Acta2008} Q. L. Yang, G. C. Zhang, A. G. Xu, Y. H. Zhao, and Y. J. Li. Acta. Physica Sinica, \textbf{57}, 940 (2008).

\bibitem{Pang-AMR2013} W. W. Pang, G. C. Zhang, A. G. Xu, P. Zhang. Adv. Mater. Research, \textbf{790}, 65 (2013).


\bibitem {Liu}J. B. Liu, D. D. Johnson. Phys. Rev. B, \textbf{79}, 134113 (2009).
%%%%%%%%%%%%%%%%%%% Xu: Grain


\bibitem{Hong-PRE2001} J. Hong, A. G. Xu. Phys. Rev. E, \textbf{63}, 061310 (2001).

\bibitem{Xu-CTP2001B} A. G. Xu, J. Hong. Commun. Theor. Phys, \textbf{36}, 1990 (2001).

\bibitem{Hong-APL2002} J. Hong, A. G. Xu. Appl. Phys. Lett, \textbf{81}, 4868 (2002).

\bibitem{Xu-CTP2001C} A. G. Xu, J. Hong. Commun. Theor. Phys, \textbf{36}, 699 (2001).


%%%%%%%%%%%%%%%%%%%%%%%%%%%%%%%%%%%%%%%%%%%%%%%%%%%%%%%%%%%%%%%%%%%%

\bibitem{Dong-CTP2013} Y. F. Dong, G. C. Zhang, A. G. Xu, Y. B. Gan. Commun. Theor. Phys, \textbf{1}, 59 (2013).



\end{thebibliography}
\end{document}